\documentclass[fleqn,usenatbib]{mnras}
\pdfoutput=1 
\PassOptionsToPackage{usenames,dvipsnames}{xcolor}
\usepackage[T1]{fontenc}
\usepackage{ae,aecompl}
\usepackage[utf8]{inputenc}
\usepackage{newtxtext,newtxmath}
\usepackage[english]{babel}
\usepackage{amsmath,amstext}
\usepackage[figure,figure*]{hypcap}
\usepackage[hang]{footmisc}
\usepackage{threeparttablex}
\usepackage{longtable}
\usepackage{inconsolata}
\usepackage{graphicx}
\usepackage{overpic}
\usepackage{multicol}
\usepackage[frozencache,cachedir=.]{minted}
\usepackage{float}

\usepackage{newtxtext,newtxmath}
\usepackage[T1]{fontenc}

\DeclareRobustCommand{\VAN}[3]{#2}
\let\VANthebibliography\thebibliography
\def\thebibliography{\DeclareRobustCommand{\VAN}[3]{##3}\VANthebibliography}

\newcommand*{\https}[1]{\href{https://#1}{\nolinkurl{#1}}}
\newcommand*{\http}[1]{\href{http://#1}{\nolinkurl{#1}}}

\DeclareUrlCommand\code{\urlstyle{tt}}

\title[OpenUniverse2024]{OpenUniverse2024: A shared, simulated view of the sky for the next generation of cosmological surveys}

\author[OpenUniverse, LSST DESC, and the Roman HLIS Cosmology, RAPID, \& Supernova Cosmology PITs]{
\parbox{\textwidth}{
{\large 
OpenUniverse,
The LSST Dark Energy Science Collaboration,
The Roman HLIS Project Infrastructure Team,
The Roman RAPID Project Infrastructure Team, 
The Roman Supernova Cosmology Project Infrastructure Team,}
{\large
A.~Alarcon,$^{1}$
L.~Aldoroty,$^{2}$
G.~Beltz-Mohrmann,$^{3}$
A.~Bera,$^{4}$
J.~Blazek,$^{5}$
J.~Bogart,$^{6}$
G.~Braeunlich,$^{7}$
A.~Broughton,$^{6}$
K.~Cao,$^{8}$
J.~Chiang,$^{6}$
N.~E.~Chisari,$^{9}$
V.~Desai,$^{10}$
Y.~Fang,$^{2}$
L.~Galbany,$^{1,11}$
A.~Hearin,$^{3}$
K.~Heitmann,$^{3}$
C.~Hirata,$^{8}$
R.~Hounsell,$^{12,13}$
B.~Jain,$^{14}$
M.~Jarvis,$^{14}$
J.~Jencson,$^{10}$
A.~Kannawadi,$^{2}$
M.~K.~Kasliwal,$^{15}$
R.~Kessler,$^{16,17}$
A.~Kiessling,$^{18}$
R.~Knop,$^{19}$
E.~Kovacs,$^{3}$
R.~Laher,$^{10}$
K.~Laliotis,$^{8}$
C.~Lin,$^{2}$
I.~Lopes,$^{20}$
E.~Macbeth,$^{8}$
A.~Mahabal,$^{15}$
R.~Mandelbaum,$^{21}$
J.~Masiero,$^{10}$
S.~Mau,$^{22}$
C.~Meehan,$^{2}$
J.~Meyers,$^{6}$
B.~Moraes,$^{20}$
R.~Paladini,$^{10}$
A.~Pearl,$^{23}$
A.~Plazas Malagon,$^{6}$
B.~Rose,$^{24}$
D.~Rubin,$^{25}$
B.~Rusholme,$^{10}$
A.~Santos,$^{26}$
N.~Šarčević,$^{27}$
D.~Scolnic,$^{2}$
J.~Singhal,$^{10}$
M.~A.~Troxel,$^{2}$\thanks{E-mail: michael.troxel@duke.edu}
N.~Van Alfen,$^{5}$
S.~Van Dyke,$^{10}$
C.~W.~Walter,$^{2}$
T.~Wu,$^{2}$
M.~Yamamoto,$^{2}$
L.~Yan,$^{28}$
and T.~Zhang$^{23}$
}
}
\\
\parbox{\textwidth}{
\begin{multicols}{2}
$^{1}$ Institute of Space Sciences (ICE, CSIC), Campus UAB, Carrer de Can Magrans, s/n, 08193 Barcelona, Spain\\
$^{2}$ Duke University, Durham, NC 27708, USA\\
$^{3}$ Argonne National Laboratory, 9700 S Cass Ave, Lemont, IL 60439, USA\\
$^{4}$ University of Texas, Dallas, 800 W Campbell Rd, Richardson, TX 75080, USA\\
$^{5}$ Northeastern University, 360 Huntington Ave., Boston, MA  02115, USA\\
$^{6}$ SLAC National Accelerator Laboratory, 2575 Sand Hill Road, Menlo Park, CA, 94025, USA\\
$^{7}$ ETH Zurich, Wolfgang-Pauli-Strasse 16, CH-8093 Zurich, Switzerland\\
$^{8}$ Ohio State University, Columbus, OH 43210, USA\\
$^{9}$ Institute for Theoretical Physics, Utrecht University, Princetonplein 5, 3584 CC, The Netherlands\\
$^{10}$ IPAC, California Institute of Technology, Pasadena, CA 91125, USA\\
$^{11}$ Institut d’Estudis Espacials de Catalunya (IEEC), 08860 Castelldefels (Barcelona), Spain\\
$^{12}$ University of Maryland, Baltimore County, Baltimore, MD 21250, USA\\
$^{13}$ NASA Goddard Space Flight Center, Greenbelt, MD 20771, USA\\
$^{14}$ University of Pennsylvania, Philadelphia, PA 19104, USA\\
$^{15}$ Division of Physics, Mathematics, and Astronomy, California Institute of Technology, Pasadena, CA 91125, USA\\
$^{16}$ Kavli Institute for Cosmological Physics, University of Chicago, Chicago, IL 60637, USA\\
$^{17}$ Department of Astronomy and Astrophysics, University of Chicago, Chicago, IL 60637, USA\\
$^{18}$ NASA Jet Propulsion Laboratory, 4800 Oak Grove Dr, Pasadena, CA 91109, USA\\
$^{19}$ Lawrence Berkeley National Laboratory, 1 Cyclotron Rd, Berkeley, CA 94551, USA\\
$^{20}$ Universidade Federal do Rio de Janeiro, Rio de Janeiro, Brazil\\
$^{21}$ Carnegie Mellon University, 5000 Forbes Ave, Pittsburgh, PA 15213, USA\\
$^{22}$ Stanford University, 450 Serra Mall, Stanford, CA 94305, USA\\
$^{23}$ University of Pittsburgh, 4200 Fifth Ave, Pittsburgh, PA 15260, USA\\
$^{24}$ Department of Physics and Astronomy, Baylor University, One Bear Place \#97316, Waco, TX 76798-7316, USA\\
$^{25}$ Department of Physics and Astronomy, University of Hawai`i at M{\=a}noa, Honolulu, Hawai`i 96822, USA\\
$^{26}$ Centro Brasileiro de Pesquisas Físicas, Rua Dr. Xavier Sigaud 150, RJ 22290-180, Rio de Janeiro, Brazil\\
$^{27}$ Newcastle University, Newcastle upon Tyne, County Durham, NE1 7RU, United Kingdom\\
$^{28}$ Caltech Optical Observatories, California Institute of Technology, Pasadena, CA 91125, USA
\end{multicols}
\vspace{-0.5cm}
}
}

\pubyear{2025}

\begin{document}
\label{firstpage}
\pagerange{\pageref{firstpage}--\pageref{lastpage}}
\maketitle

\begin{abstract}
The OpenUniverse2024 simulation suite is a cross-collaboration effort to produce matched simulated imaging for multiple surveys as they would observe a common simulated sky. Both the simulated data and associated tools used to produce it are intended to uniquely enable a wide range of studies to maximize the science potential of the next generation of cosmological surveys. We have produced simulated imaging for approximately 70 deg$^2$ of the Vera C. Rubin Observatory Legacy Survey of Space and Time (LSST) Wide-Fast-Deep survey and the Nancy Grace Roman Space Telescope High-Latitude Wide-Area Survey, as well as overlapping versions of the ELAIS-S1 Deep-Drilling Field for LSST and the High-Latitude Time-Domain Survey for Roman. OpenUniverse2024 includes i) an early version of the updated extragalactic model called {\tt Diffsky}, which substantially improves the realism of optical and infrared photometry of objects, compared to previous versions of these models; ii) updated transient models that extend through the wavelength range probed by Roman and Rubin; and iii) improved survey, telescope, and instrument realism based on up-to-date survey plans and known properties of the instruments. It is built on a new and updated suite of simulation tools that improves the ease of consistently simulating multiple observatories viewing the same sky. The approximately 400\,TB of synthetic survey imaging and simulated universe catalogs are publicly available, and we preview some scientific uses of the simulations. 
\end{abstract}

\begin{keywords}
transients: supernovae -- large-scale structure of Universe -- software: simulations
\vspace{-.9cm}
\end{keywords}


\section{Introduction}

With the next generation of wide-area ground- and space-based photometric surveys like the Vera C.\ Rubin Observatory Legacy Survey of Space and Time (LSST\footnote{\url{https://rubinobservatory.org}}, \citealt{2019ApJ...873..111I}), the Euclid mission\footnote{\url{https://www.cosmos.esa.int/web/euclid}} \citep{2011arXiv1110.3193L,2022A&A...662A.112E}, and the Nancy Grace Roman Space Telescope\footnote{\url{https://roman.gsfc.nasa.gov}} taking data over the next decade, we are poised to fundamentally change the way we study the Universe. With thousands of square degrees of overlapping wide-area imaging from ground and space, coupled with tens of square degrees of overlapping ultra-deep fields, there is an unprecedented need for expanding the limits of our computational techniques, analysis methods, and fundamental understanding of the Universe. Work preparing for this data is ongoing in many collaborations, and there is a critical need for realistic, large simulated data sets with which to explore joint calibration, analysis, and other scientific and technical possibilities that emerge with the combination of these new data sets \citep[e.g.,][]{2022zndo...5836022G}. With the new simulation campaign described in this paper, we provide a snapshot of that future by simulating hundreds of terabytes of overlapping images from LSST and the Roman Space Telescope. 

\begin{figure*}
\includegraphics[width=\columnwidth]{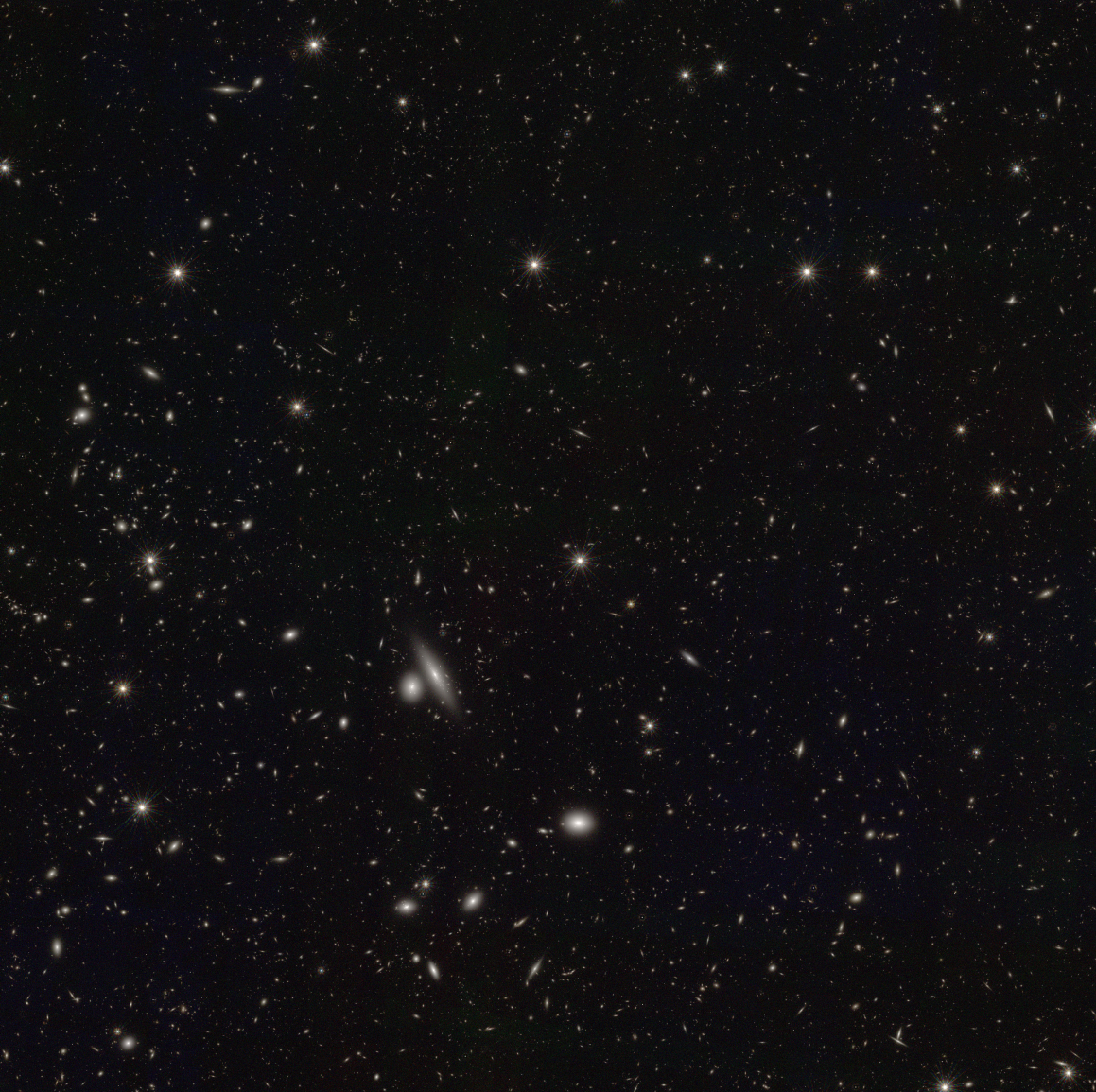}
\includegraphics[width=\columnwidth]{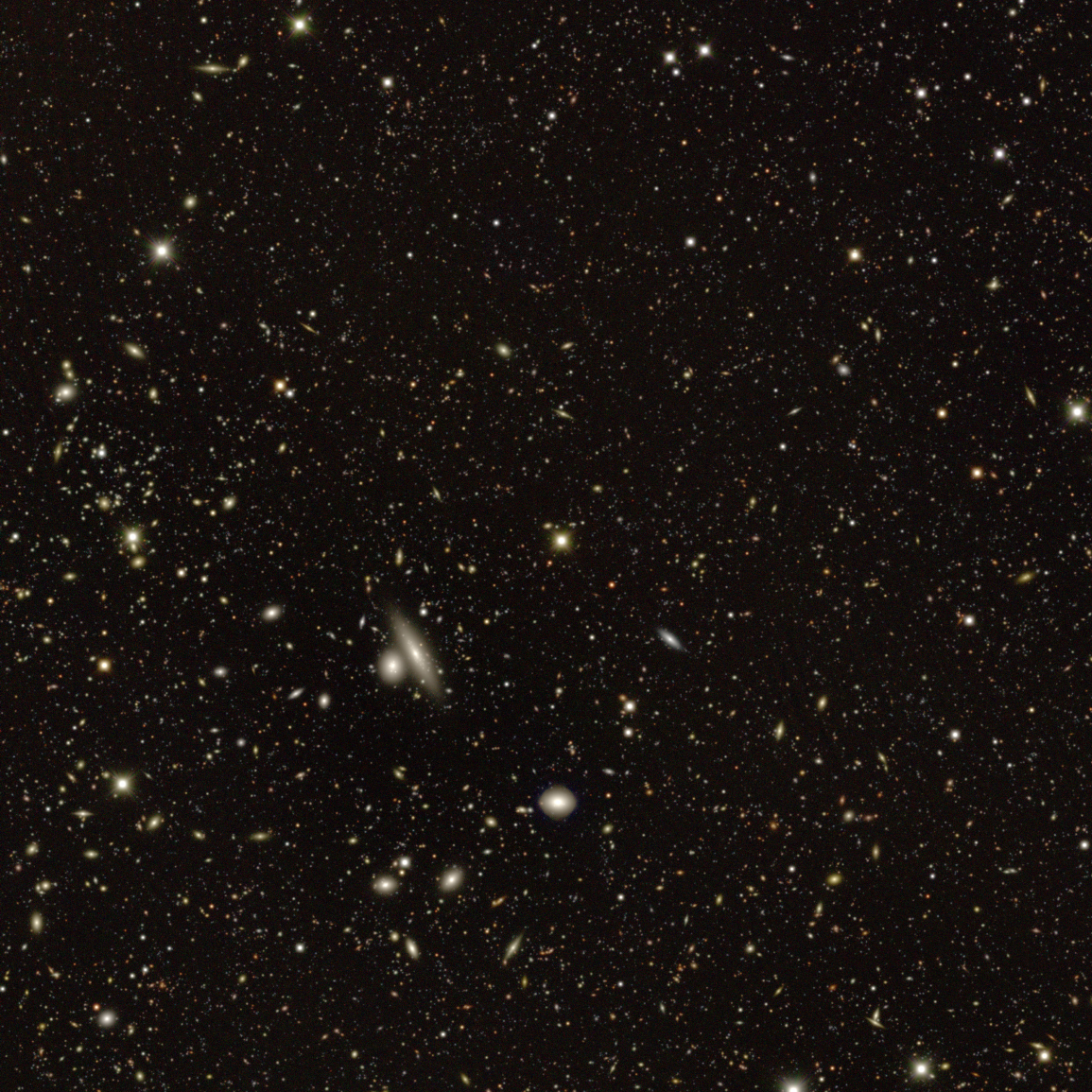}
\caption{\label{fig:coaddimages}A comparison of color coadd images from the Roman HLWAS (left; Y106/J129/H158 color composite) and LSST WFD (right; g/r/i color composite) overlapping a single LSST patch of the synthetic sky with central coordinates $(\textrm{RA}, \textrm{Dec})=(9.55^{\circ}, 44.1^{\circ})$, $\Delta\textrm{RA}=0.325^{\circ}$, and $\Delta\textrm{Dec}=0.233^{\circ}$. Both coadds are full simulated-survey depth -- the full WAS-depth for Roman and the first five years of the simulated observing sequence for LSST WFD. The LSST coadd  matches the native pixel scale of 0.2 arcsec, while the Roman \textsc{imcom} coadd is significantly oversampled from the native Roman pixel scale of 0.11 arcsec to a coadd pixel scale of 0.039 arcsec, in order to achieve a Nyquist sampled coadd image. Note that the images are substantially degraded in quality for rendition in this document.}
\end{figure*}

In particular, ground- and space-based imaging from these surveys provides highly complementary and synergistic information that can be used together to substantially mitigate systematics and further improve our knowledge of the Universe (e.~g.,~\citealt{2017ApJS..233...21R}). LSST will provide exquisite optical photometry from the ground that will strongly benefit photo-z inference for Euclid and Roman, while Euclid and Roman will provide highly resolved space-based imaging to help characterize or deblend LSST images \citep[e.g,][]{2021arXiv210706984J,2021NatRP...3..712M} that are degraded by an atmospheric point-spread function (PSF). Roman will also provide near-infrared measurements of galaxy shapes for weak gravitational lensing to complement optical measurements from LSST and Euclid. Both Euclid and Roman will have an overlapping grism (and prism for Roman) spectroscopic survey in some fields. 

These new simulations builds on previous work to jointly simulate the same sky viewed by multiple telescopes, as described in \cite{troxel23}. The current simulation takes advantage of new tools that make these joint simulations easier to achieve, as well as improved galaxy models and a range of updates to the telescope and survey models that make these simulations more realistic. These updates overcome some significant limitations in realism in previous simulations, in particular the fidelity of galaxy colors and other properties. 

The simulations described here provides realistic imaging data for the LSST ELAIS Deep Drilling Field (DDF) and an approx.~70 deg$^2$ overlapping region of the LSST Wide-Fast-Deep (WFD) survey. It also provides imaging for a potential complete Roman Reference High-Latitude Time-Domain Survey (HLTDS) and approx.~70 deg$^2$ of the Reference High-Latitude Wide-Area Survey (HLWAS), each also overlapping the ELAIS field. Figure \ref{fig:coaddimages} compares color coadded images for the LSST WFD and Roman HLWAS survey imaging.

These simulations were produced as part of a collaboration between NASA's OpenUniverse team, the LSST Dark Energy Science Collaboration (DESC), the Roman High-Latitude Imaging Survey (HLIS) and Supernova Project Infrastructure Teams (PITs), and others in the Roman community. They were made possible through a special Argonne Leadership Computing Facility (ALCF) Discretionary award using the entire Theta supercomputer\footnote{\url{https://www.alcf.anl.gov/alcf-resources/theta}} shortly before the system was retired at the start of 2024. To maximize the scientific outcomes from these simulated datasets, all of the products (both the resulting simulations and the code packages to produce them) are being made publicly available with this paper for exploration by the scientific community beyond these collaborations.

In this paper, we document details of the simulations and data products, summarize validation of the simulations, and provide other information necessary to utilize the simulated data. We summarize the OpenUniverse2024 framework in Sec.~\ref{sec:introsummary}. More detailed information on the mock universe is described in Sec.~\ref{sec:features}, the simulated surveys in Sec.~\ref{sec:surveys}, and the simulated imaging in Sec.~\ref{sec:imagesims}. The available data products are described in Sec.~\ref{sec:products}. Section \ref{sec:science} describes some preliminary science applications, and we conclude with a summary and outlook in Sec.~\ref{sec:outlook}.

\section{OpenUniverse2024 simulation framework}\label{sec:introsummary}

The OpenUniverse2024 simulation suite is weaved together from 1) several stages of cosmological, galactic, and extragalactic simulations, 2) current proposals for the Rubin LSST and Roman survey strategies, 3) simulations of the Rubin and Roman observatory, optics, and sensors, and finally 4) processed through existing calibration and measurement pipelines for each mission. This effort bridged expertise across the time-domain and static science cases of the Rubin Observatory and Roman Space Telescope, benefited from resources and personnel spanning the US Department of Energy (DOE) and National Aeronautics and Space Administration (NASA) labs and research communities, and combined effort from NASA's OpenUniverse team, the LSST DESC, the Roman High-Latitude Imaging Survey (HLIS) and Supernova Project Infrastructure Teams (PITs), and others in the Roman community. Taking advantage of a short-fuse opportunity provided by the ALCF in 2023, the teams had about six months to put together a scientific plan and goals, and to complete necessary development and validation of a large range of updates to these simulation components. These updates and simulation components are described in much more detail in the following sections, but we provide a general summary of the ingredients to the simulation and its science goals here.

The overall science goals of the simulations were to 1) improve some known deficiencies in the previous wide-area LSST DESC DESC Data Challenge 2 (DC2)-based \citep{korytov2019,dc2a,dc2b} joint Roman--Rubin simulations \citep{troxel23}, such as the realism of infrared and optical--infrared colors of galaxies, 2) expand the utility of these joint survey simulations to the time-domain community by simulating full time-domain fields in both surveys, and 3) provide an updated image simulation to account for improvements in a) optical and sensor modelling, b) realism of the simulated surveys, particularly the change since LSST DESC DC2 to a ``rolling'' cadence for LSST and c) to include more options in bandpass coverage for Roman imaging to support the Core Community Survey selection process in 2024.

\begin{figure*}
\includegraphics[width=\textwidth]{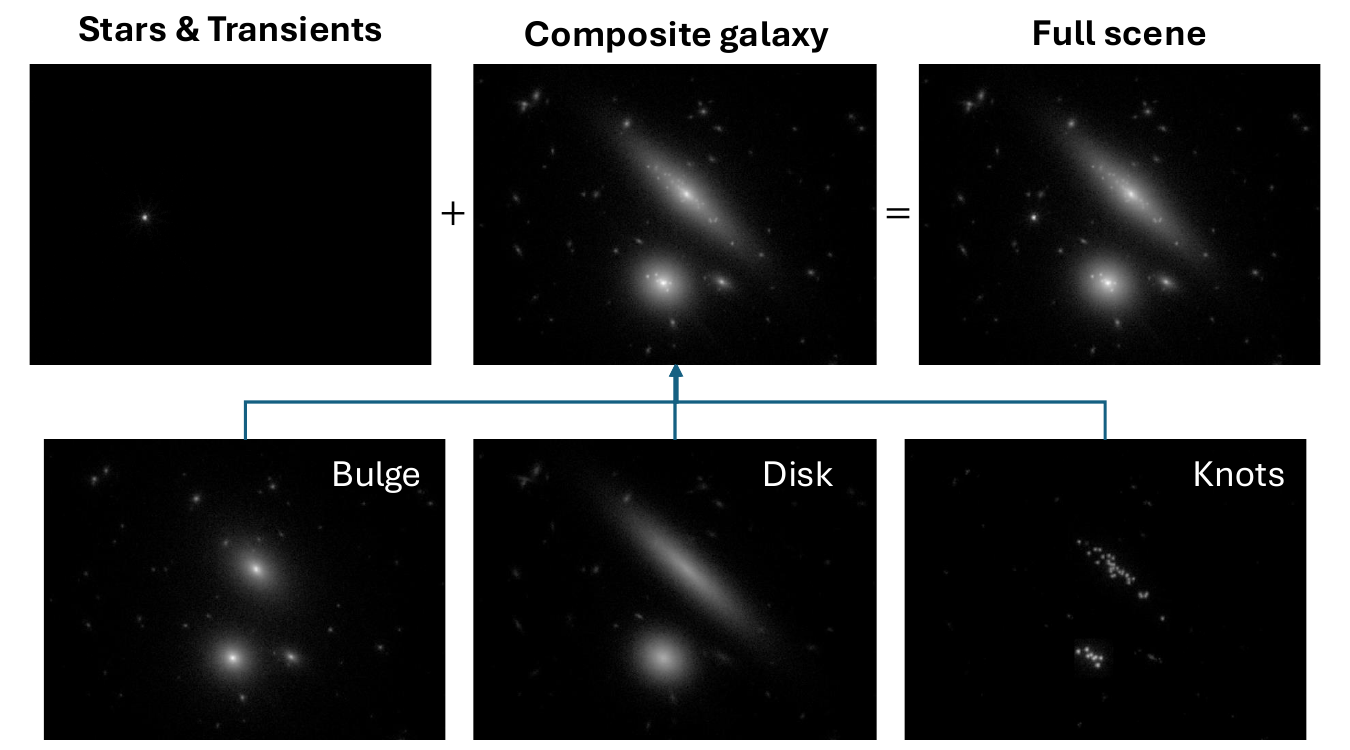}
\caption{\label{fig:galmodel}A demonstration of how the image scene is built. Each panel is approx.~50 arcsec across. Each unsaturated star, transient, and galaxy is built in the image scene photon-by-photon. Galaxies are built from a composite morphological model consisting of a bulge, disk, and star-forming knot regions. The SED and flux fraction of each component of the galaxy is self-consistently generated in the {\tt Diffsky} model to represent the appropriate populations of stars present in each component of the galaxy model.}
\end{figure*}

While the end product of OpenUniverse2024 is two sets of synthetic imaging for wide-area and time-domain surveys for both Roman and Rubin LSST, including over 300TB of 4 million individual images, the process to get to that simulated imaging data is quite complex. We begin with a cosmological gravity-only N-body simulation called Outer Rim \citep{heitmann2019_outerrim} that is evolved from initial conditions informed by the Cosmic Microwave Background, and from which dark matter halos are identified. Galaxies are assigned to with halos and subhalos, and a complex model called {\tt Diffsky} is applied that converges to match a wide range of existing observations of galaxies across redshift, magnitude, and other observable properties. This process is described in Sec.~\ref{sec:diffsky}. 

Ten classes of transient phenomena, including supernovae Type Ia, Ib, Ic, \& II, tidal disruption events, pair instability supernovae, and kilonovae, are modelled based on observations of similar transients and placed within the galaxies of OpenUniverse2024 according to their expected frequency in each galactic environment. These models and the assignment to galaxies are described in Sec.~\ref{sec:transients}. 

All objects are simulated chromatically with realistic SEDs.\footnote{Objects that would be drawn in an image with fewer than 40 photons, and thus just contribute to correlated background noise in the image, are instead drawn with a flat SED model for computational efficiency.} For the galaxies, this is a composite SED built from bulge, disk, and star-forming regions of the galaxies with their own self-consistent star formation history. The paths of the light from objects are simulated via ray tracing through a moderate-resolution map of the N-body mass distribution of the N-body mass distribution to infer properties for each object like gravitational lensing, which both distorts the shapes of extended objects like galaxies and magnifies the observed flux of all objects. We also include models of intrinsic dust for each galaxy and the Milky Way that modulates the observed colors of objects. The Milky Way itself is also modelled \citep{2008ApJ...673..864J}, and we simulate the distribution of stars we would observe in this region of the sky. 

Once we have defined all of these classes of objects and their properties to represent the contents of our synthetic universe, we use detailed simulations from each mission for the observing sequences for each telescope and survey to define what images we must simulate. These simulated surveys are described in Sec.~\ref{sec:surveys}. 

The synthetic images themselves are produced using a complex simulation of each observatory, its optics system, and properties of the sensors and read-out electronics, which is described in Sec.~\ref{sec:imagesims}. We build each scene of the sky from our mock universe photon-by-photon, a process summarized in Fig.~\ref{fig:galmodel}. The images are saved and later processed through available science pipelines for each survey, which results in final processed coadd images like shown in Fig.~\ref{fig:coaddimages}.

\section{Astrophysical \& Cosmological Features of the simulations}
\label{sec:features}

The new simulation suite makes use of some existing resources from the LSST DC2 simulations \citep{korytov2019,dc2a,dc2b,troxel23}, while supplementing them with updates to mitigate known limitations of the realism of some aspects of the DC2 simulations and improve the realism of the modelling of the telescopes and surveys. In particular, we utilize the same N-body simulation Outer Rim \citep{heitmann2019_outerrim}, Milky Way simulation Galfast \citep{2008ApJ...673..864J}, and renormalized Milky Way dust models \citep{2005AJ....130..659A,1998ApJ...500..525S} used in the previous set of LSST DESC DC2 simulations, which can be referred to for more details.
The major updates to the astrophysical inputs to the simulations include the modelling of galaxies and a variety of transient objects.

In total, the region of the simulated universe used for the simulated Rubin and Roman surveys contains around 1.3 million Milky Way stars, 117 million galaxies, and 1.4 million transient objects.

\subsection{Extragalactic catalog}\label{sec:diffsky}

The production of the extragalactic catalog uses the GalSampler technique \citep{hearin2020_galsampler} to populate host halos in the Outer Rim simulation with central and satellite galaxies. First, a high-resolution N-body simulation with merger trees (the {\em source simulation}) is populated with a model of the galaxy--halo connection; second, the synthetic galaxy population is transferred to a host halo catalog in a larger-volume simulation (the {\em target simulation}) via a halo-to-halo correspondence; finally, galaxies in the target simulation are supplemented with additional properties and modifications according to the specifications set by the downstream scientific applications targeted by the synthetic OpenUniverse2024 surveys.

The same GalSampler-based methodology was used to create the cosmoDC2 extragalactic catalog \citep{korytov2019}; here we have used newly developed models in the third step to bestow additional properties onto the galaxy population. Most of these new models are prototypes from ongoing work on {\tt Diffsky}, a differentiable and probabilistic model of the galaxy--halo connection.\footnote{\url{https://diffsky.readthedocs.io/en/latest/}} In this section, we outline the key functional forms used to compute SEDs and photometry in the extragalactic catalog. Parameter values and implementation details can be found in the publicly available source code \href{https://lsstdesc-diffsky.readthedocs.io/en/latest/}{lsstdesc-diffsky}, but we defer a more comprehensive presentation of the final {\tt Diffsky} model to a future publication. 

\subsubsection{Galaxies in the source simulation}
As in \cite{korytov2019}, the starting point of our catalog is the Small MultiDark Planck(SMDPL) simulation \citep{SMDPL2} populated with galaxies based on the UniverseMachine model \citep{behroozi2019UM}. SMDPL is a high-resolution 
N-body box with $L_{\rm box}=400 {\rm Mpc}/h$ and particle mass $m_{\rm p}=9.63\times 10^7M_{\odot}/h$ in which halos and subhalos have been identified with Rockstar \citep{behroozi_etal13_rockstar}, and merger trees have been identified with Consistent Trees \citep{behroozi_etal13_consistent_trees}. The UniverseMachine model paints a star formation history (SFH) onto each (sub)halo in the box in a manner that results in broad agreement between the mock catalog and a wide range of observational measurements such as stellar mass functions, cosmic star formation rates, and two-point clustering.

Before transferring synthetic galaxies from source to target simulation, we replace each UniverseMachine SFH with a smooth approximation based on the Diffstar model \citep{alarcon2023_diffstar}. Diffstar models SFH by parametrizing basic features of galaxy formation physics such as star formation efficiency, a gas consumption timescale, quenching, and rejuvenation. In replacing the UniverseMachine SFHs with Diffstar, we first fit the mass assembly history of the main progenitor (MAH) of each simulated merger tree with the Diffmah model  \citep{hearin2021_diffmah}, and then find best-fitting Diffstar parameters of each simulated SFH. In developing the {\tt Diffsky} modeling framework, we have found that using parametric approximations to the MAHs helps ensure well-behaved SFHs even for galaxies residing in halos that are only well-resolved for a handful of snapshots; moreover, replacing the full MAH with an analytic approximation reduces the memory footprint of the computation by $\sim5$x, significantly improving the capability to leverage GPU compute resources.

A significant fraction of low-mass (sub)halos in SMDPL have too few particles to have a physically meaningful MAH or SFH; for such objects, we first paint a synthetic MAH based on DiffmahPop, a statistical model that connects (sub)halo mass to a point in Diffmah parameter space \citep{hearin2021_diffmah}; we then paint a synthetic SFH based on DiffstarPop, a separate statistical model that connects (sub)halo mass to a point in Diffstar parameter space.

For every halo and subhalo in SMDPL, this methodology produces a mock in which each individual galaxy has a unique SFH deriving from the parametric approximation to its assembly history.

\subsubsection{Transferring galaxies from source to target simulation}
In the next step of our pipeline, we transfer galaxies from SMDPL into a lightcone of host halos in the Outer Rim simulation \citep{heitmann2019_outerrim}. The resampling technique uses a KD-tree search to identify matches between host halo masses in the Outer Rim and SMDPL simulations. Redshift by redshift, host halos from the two simulations are matched, and the central and satellite galaxy content of each SMDPL host halo is transferred to its corresponding Outer Rim host halo. The position and velocity of the central galaxy is set equal to that of the Outer Rim host halo. We adopt an ellipsoidal NFW profile for the spatial distribution of satellite galaxies, aligning the major axes of the ellipse with the triaxial shape of the target halo when available, and using randomly selected axes when shape information is not available in the target halo (e.g., due to it being too poorly resolved in particle density). The velocities of satellite galaxies are set by solving the Jeans equation under the assumption of spherical symmetry. 

At this stage of our pipeline, every host halo in the Outer Rim lightcone contains a population of central and satellite galaxies with lightcone coordinates and star formation histories.

\subsubsection{SEDs and photometry}
We use additional models of stellar population synthesis (SPS) to transform our mock catalog of SFHs into a synthetic universe of galaxies with spectral energy distributions (SEDs) and photometry. In particular, we use new population-level models of stellar metallicity, dust attenuation, and burstiness to map additional properties onto our synthetic SFHs, and then use the DSPS library \citep{hearin_etal23_dsps} to perform the SPS computations needed to produce an SED for each object.

In SPS, the composite SED of a galaxy is a PDF-weighted combination of the SEDs of a collection of Simple Stellar Populations (SSPs). Our SSP template SEDs are taken from the MILES library \citep{falcon_barroso_etal11_miles_update}, extracted using {\tt python-fsps} wrapping FSPS v3.1 with default settings for nebular emission. To calculate the SED of a galaxy, the SSP templates are convolved against the joint PDF of stellar age and metallicity, and normalized by the total stellar mass formed. We now outline the modeling ingredients used to map an SED onto each galaxy in the catalog.

We map stellar metallicity, $Z$, onto our mock galaxies according to a scaling relation with stellar mass, $M_{\star}.$ For the basic form of this scaling relation, we assume a power law with a rolling index, $\langle \log Z\ \vert\ x\rangle\propto C_1 + \alpha(x)\cdot (x-x_1),$ where $x\equiv\log_{10}M_{\star},$ $x_1=12,$ and $\alpha(x)$ is a sigmoid function:
\begin{equation}
\label{eq:sigmoid}
\alpha(x) = y_{\rm min} + \frac{y_{\rm max}-y_{\rm min}}{1 + \exp(-k(x-x_0))}.
\end{equation}
Time evolution of this scaling relation is captured by allowing the parameters $C_1$ and $y_{\rm min}$ to vary with redshift. When computing the SED of the galaxy, we assume a lognormal distribution of stellar metallicity centered at the value defined by the above scaling relations with the stellar mass and redshift.

The effect of dust on galaxy SED is captured by mapping an attenuation curve, $A(\lambda),$ onto each synthetic galaxy, so that the observed SED, $L_{\rm obs}(\lambda),$ is given by the product of the intrinsic SED, $L_{\rm em}(\lambda),$ with a transmission function:
\begin{eqnarray}
\label{eq:attenuation_Funo}
L_{\rm obs}(\lambda) &=& L_{\rm em}(\lambda)\cdot F_{\rm trans}(\lambda) \\
F_{\rm trans}(\lambda) &=& F_{\rm uno} + (1-F_{\rm uno})10^{-0.4A(\lambda)}.
\end{eqnarray}
The quantity $F_{\rm uno}$ was introduced in \citet{lower_etal22} to capture the effect of sightlines that are unobscured by dust. Each attenuation curve in the mock is specified by the same parametric model used in \citet{noll_etal09}:
\begin{equation}
\label{eq:attenuation}
A(\lambda)/A_V = \left[k(\lambda) + D_{\rm UV}(\lambda)\right]\left(\lambda/\lambda_V\right)^{\delta}.
\end{equation}
In Eq.~\ref{eq:attenuation}, $k(\lambda)$ is fitting function that has been tuned to approximate the same reddening curve used in \citet{noll_etal09}, which at wavelengths longer than 0.15 microns closely agrees with the attenuation curve in \citet{calzetti_2000}, and at shorter wavelengths closely agrees with \citet{leitherer_etal2002}. The function $D_{\rm UV}(\lambda)$ is a Drude profile used to approximate the UV bump centered at $2175\mathring{\rm A}.$

For each individual galaxy in the mock, the dust transmission function, $F_{\rm trans}(\lambda),$ is therefore determined by the following free parameters: $F_{\rm uno}, A_V, $ and $\delta.$ In our model for the dust attenuation of galaxy populations, each of these free parameters is allowed to vary according to the stellar mass and SFR of the galaxy. For the case of $F_{\rm uno}$ and $A_V,$ we capture an additional dependence upon the age of the emitting stellar population, such that younger stars have a distinct unobscured fraction and dust normalization relative to older stars within the same galaxy. Each of these dependencies are encoded with a two-dimensional sigmoid function that jointly varies with $\log M_{\star}$ and ${\rm sSFR}\equiv \log \dot{M}_{\star}/M_{\star}.$ We furthermore allow the relationship between the height of the UV bump and $\delta$ to vary according to the same relationship reported in \citet{kriek_conroy13}.

The Diffstar model describes the {\em smooth} component of galaxy SFH, and we implement a new model of burstiness to capture the effect of short-timescale SFR on the galaxy SED. Briefly, using $\tau$ to denote stellar age, we decompose the distribution of stellar ages as follows:
\begin{equation}
\label{eq:burstiness}
P(\tau\vert t_{\rm obs}) = F_{\rm burst}\cdot P_{\rm burst}(\tau\vert\theta_{\rm burst}) + (1-F_{\rm burst})\cdot P(\tau\vert\theta_{\rm SFH}, t_{\rm obs}).
\end{equation}
On the RHS of Eq.~\ref{eq:burstiness}, the first term encodes the stellar population formed in a recent burst, the second term is calculated by numerically integrating the Diffstar SFH, and $F_{\rm burst}$ controls the fractional contribution of the bursting population to the total stellar mass at the time of observation. The quantity $P_{\rm burst}(\tau)$ is a parametric family of PDFs controlled by parameters $\theta_{\rm burst}.$ This family is defined by a triweight Gaussian: a degree-6 polynomial with rational coefficients tuned to approximate a Gaussian distribution that vanishes at $3\sigma$ with vanishing derivatives (see Appendix E of \citet{hearin_etal21_shamnet} for details). The shape of this family of PDFs is controlled by two parameters: $\tau_{\rm peak}$ defines the centroid, and $\tau_{\rm max}$ defines the upper bound of the support. In this model, a burst of SFH is described by three parameters: $\tau_{\rm peak}$ and $\tau_{\rm max}$ control differences in the distribution of star-formation timescales in the recent history of the galaxy, and $F_{\rm burst}$ controls the total mass in the bursting population. For the population-level model that maps values of $\theta_{\rm burst}$ onto simulated halos, we use sigmoid-based scaling relations between each of our three parameters and $M_{\star}$ and sSFR. We allow $F_{\rm burst}$ and $\tau_{\rm peak}$ to vary in sigmoid fashion independently in $M_{\star}$ and sSFR, whereas $\tau_{\rm max}$ only scales with $M_{\star}$.

\subsubsection{Morphology}
We use a disk-bulge decomposition with separate stellar populations in the disk and bulge. The disk is furthermore comprised of a smooth distribution, and star-forming knots. To calculate the stellar populations in the bulge, we convolve the Diffstar SFH with the bulge efficiency function, $F_{\rm b,eff}(t),$ which defines the fraction of $\dot{M}_{\star}$ that forms in the bulge relative to the disk. For the shape of $F_{\rm b,eff}(t),$ we adopt a sigmoid function of cosmic time; when this sigmoid tends to have larger values at earlier times, this corresponds to the bulge of the galaxy forming relatively earlier than the disk, producing an older stellar population in the bulge relative to the disk. Our population-level model of the disk and bulge is then a mapping between {\tt Diffsky} galaxies and the parameters regulating the sigmoid-type shape of $F_{\rm b,eff}(t)$. We constructed this model to produce more bulge-dominated galaxies at high mass, and more disk-dominated galaxies at low mass, while ensuring older stellar populations in bulges relative to disks, and a generally broad diversity of morphologies at all mass and redshift.

For all remaining properties of morphology, such as ellipticity and black hole properties, we use the same models implemented in \cite{korytov2019} for cosmoDC2. Briefly, the model for the scaling relations used for galaxy size was calibrated against \cite{zhang_wang17}; the models for black hole mass and accretion rate were calibrated against \cite{kormendy_ho13} and \cite{aird_etal17a}, respectively; and the model for ellipticity was calibrated against \cite{Joachimi2013}. We refer the reader to the cosmoDC2 paper for details.

Within the smooth disk profile, we model a random set of star-forming ``knots'' that contain part of the $F_{\textrm{burst}}$ population of stars in the disk. The mass of each galaxy's knot population is determined by  $F_{\textrm{knot}}$, the fraction of the disk mass bound up in star-forming knots. Each galaxy is assigned its own value of  $F_{\textrm{knot}}$ drawn from a uniform random distribution spanning $(0, 0.2),$ so that $M_{\star}^{\rm knot}\equiv F_{\textrm{knot}}\cdot M_{\star}^{\rm disk}.$ For each galaxy, if $M_{\star}^{\rm knot}>M_{\star}^{\rm burst},$ then the entirety of the galaxy's burst population is found in the knots, and the remaining portion of $M_{\star}^{\rm knot}$ is made up of the same stellar population as the smooth disk; if $M_{\star}^{\rm knot}<M_{\star}^{\rm burst},$ then the entirety of the mass in the knot is composed of the bursting population, and the remaining mass $(M_{\star}^{\rm burst}-M_{\star}^{\rm knot})$ is assigned to the smooth disk. In this way, the SED of the star-forming knots is bluer in color and has stronger emission lines relative to the smooth disk, but with significant variance across the galaxy population. Knots are a fixed physical size of 250 pc over time, which is translated to an observed size in arcsec based on the angular diameter distance to the redshift of the galaxy. The number of knots modelled in each galaxy is random, with the maximum number being a linear function of $\log$ stellar mass between 3 and 50 knots. The placement of the knots are drawn randomly based on the exponential profile of the disk for each galaxy.

\begin{figure}
    \includegraphics[width=\columnwidth]{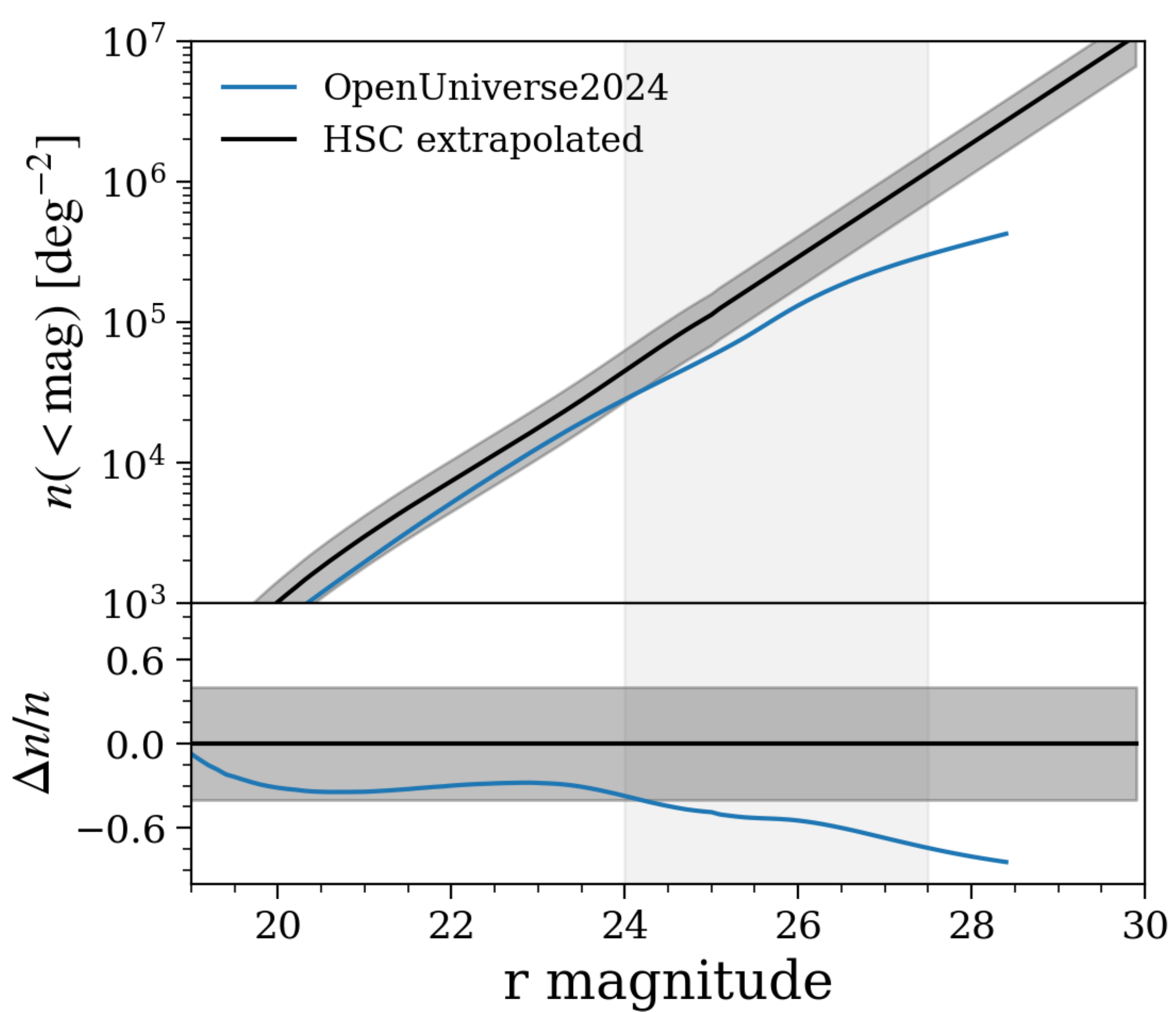}
    \vspace{-0.2cm}
    \caption{\label{fig:validation_romanrubin_catalog1} LSST $r$-band cumulative magnitude distribution for the extragalactic catalog. This is compared to an observed cumulative magnitude distribution from HSC Deep survey data, which is extrapolated to magnitudes fainter than 25. The two begin to diverge in the shaded column at faint magnitudes where the simulation is incomplete. The dark gray band represents uncertainty on the HSC extrapolation.}
\end{figure}

\begin{figure}
    \includegraphics[width=\columnwidth]{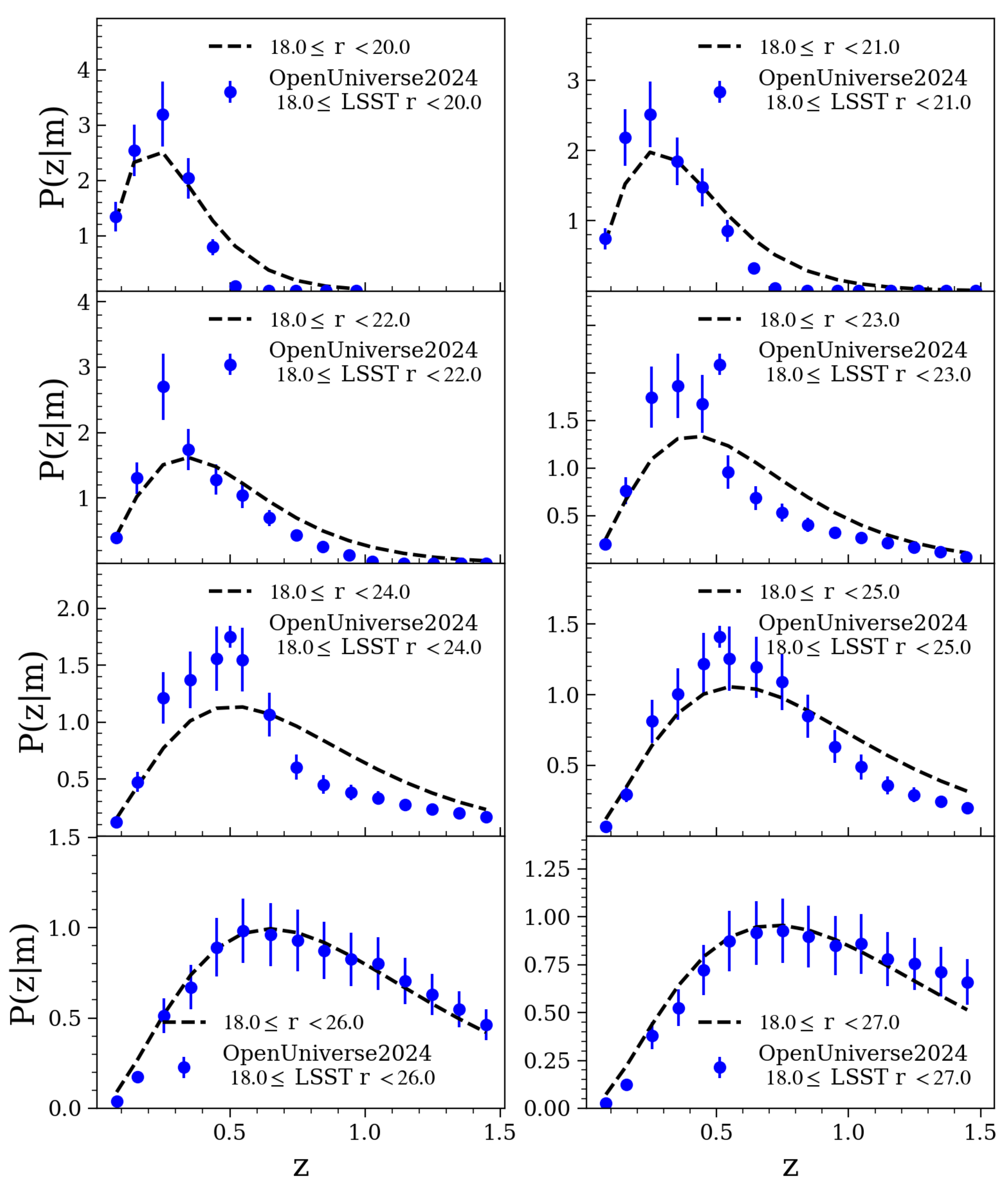}
    \caption{\label{fig:validation_romanrubin_catalog2} LSST redshift distributions and uncertainties for selected ranges of the LSST $r$-band magnitude for the extragalactic catalog. These measurements from the simulation are compared to predicted LSST redshift distributions from \protect\cite{coil2004}.}
\end{figure}

\begin{figure*}
    \label{fig:color_redshift_scatter}
    \includegraphics[width=1.65in]{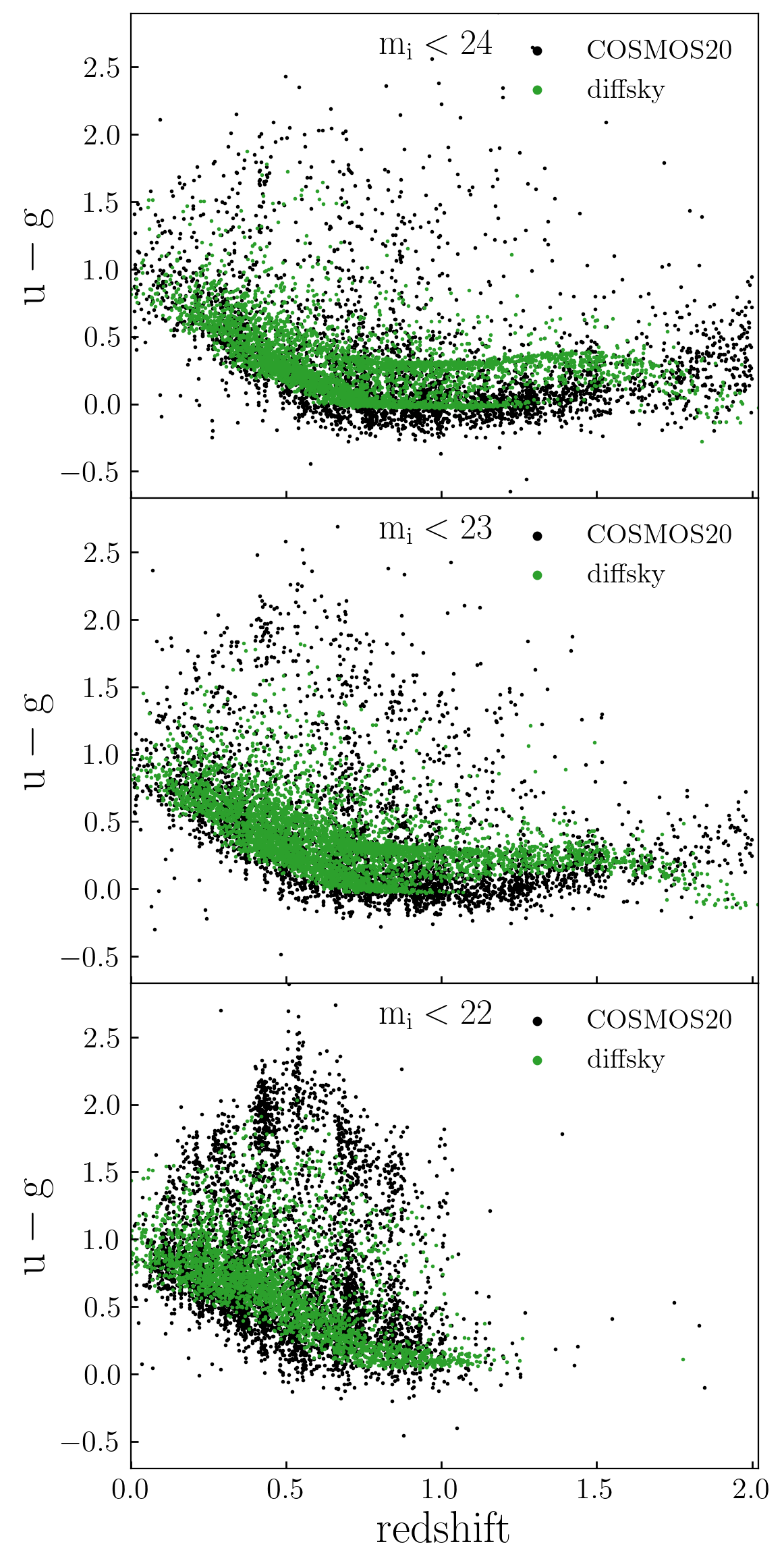}
    \includegraphics[width=1.65in]{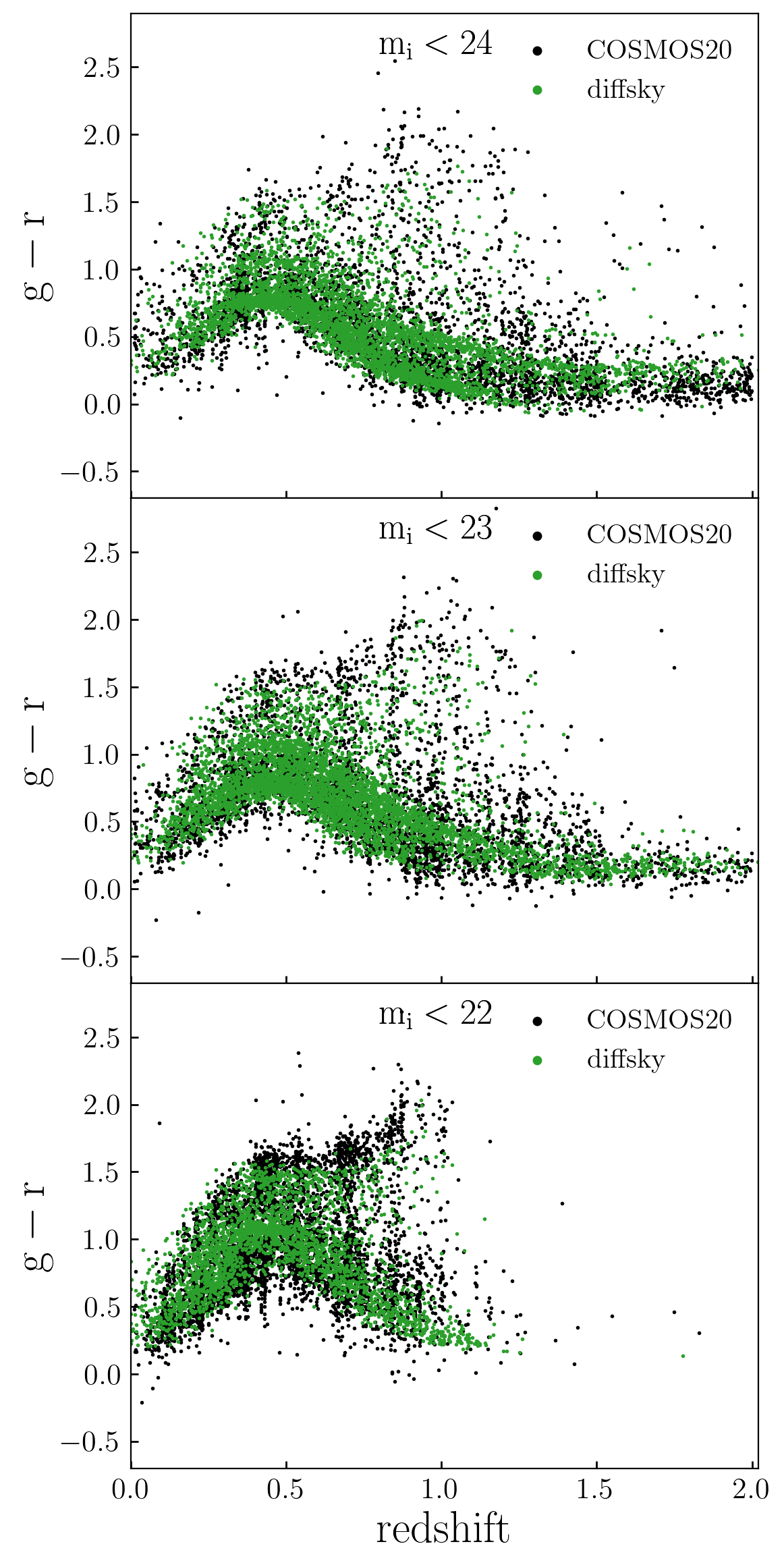}
    \includegraphics[width=1.65in]{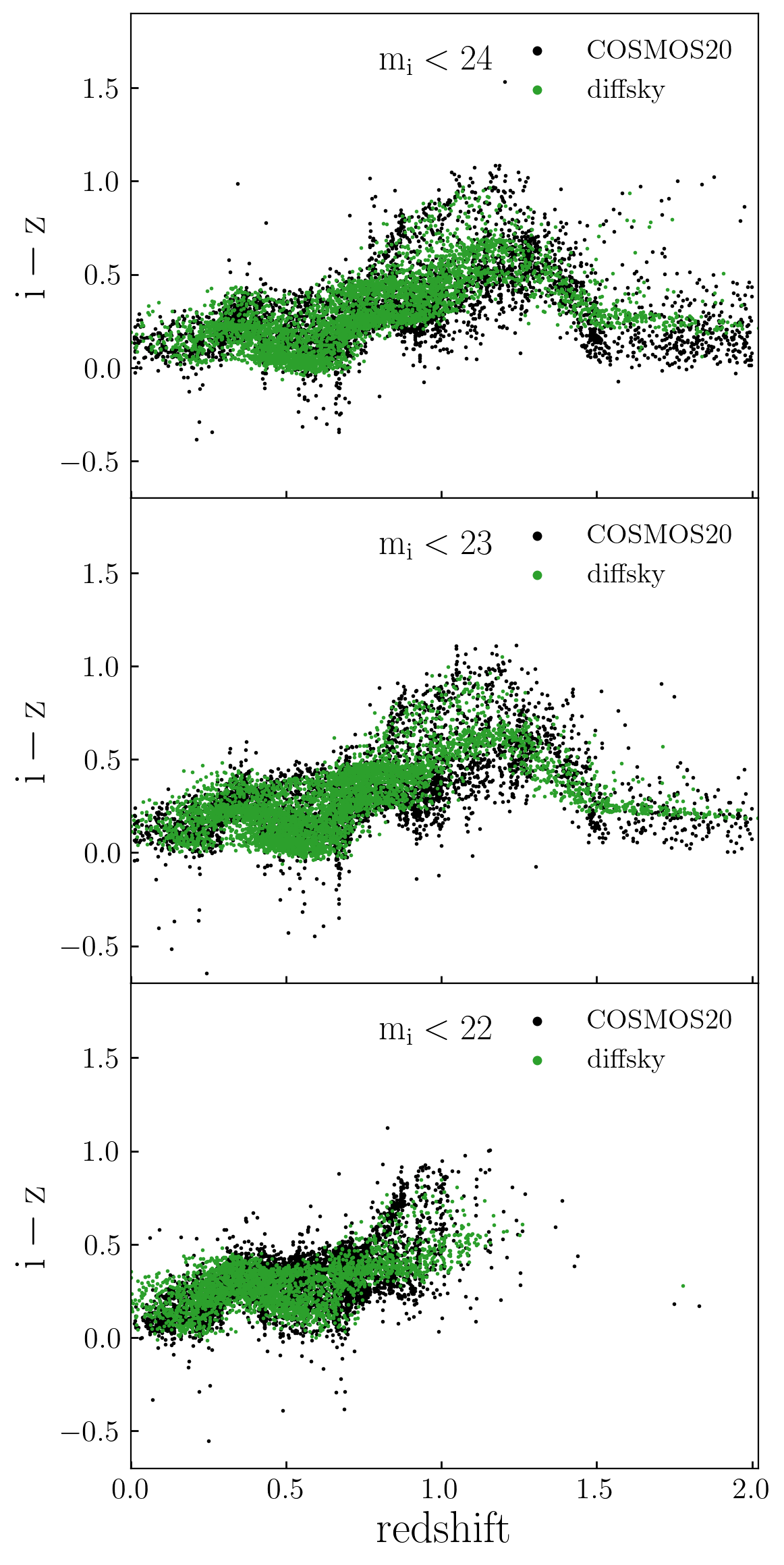}
    \includegraphics[width=1.65in]{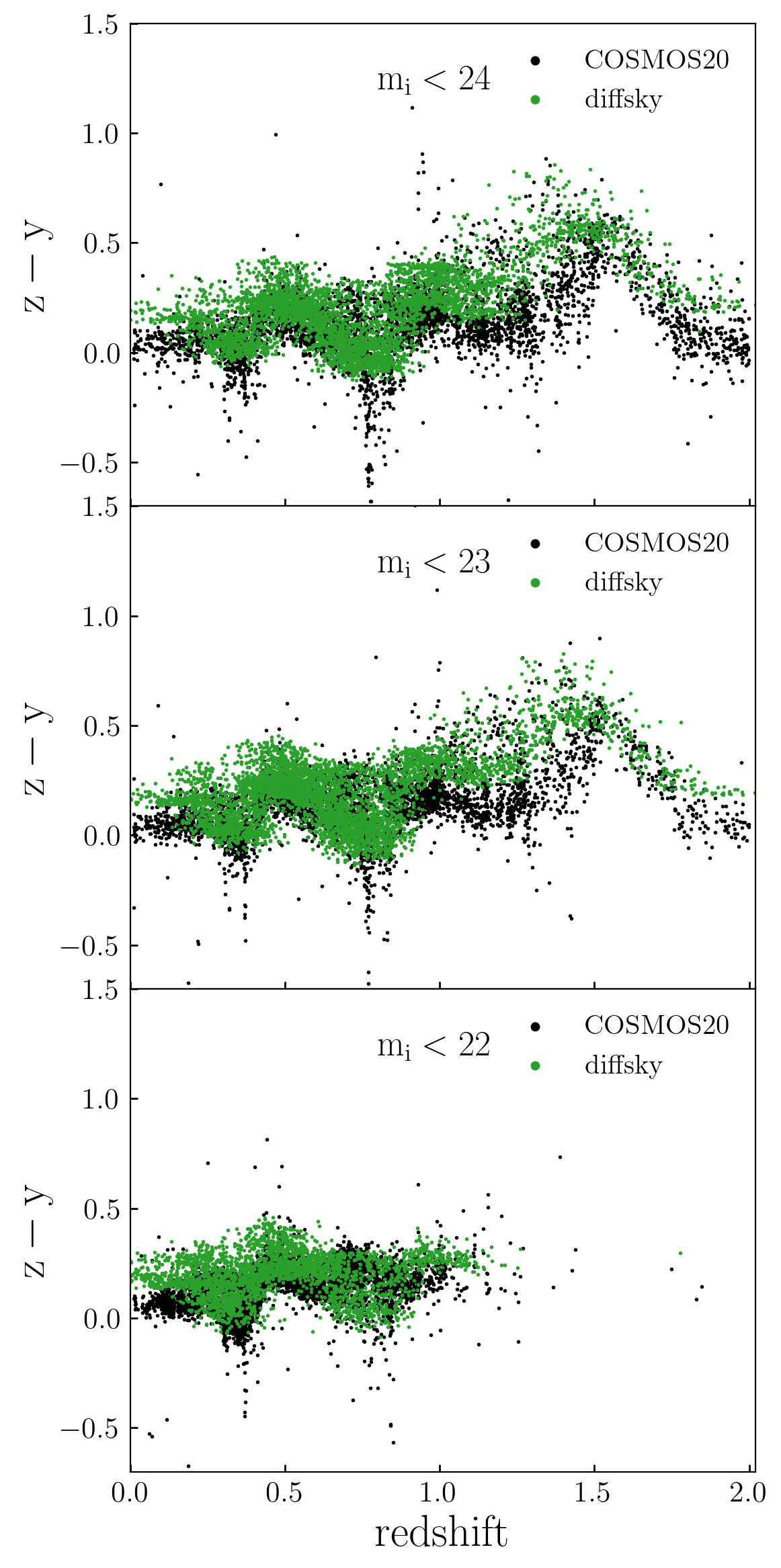}
    \caption{Redshift evolution of broadband optical galaxy colors in the {\tt Diffsky} mock catalog (green) compared with equivalent samples of galaxy colors from COSMOS20 (black). Each panel contains a different subsample of objects drawn from each limiting-magnitude selection. Figure is broken into: (left to right) u-g, g-r, i-z, and z-y colors, and (top to bottom) several limiting magnitude ranges. }
\end{figure*}

\begin{figure}
    \label{fig:color_redshift_scatter_nir}
    \includegraphics[width=1.5in]{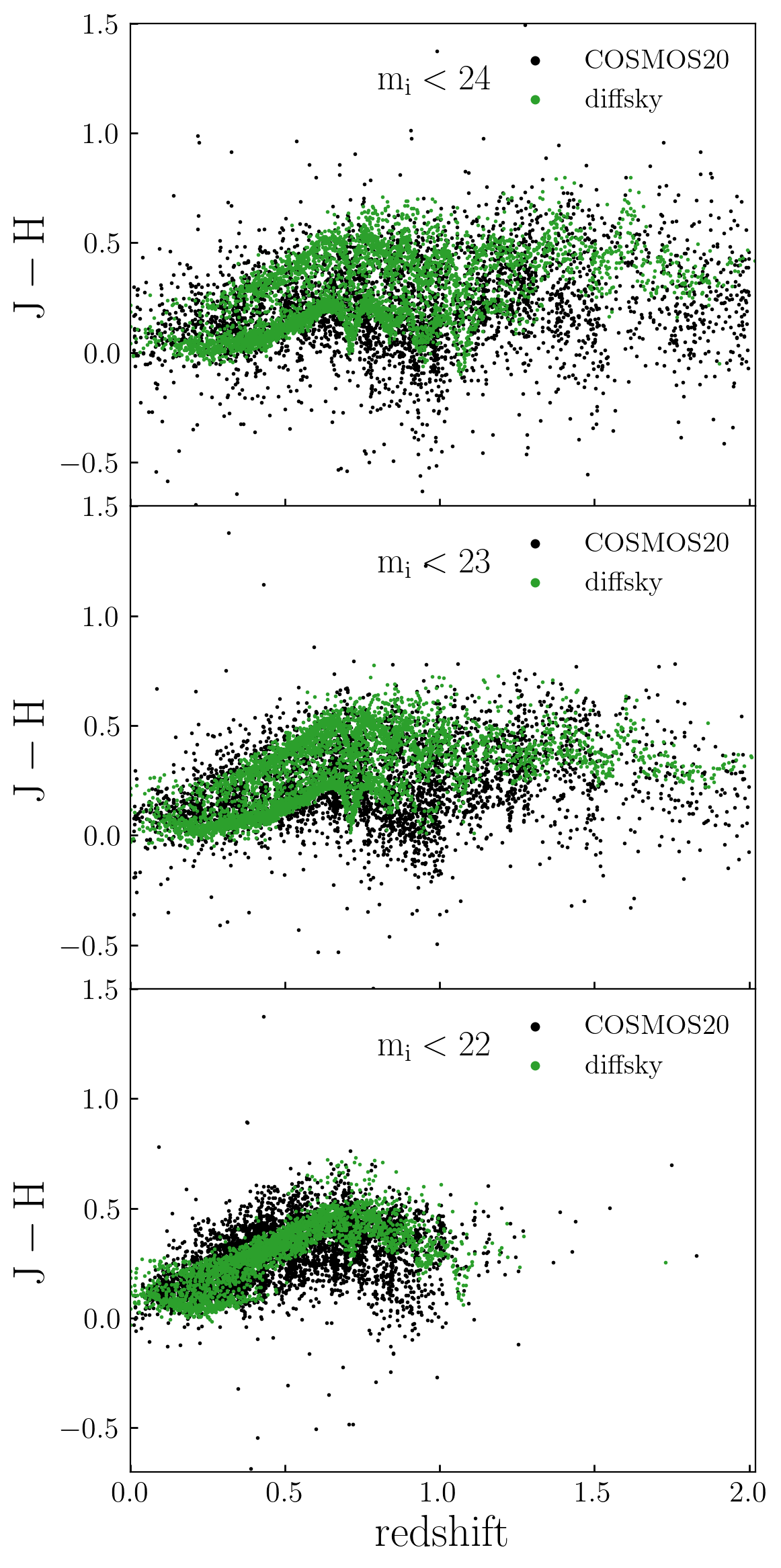}
    \includegraphics[width=1.5in]{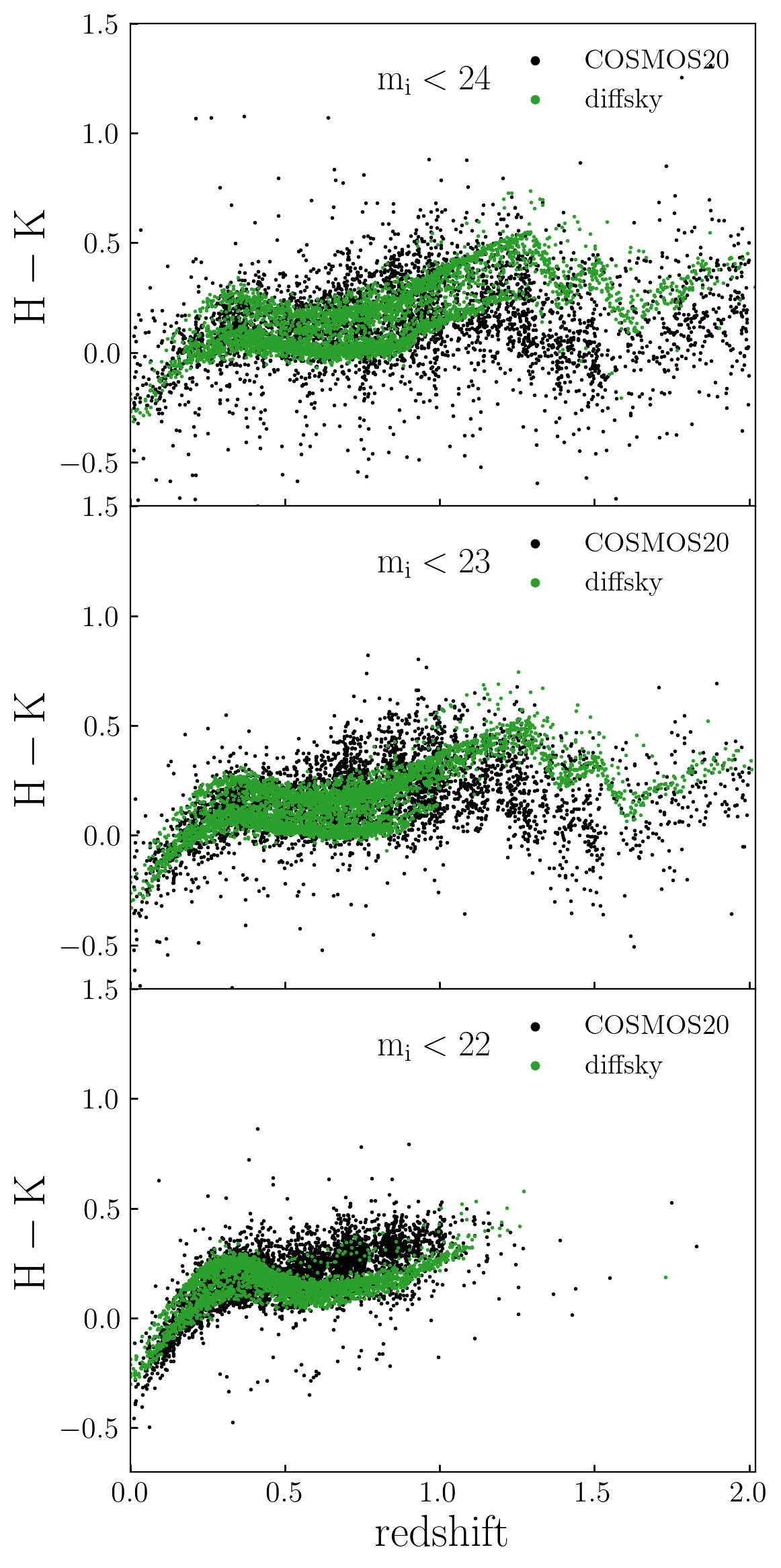}
    \caption{Same as Figure~\ref{fig:color_redshift_scatter}, but for Roman's bandpass colors.}
\end{figure}

\subsubsection{Validation results}

In Figs.~\ref{fig:validation_romanrubin_catalog1} \& \ref{fig:validation_romanrubin_catalog2}, we show selected validation tests for the extragalactic OpenUniverse2024 catalog. These tests were developed to validate the cosmoDC2 catalog and are described in \cite{korytov2019} and \cite{kovacs2022}. The upper panel of Fig.~\ref{fig:validation_romanrubin_catalog1} shows the cumulative galaxy number count per square degree for galaxies with  LSST $r$-band magnitudes less than a specific value, as a function of that value. The solid blue and black lines show the catalog result and the cumulative count extrapolated\footnote{See \cite{kovacs2022} for a discussion of the extrapolation procedure.} from HSC Deep survey
data \citep{aihara2018}, respectively.  The dark shaded band shows a $\pm40$\% fractional difference in the number counts around the HSC result. The light shaded vertical band shows the validation region for the cosmoDC2 catalog, which was designed to check number counts for weak lensing analyses. The lower panel shows the fractional difference between the HSC extrapolation and the catalog result. This catalog falls slightly below the validation criterion in the validation region because we did not inject additional synthetic ultra-faint galaxies into the catalog to compensate for the effects of the finite mass resolution of the underlying simulation.  Figure~\ref{fig:validation_romanrubin_catalog2}  shows the normalized redshift distribution for 8 bins of $r$-band magnitude ranging from 18 to 27, as given in the legend for each panel. The magnitude bins become progressively wider, but are dominated by the faintest magnitudes in the bin for each case. The dashed lines show fits to the observational data from \cite{coil2004}, extrapolated to fainter magnitudes than were observed in that dataset. The distributions agree well for faint magnitudes, but for brighter magnitudes there are too many (few) galaxies below (above) a redshift of $\approx$0.5. Compared to cosmoDC2, this catalog shows a similar excess at low redshifts of bright galaxies with $r<21$, a worse excess at low redshift for galaxies with $r< 24-25$, and better agreement for galaxies with $r<27$.

In Figs.~\ref{fig:color_redshift_scatter} and \ref{fig:color_redshift_scatter_nir}, we compare the redshift evolution of broadband colors in COSMOS-20 to the mock catalog. Redshift appears on the x-axis of each panel, and color on the y-axis. Different colors are plotted in different columns; within each column, different rows show results for galaxy samples selected with different magnitude cuts. Optical colors are shown in Fig.~\ref{fig:color_redshift_scatter}, NIR colors in Fig.~\ref{fig:color_redshift_scatter_nir}.

As discussed in the introduction, these simulations were made possible by a limited-time opportunity to run image simulations on the Theta supercomputer before its decommissioning. The {\tt Diffsky} model is a long-term project to forward-model galaxy SEDs based on high-resolution N-body simulations, and the {\tt Diffsky} extragalactic catalog simulated here is a coarse calibration of a prototype version of the model. As shown in the validation figures, the broadband optical--NIR colors are broadly representative of the real universe, but the observational data used as constraints are not recovered with high precision. The SEDs and colors in the synthetic catalog should be sufficient for a wide range of scientific applications, although downstream projects in need a high-fidelity representation of galaxy SED evolution will likely require a higher-precision calibration than was possible to achieve before Theta was decommissioned. In particular, the NIR colors shown in Fig.~\ref{fig:color_redshift_scatter_nir} still have narrower distributions than found in real data, though those distributions follows observed trends very well.

\begin{figure*}
\includegraphics[width=0.33\textwidth]{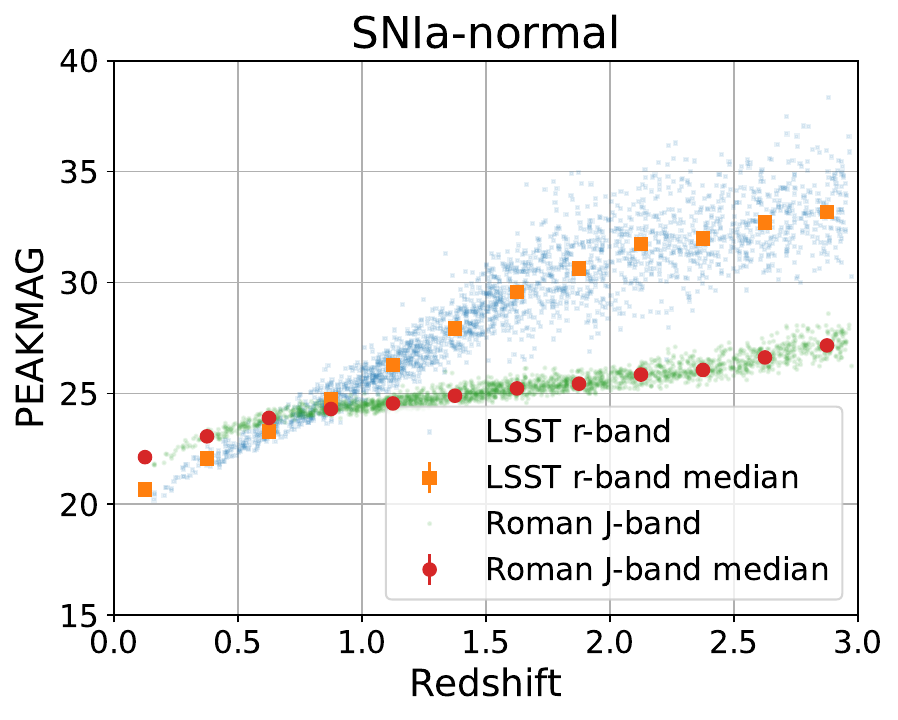}
\includegraphics[width=0.33\textwidth]{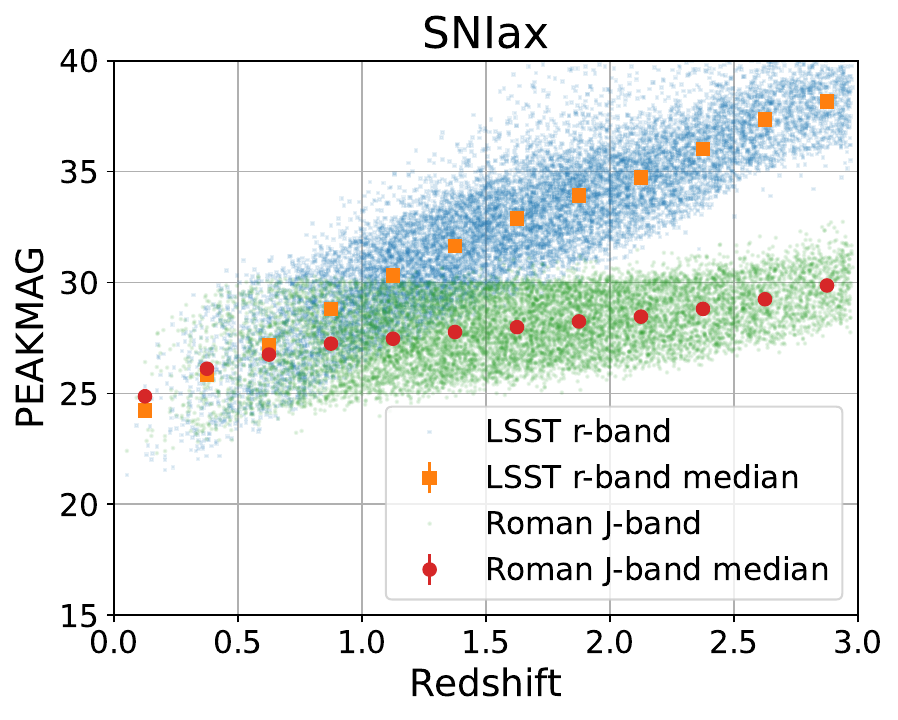}
\includegraphics[width=0.33\textwidth]{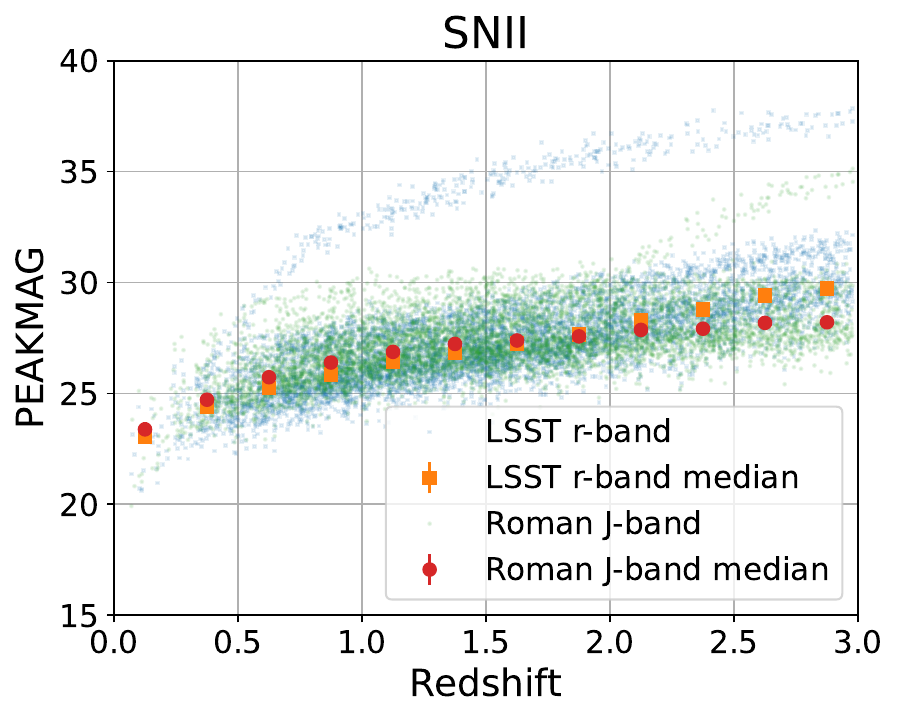}
\includegraphics[width=0.33\textwidth]{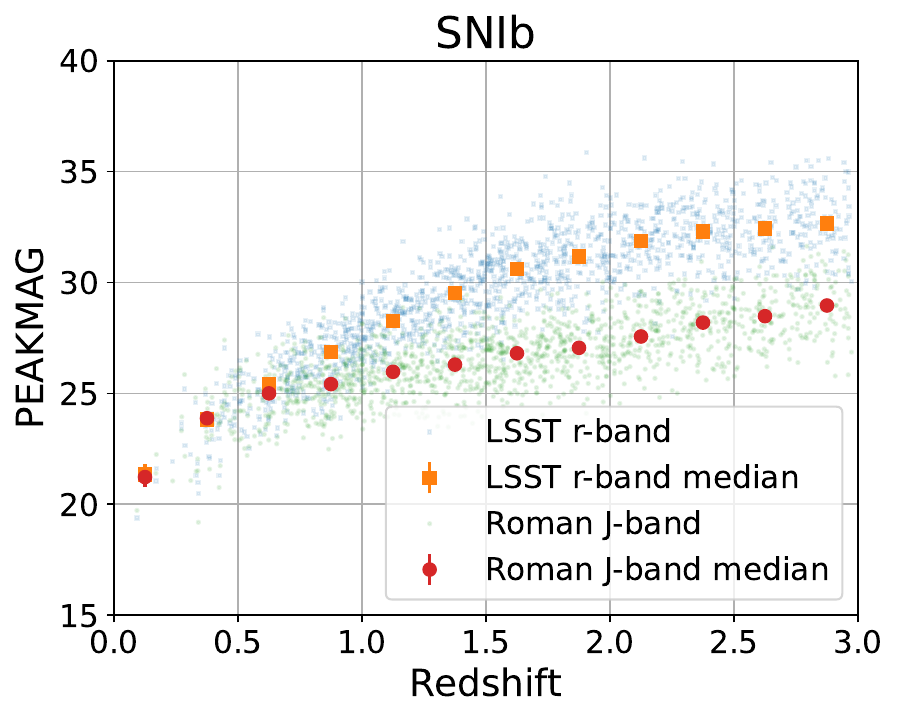}
\includegraphics[width=0.33\textwidth]{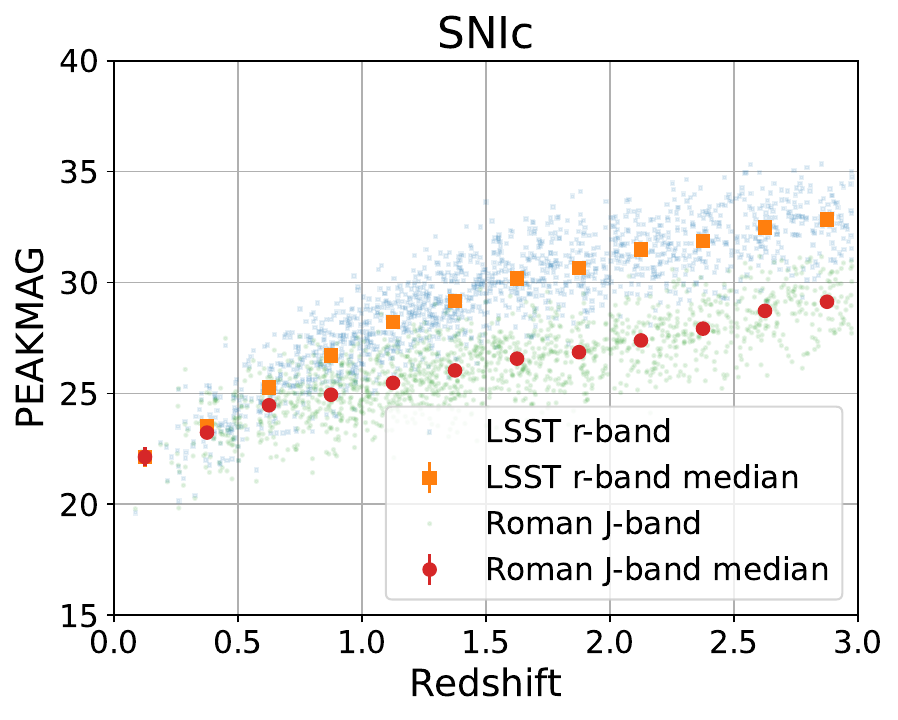}
\includegraphics[width=0.33\textwidth]{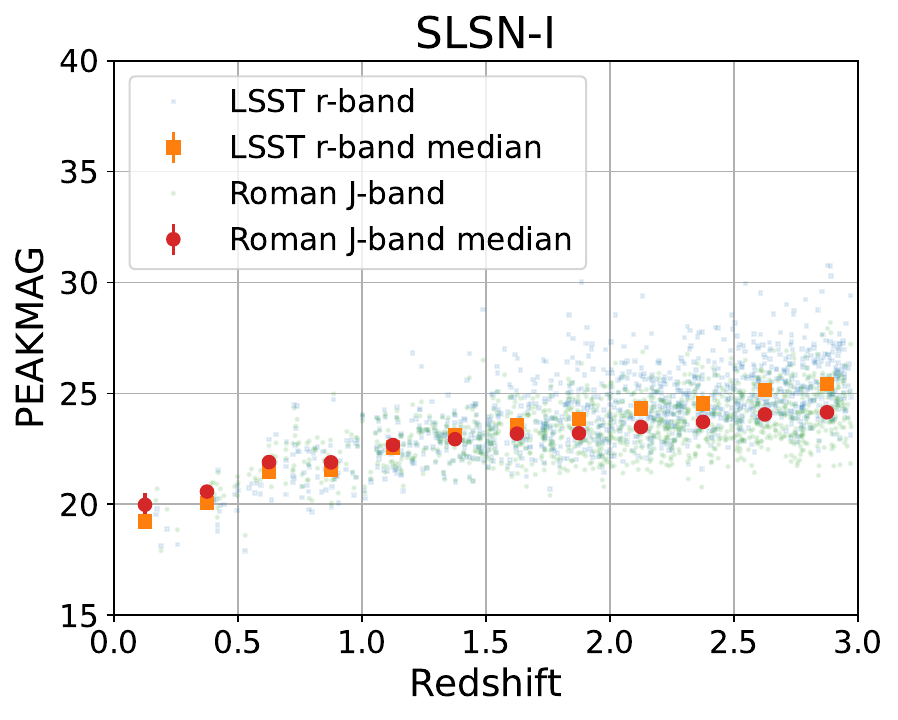}
\includegraphics[width=0.33\textwidth]{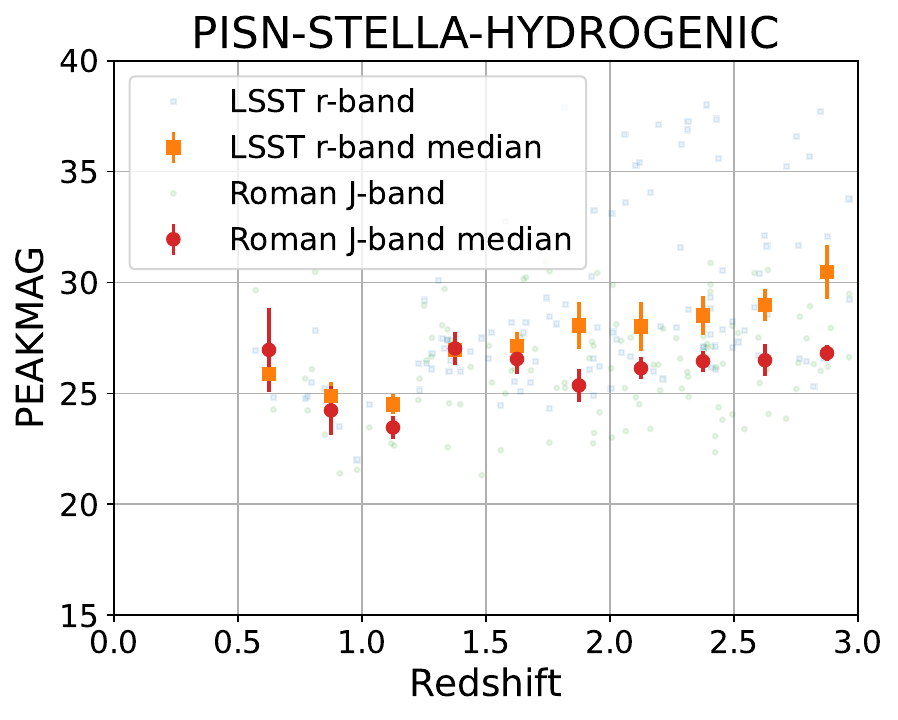}
\includegraphics[width=0.33\textwidth]{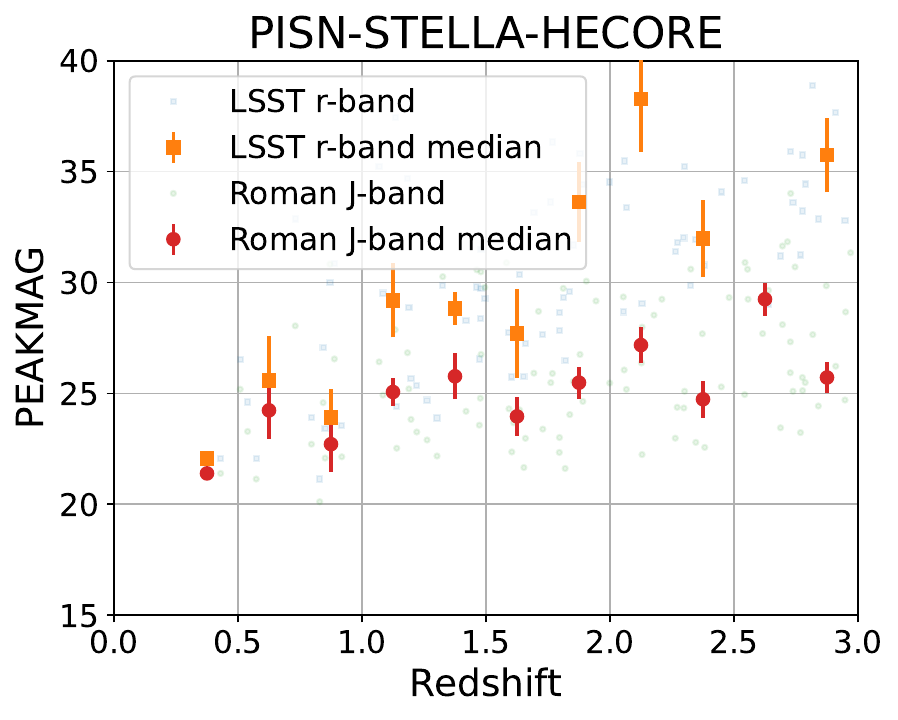}
\includegraphics[width=0.33\textwidth]{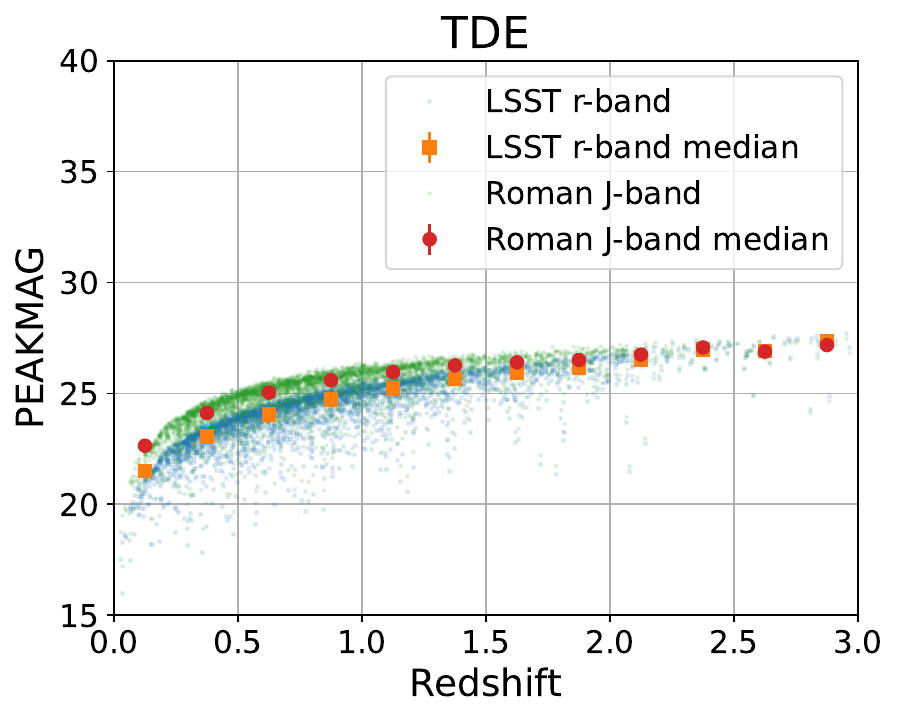}
\includegraphics[width=0.33\textwidth]{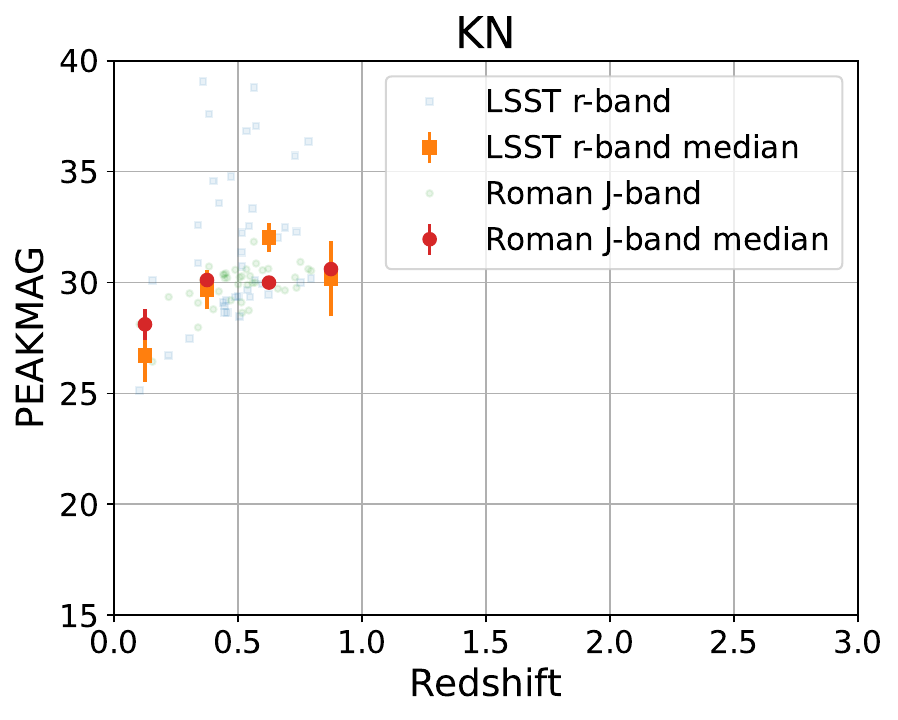}
\caption{\label{fig:peakmag_vs_z} 
For each transient model, peakmag vs. redshift is shown for 
    LSST $r$-band (blue points, orange squares for median) and 
    Roman $J$-band (green points, red circles for median).
    All events are shown for the rare transients (SLSN-I, PISN, TDE, KN);
    the other models are prescaled by 100 to avoid saturating the plot.
    }
\end{figure*}

\begin{figure*}
\includegraphics[width=0.31\textwidth]{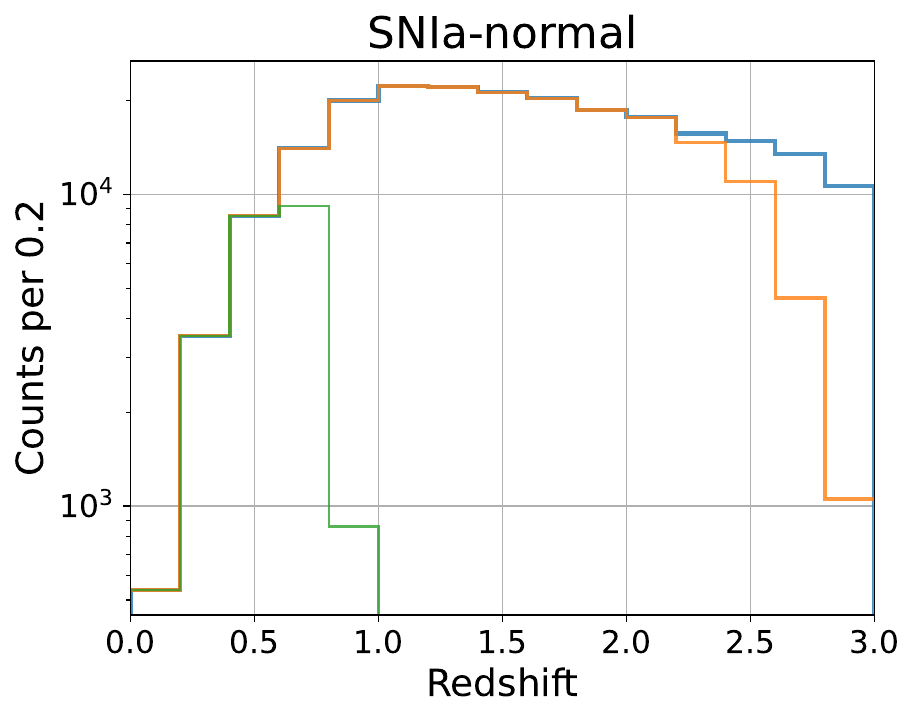}
\includegraphics[width=0.31\textwidth]{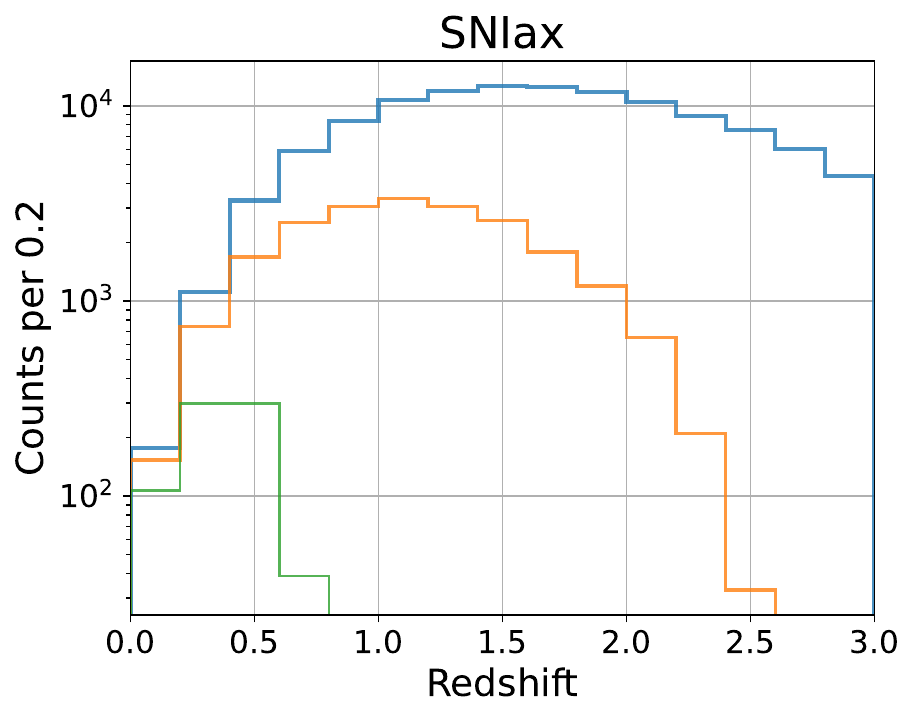}
\includegraphics[width=0.31\textwidth]{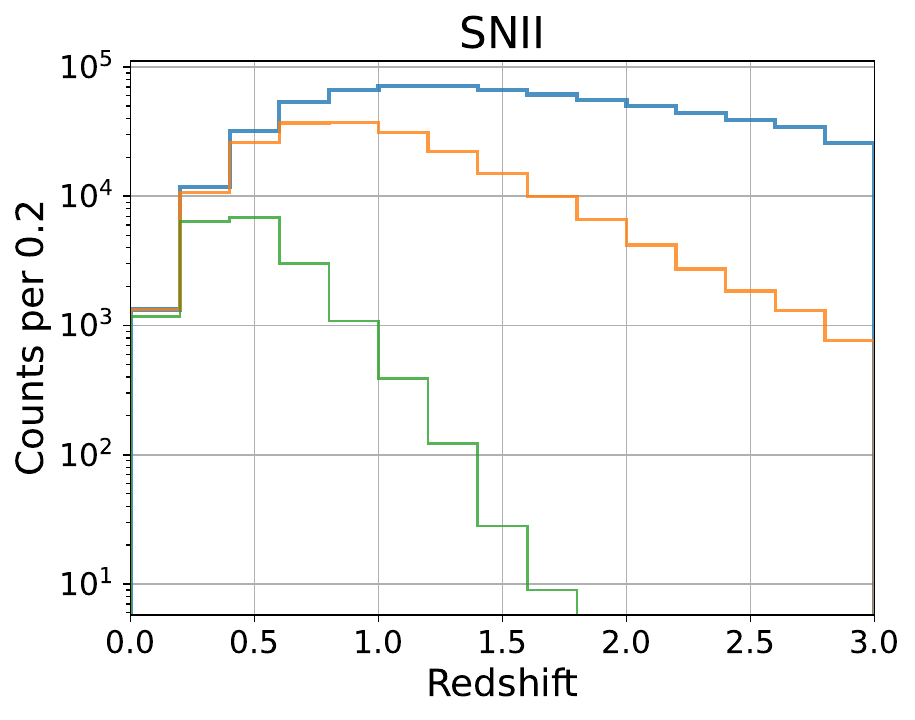}
\includegraphics[width=0.31\textwidth]{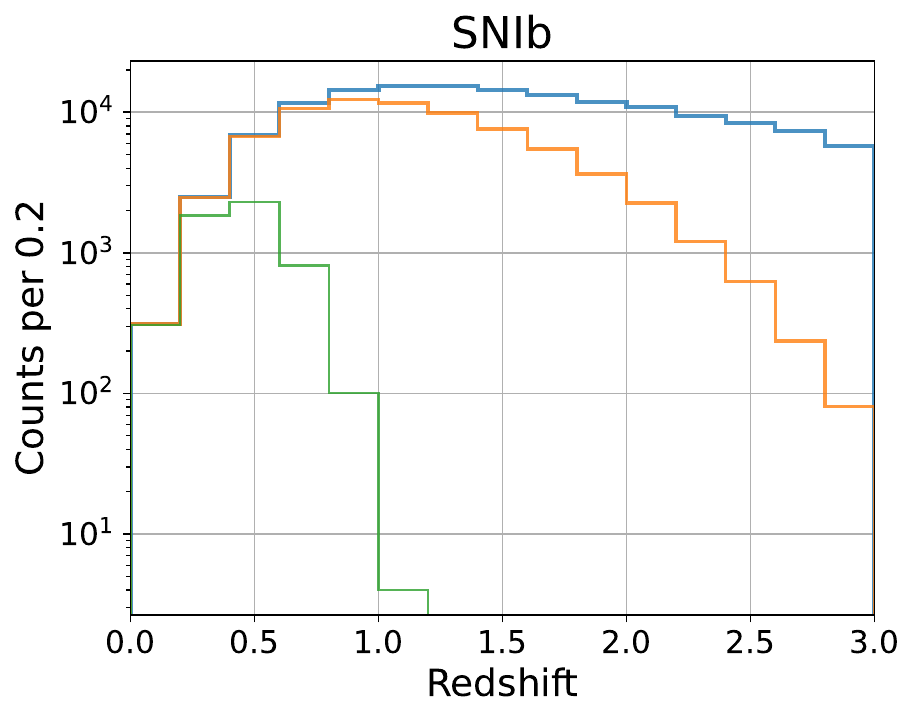}
\includegraphics[width=0.31\textwidth]{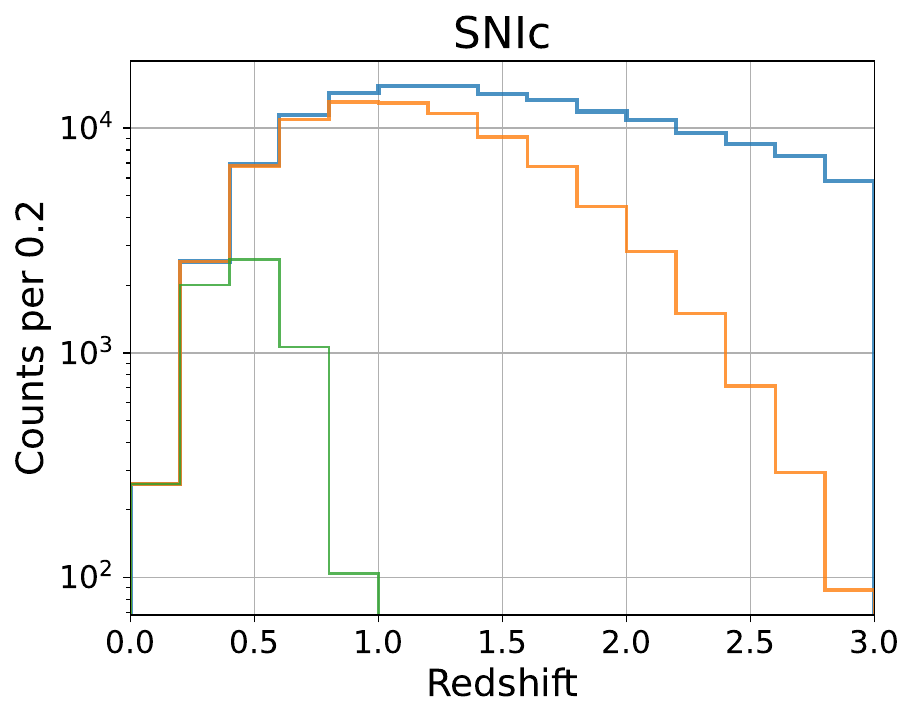}
\includegraphics[width=0.31\textwidth]{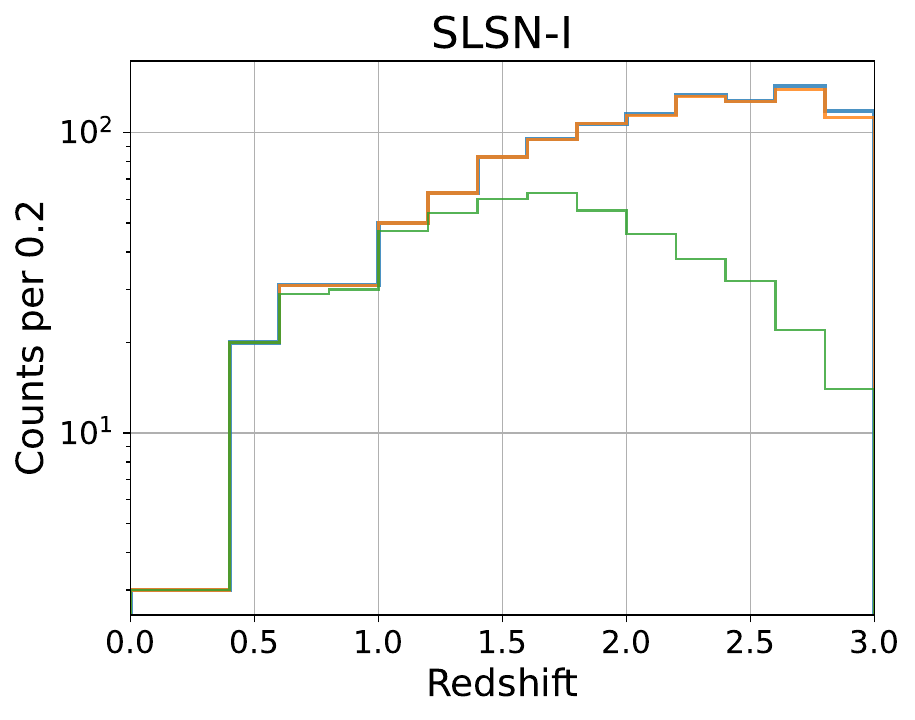}
\includegraphics[width=0.31\textwidth]{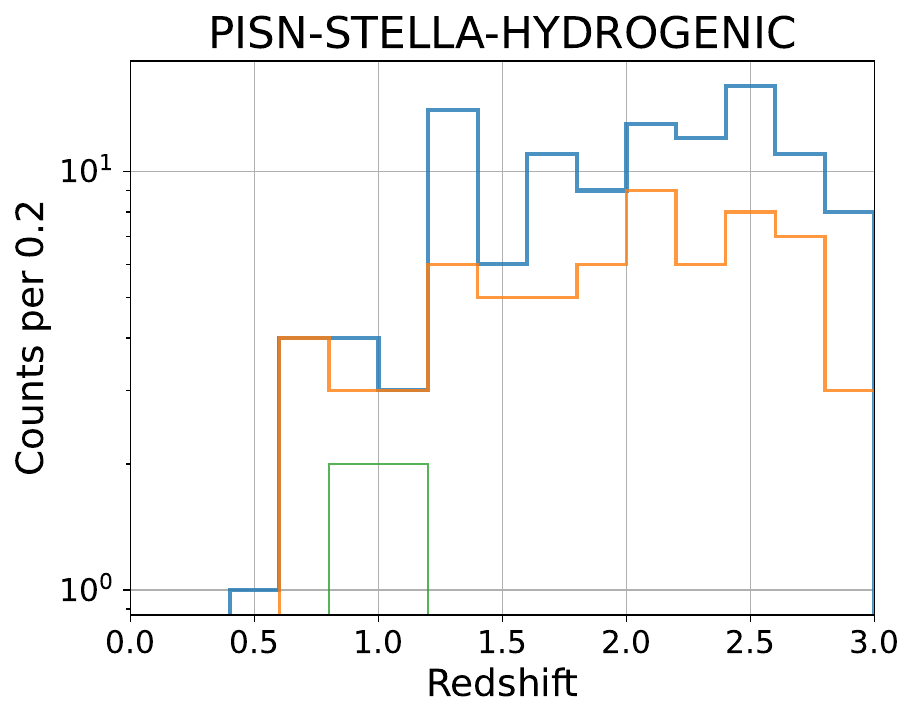}
\includegraphics[width=0.31\textwidth]{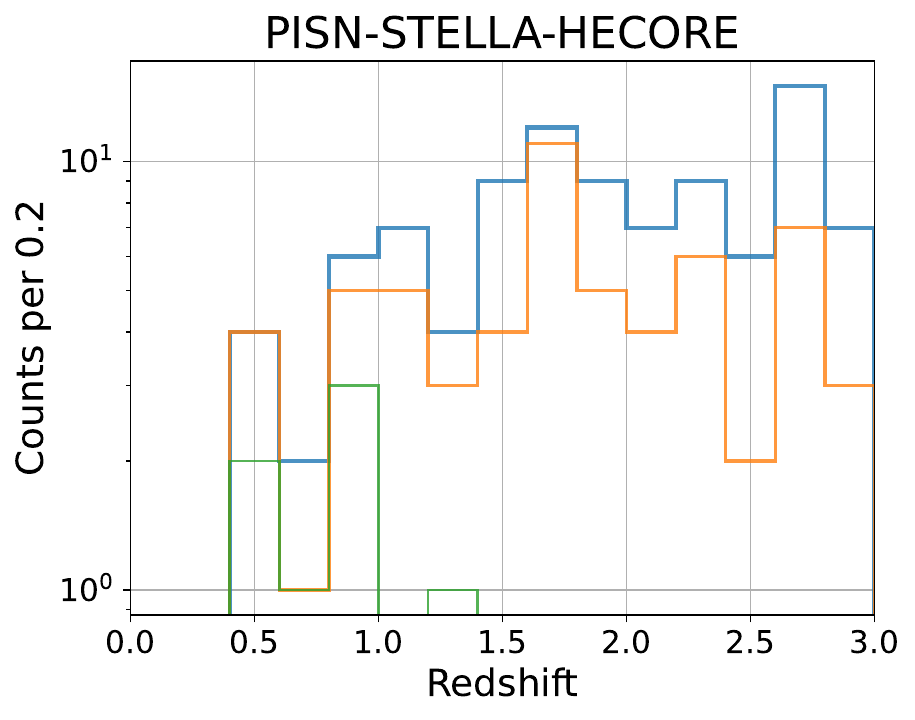}
\includegraphics[width=0.31\textwidth]{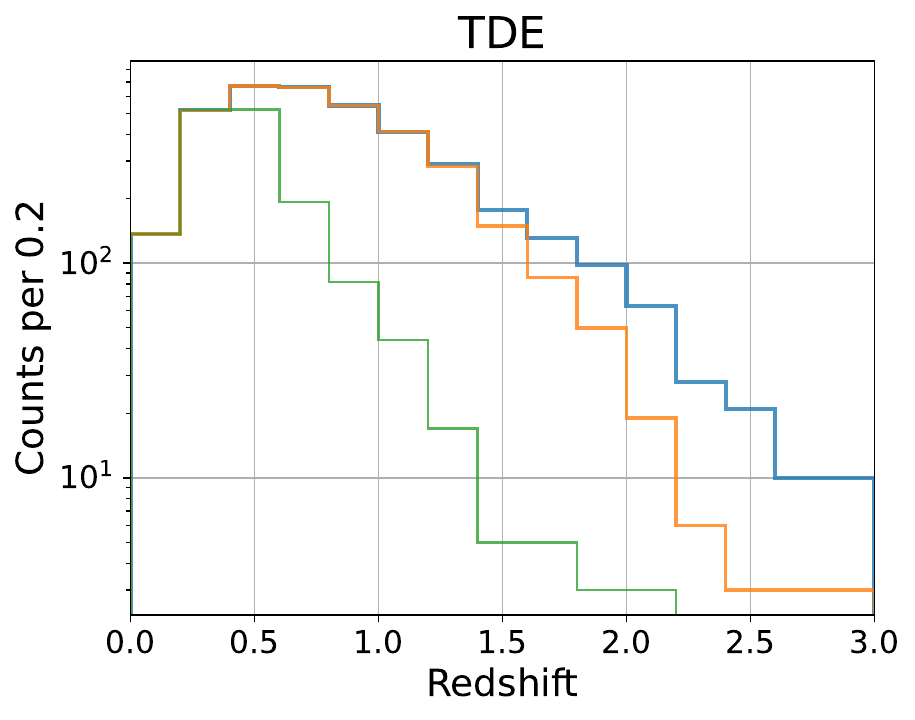}
\includegraphics[width=0.40\textwidth]{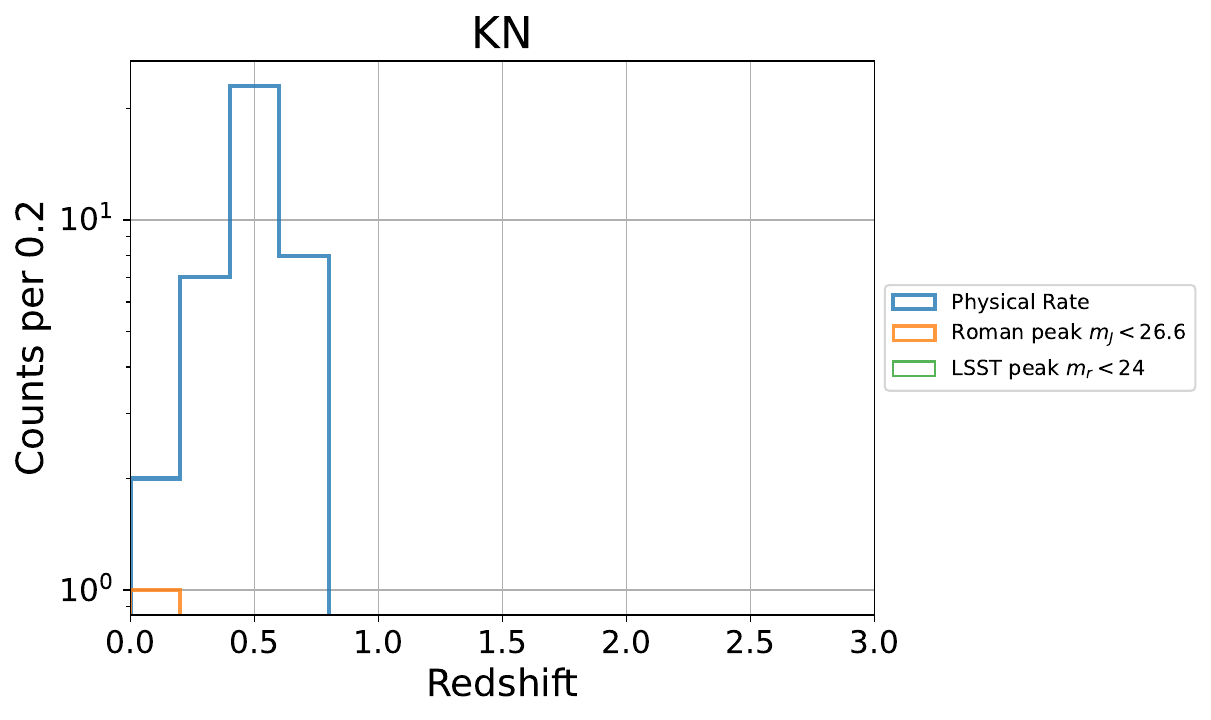}
\caption{\label{fig:transient_zov} 
For each transient model, the redshift distributions are overlaid for 
physical rate (blue), and single-visit $5\sigma$ depth for
Roman $J$-band with 160~s exposure time ($m_J<26.6$; orange); 
LSST $r$-band with 30~s exposure time ($m_r<24$; green).
    }   
\end{figure*}

\subsection{Transient catalog}\label{sec:transients}

\newcommand{\projectName}{OpenUniverse2024} 
\newcommand{\plasticc}{PLAsTiCC}
\newcommand{\elasticc}{ELAsTiCC}
\newcommand{\URLSNANA}{\url{https://github.com/RickKessler/SNANA}}
\newcommand{\URLIax}{\url{https://github.com/RutgersSN/SNIax-PLAsTiCC}}
\newcommand{\rateunit}{Mpc$^{-3}$yr$^{-1}$ }
\newcommand{\snana}{{\tt SNANA}}
\newcommand{\hostlib}{{\tt HOSTLIB}}
\newcommand{\Trest}{T_{\rm rest}}

\newcommand{\NtransientLC}{1.39~million}
\newcommand{\NtransientSED}{312~million}
\newcommand{\NtransientClass}{10}
\newcommand{\URLSNANAMODELS}{\url{https://zenodo.org/records/14749318}}

A library of SED-based transient models 
originally developed for a public classification challenge known as
\plasticc\ :  
``Photometric LSST Astronomical Time-series Classification Challenge"
\citep{PLASTICC2018,Hlozek2023_PLASTICC} was leveraged, extended, and updated for our simulations, in particular extending models to fully cover the IR wavelengths of Roman.
The simulation code for these models is part of the ``Supernova Analysis" (\snana) 
software package \citep{SNANA},\footnote{\URLSNANA}   
and details of the simulation and models are described in 
\citet[K19]{Kessler2019_PLASTICC}.
Here we leverage this catalog-level simulation infrastructure to
include the $\NtransientClass$ extragalactic transient models shown in 
Table~\ref{tab:transients} and Figs.~\ref{fig:peakmag_vs_z}-\ref{fig:transient_zov}.

While Galactic transients were included in \plasticc, they are
not included here.
The total number of generated events is \NtransientLC, which includes
\NtransientSED\ SEDs on a Modified Julian Date (MJD) grid extending from 61444 to 63269
and a redshift range extending out to $z=3$.
To include light curves that are not fully contained in the survey MJD range, 
the MJD at peak brightness was generated over a wider range: 61374 to 63299.
The only selection requirement is that the peak magnitude must be brighter 
than 30 in at least one band,
and therefore a difference-imaging analysis will select a subset
corresponding to the survey detection limit.

\begin{table} 
\begin{center}
\caption{\label{tab:transients} List of transient models and summary information for each model, including number of events, rest-frame phase $\Trest$ (days) with $\Trest=0$ at peak brightness, and artificial rate multiplier.} 
\begin{tabular}{l | r r | c c c  }
\hline
\hline
      &            &                             & $\Trest$ &          & rate    \\
      & \multicolumn{2}{c}{Number of generated:} & range    & integer  & multi-  \\
model &  events    & SEDs                        & (days)   & type    &  plier  \\ 
\hline
SN~Ia     & 224,575  & 67,070,894  & $-20, +300$ & 10 & 1     \\
SNIax     & 165,703  & 31,978,600  & $-50, +300$ & 12 & 1    \\
SNII      & 694,064  & 144,330,844 & $-50, +200$ & 32 & 1    \\
SNIb      & 148,722  & 33,604,658  & $-50, +200$ & 21 & 1    \\
SNIc      & 148,724  & 33,753,628  & $-50, +200$ & 26 & 1    \\
SLSN-I    & 1,128    & 398,547     & $-50, +300$ & 40 & 1.5  \\
TDE       & 3,784    & 583,063     & $-100, +250$ & 42 & 10  \\
PISN-H    & 113      & 33,694      & $-50, +200$  & 57 & 4   \\
PISN-He   & 112      & 35,318      & $-50, +200$  & 58 & 4    \\
KN        & 53       & 12,785      & $-5, +30$    & 50  & 100  \\ 
RanMag    & 27,884   & 138,855     & $-100, +300$ & 99  & --  \\
\hline
Total     & 1,386,978 & 311,802,031 & -- & -- & -- \\
\hline
\end{tabular}
\end{center}
\end{table}

Most of the \plasticc\ model SEDs extend only to 11,000\AA\ in the rest-frame,
well below the 25,000\AA\ limit of the Roman $K$-band.
We therefore use updated models that are based on optical + NIR data to
accommodate the Roman $K$-band at low redshifts. 
The SEDs are modeled on an MJD grid spanning the 5~year
survey, and each SED is transformed to the observer frame accounting for
cosmic expansion. 
To ensure uniform modeling of Galactic extinction, weak lensing, 
and peculiar velocities, these effects are {\it not} modeled by
\snana, and instead they are incorporated in an identical way to stars and galaxies within the SkyCatalog framework, described in more detail below.

\subsubsection{Classes of transients} 
A brief technical summary of each model is given below, 
where an "SED template" refers to a time series of rest-frame
SED on a roughly 1-day grid.
More detailed astrophysical descriptions can be found in Sec.~4 of K19
and references within,
and the model libraries are also available to download from zenodo.\footnote{\url{https://zenodo.org/records/14749318}}

\medskip\noindent{\bf Type Ia Supernova (SN~Ia):}  \\
\plasticc\ modeled SN~Ia SEDs using the the semi-analytical SALT2 model from \citet{Guy2010_SALT2,Betoule2014_JLA}.
Here we use an updated SALT3 model trained with the SALTshaker code described in \citet{Kenworthy2021_SALT3}. 
The original SALT3 SED model extended to 11,000\AA,
and \citet{Pierel2022_SALT3NIR} extended this model to 20,000\AA\ by
expanding the SN~Ia training set to include NIR data from 
ground base surveys and from the Hubble Space Telescope
(see list of surveys and references in Sec 3.2 in \citet{Pierel2022_SALT3NIR}). 
For this simulation project, the SALT3 model was further extended to 25,000\AA\ using the
methods in \citet{Pierel2018_SEDextend}.

The stretch and color populations are from \citet{SK2016},
and the intrinsic scatter model is the ``G10'' model 
from \citet{Kessler2013}.
The volumetric rate model is the same as in \plasticc:
see Eqs.~1-2 in K19.

\medskip\noindent{\bf Peculiar Type Iax Supernova (SNIax):} \\
We use 1,000 SED templates based on the original model\footnote{\URLIax}  
used in \plasticc.
This model is defined to 25,000~\AA, and thus does not need to 
be extended in wavelength.
We use the updated model that includes host-galaxy extinction 
(See Fig.~4 in \citet{Vincenzi2021_CCcontam}).
The volumetric rate model follows the star formation rate;
see Sec 4.3.2 in K19.

\medskip\noindent{\bf Type II/Ib/Ic Core Collapse (CC) Supernova (SNII/SNIb/SNIc):} \\
For SNII/Ib/Ic, we use 24/13/7 SED templates from 
\citet{Vincenzi2019_CCSED}.
The SED wavelength range was extended from 11,000\AA\ to 25,000\AA\
using methods in \citet{Pierel2018_SEDextend}. 
The luminosity function (LF) is approximated by a coherent Gaussian smear
at all phases and wavelengths: 
$\sigma=0.40,0.65$~mag for SNIIP and SNIIL, and
$\sigma=1.12,1.18$~mag for SNIb and SNIc.
The core collapse volumetric rate is from 
\citet{Strolger2015_CCrate} (green curve in Fig.~6).
The CC rate fractions are 0.70, 0.15, 0.15 for SNII, SNIb, SNIc, respectively.

\medskip\noindent{\bf Type I Superluminous Supernova (SLSN-I):} \\
Rather than extending the wavelength range of the original SLSN-I model from PLAsTiCC, 
we adopt the SED time sequence derived from the observed light curves of Gaia16apd, 
one of the closest SLSNe-I ever discovered. 
At $z=0.102$, Gaia16apd was extensively followed up over a wide range of wavelength 
and time phase \citep{Yan2017_SLSN, Kangas2017_SLSN}. 
Most importantly, Gaia16apd is one of a few SLSNe-I which have {\it HST} far and near-UV spectra 
as well as {\it SWIFT} UV light curves at early photospheric phases \citep{Yan2017_SLSN}. 
These UV data enable us to derive the reliable SED time sequence through black-body fitting 
and cover both far-UV and near-IR wavelengths spanned by the Roman photometric bands.
Host galaxy extinction is modeled using Eq.~2 in \citet{WV07_HOSTXT}.

At $z\sim0.2$, the SLSN-I volumetric rate has been estimated by
\citet{Cooke2012_SLSN}, based on the data from the 
Robotic Optical Transient Search Experiment-IIIb \citep{Quimby2013_SLSN}
and the Zwicky Transient Facility \citep{Perley2020_ZTF}.
At $z\sim1$, the rate has been measured by the 
Canada-France-Hawaii Telescope Supernova Legacy Survey (SNLS) \citep{Prajs2017_SLSN}.
We adopt the model fit to the estimated volumetric rates as a function of redshift in the form of $3\times10^{-8}(1+z)^{2.2}$~\rateunit\ \citep{Prajs2017_SLSN}. 

The SLSN-I luminosity function is approximated by a coherent Gaussian smear with 
$\sigma = 1$\,mag at all phases and wavelengths. 
This approximation is clearly too simple compared to the diversity demonstrated 
by the large sample of low-$z$ SLSN-I from ZTF \citep{Chen2023_SLSN}. 
Building more realistic diversity in light curve morphologies is a longer-term simulation effort, not achievable on the time-frame of these current simulations.

\medskip\noindent{\bf Tidal Disruption Event (TDE):} \\
Rather than extending the wavelength range of the original 
TDE model from \plasticc, 
we extracted an SED time sequence 
from the multi-wavelength data collected for the closest TDE event: AT2019qiz at $z=0.0151$. 
The unique advantage of this event is its complete UV photometry \&\ spectral coverage at both 
the early and the peak phases of its light curve evolution \citep{Hung2021_TDE,Nicholl2020_TDE}. 
Applying a blackbody emission model and interpolation to the UV-optical light curves, 
we construct a grid of rest-frame SED templates covering wide ranges of phase 
from early to late times as well as wavelength from UV to infrared.
The same SED time series is used for each simulated event, and we apply a
luminosity function (LF) estimated from a large sample of TDEs discovered by ZTF \citep{Yao2023_TDE_LF}, and the LF is assumed to include host-galaxy extinction.
The local TDE volumetric rate is estimated to be roughly 
$1\times10^{-6}$~\rateunit\ \citep{Velzen2018_TDE_RATE}.
To enhance the statistics, we artificially multiplied this rate by a factor of ten.

\newcommand{\URLELASTICC}{\url{https://portal.nersc.gov/cfs/lsst/DESC_TD_PUBLIC/ELASTICC}}

\medskip\noindent{\bf Pair Instability Supernova (PISN):} \\
The original PISN model for \plasticc\ was based on MOSFIT \citep{MOSFIT2018},
but the maximum SED wavelength of 11,000~\AA\ is not sufficient for
Roman sims. Instead, we use two SED models developed as part of
\elasticc\footnote{\URLELASTICC}, a data challenge to test 
real-time transient classification within LSST-DESC.

The first PISN model set is based on the massive-star progenitor models of
\citet{Gilmer2017_PISN}, which retain part of their initial H-rich envelope 
(except for the most massive 250 solar-mass progenitor). 
The FLASH hydrodynamics code \citep{FLASH2000,FLASH2009}
was used to generate the explosion, whose output was post-processed with the multi-group radiation-hydrodynamics code STELLA 
\citep{Blin2006_STELLA,Blin2011_STELLA} 
to compute SEDs and broad-band light curves 
\citep{Kozy2014_PISN,Kozy2017_PISN}.
The second PISN model set is based on the He-core progenitor models of
\citep{HegarWoosley2002_PISN},
whose explosion was generated with the hydrodynamics code 
KEPLER \citep{Weaver1978,Woosley2002}.
The synthetic SEDs were generated with same STELLA code that
was used for the first model set.
Host galaxy extinction is modeled using Eq.~2 in \citet{WV07_HOSTXT}.

The same number of events were generated from each model,
and the volumetric rate vs. redshift is from \citet{PLK2012_PISN_RATE}.
We have enhanced the PISN sample by artificially increasing this rate
by a factor of four.

\medskip\noindent{\bf Kilonova (KN):} \\
We model KNe using 329 SED time series templates from
\citet{Kasen2017_KN}, the same model used in the original \plasticc\ challenge,
and add host extinction using Eq.~2 in \citet{WV07_HOSTXT}.
This SED wavelength range extends to 30,000\AA, and therefore 
no SED modifications were applied. 
We enhanced this sample using a volumetric rate of
$1\times 10^{-7}$~{\rateunit}, which is roughly 2 orders of magnitude
larger than the expected rate.

\medskip\noindent{\bf Bright Random Magnitudes (RanMag):} \\
To enable more detailed difference-imaging studies,
we enhanced the sample of bright sources near bright galaxies 
by generating events with a random absolute mag between $-22$ and $-17$, 
and random redshift between 0 and 2. These apparently bright, low-$z$ transients are rare in the physical rate model, due to volume effects at lower redshift, and so this class of models artificially boosts their number in the simulation to enable studying the recovery of these objects. The model SED is an AB spectrum.
To help avoid confusion between RanMag and astrophysical transients,
the RanMag sources begin 280 days after the last MJD for astrophysical transients.

\subsubsection{Host galaxies} 
The galaxies in the simulation 
were used to construct a host library (\hostlib) for the 
\snana\ simulation. The \hostlib\ includes a bulge+disk Sersic profile
that is used to place transients near a galaxy with a weight proportional
to the local surface brightness. The \hostlib\ also includes 
estimates of stellar mass and star formation rate, which are used to 
model transient-host correlations such as core collapse SNe being found
only in star forming galaxies.

\newcommand{\snmagExact}{m_{10\text{\AA}}}
\newcommand{\snmagApprox}{m_{100\text{\AA}}}
\newcommand{\snmagBinCor}{m_{\rm binCor}}

\subsubsection{Rebinning to reduce SED output size}\label{magcor}
For the \snana\ simulation, the default wavelength bin size is 10~\AA\
for integrating broadband fluxes. Since the \projectName\ simulation requires
writing the model SED in 1~day bins, we reduce this output by writing
the SED in courser 100~\AA\ bins. This courser bin size results in 
slight errors in the integrated broadband flux, and we therefore include a 
correction for each SED using
\begin{equation}
   \snmagBinCor = \snmagExact - \snmagApprox
\end{equation}
where 
$\snmagExact$ is the precise synthetic magnitude computed with 10~\AA\ bin SEDs, and
$\snmagApprox$ is the synthetic magnitude computed with 100~\AA\ bins.
When constructing the Python object representing the transient within SkyCatalogs,  
each SED flux is corrected by $10^{-0.4\snmagBinCor}$.

In the time dimension,
the SEDs are written in 1~day MJD bins for $\Trest < 40$~days,
where $\Trest$ is the rest-frame phase (days) with respect to the time of peak brightness (i.e., $\Trest=0$ at peak brightness).
To reduce output, we use 2~day bins for $40<\Trest<80$~days,
and 4-day bins for $\Trest > 80$~days.
The SED flux at each observed MJD is linearly interpolated between the closest MJDs
for which an SED is written.
Since the Kilonova model evolves much more rapidly than other models,
the SEDs for this model are written in 0.1~day MJD bins.

\subsubsection{Known issues}
First, for the core collapse models (SNII, SNIb, SNIc), the wavelength range was extended for the set of templates that had been corrected for host-galaxy extinction. However, host extinction was not enabled in the simulation because we had intended to extend the templates without host extinction corrections. Host-transient correlations were also incorrectly modeled, resulting in an average stellar mass about 1~dex too low. Thus the Poisson noise contribution from the host is underestimated.

\section{Simulated Survey Features}\label{sec:surveys}

\begin{figure}
\hspace{-0.2cm}\includegraphics[width=\columnwidth]{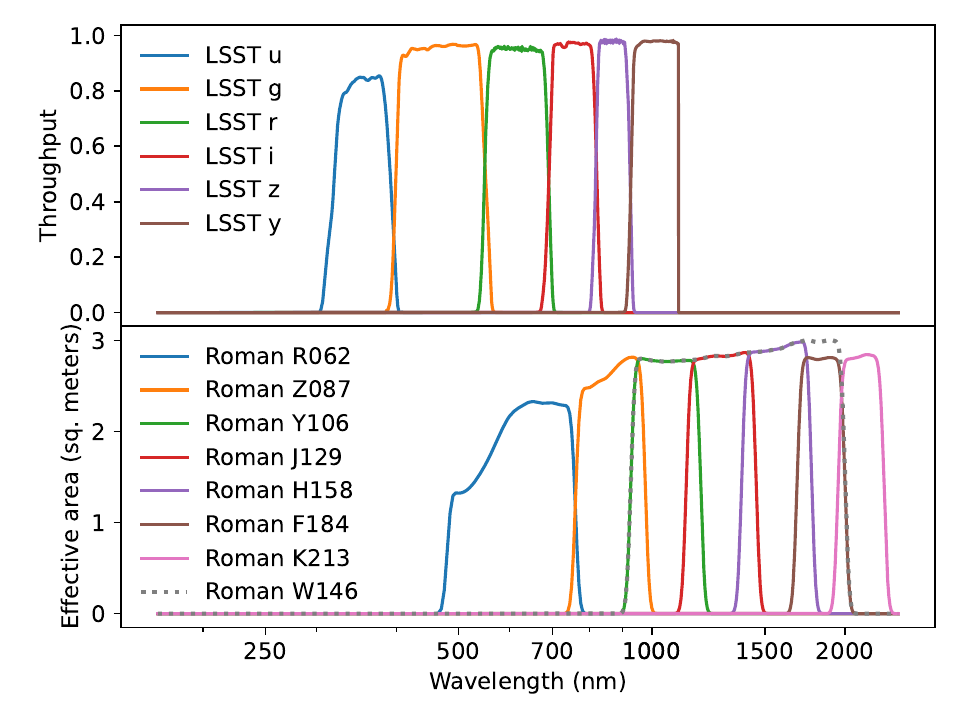}
\vspace{-0.2cm}
\caption{\label{fig:bpass}Comparison of the LSST and Roman bandpass wavelength ranges and throughput used in the simulation suite.}
\end{figure}

\begin{figure*}
\vspace{-1cm}
\includegraphics[width=\columnwidth]{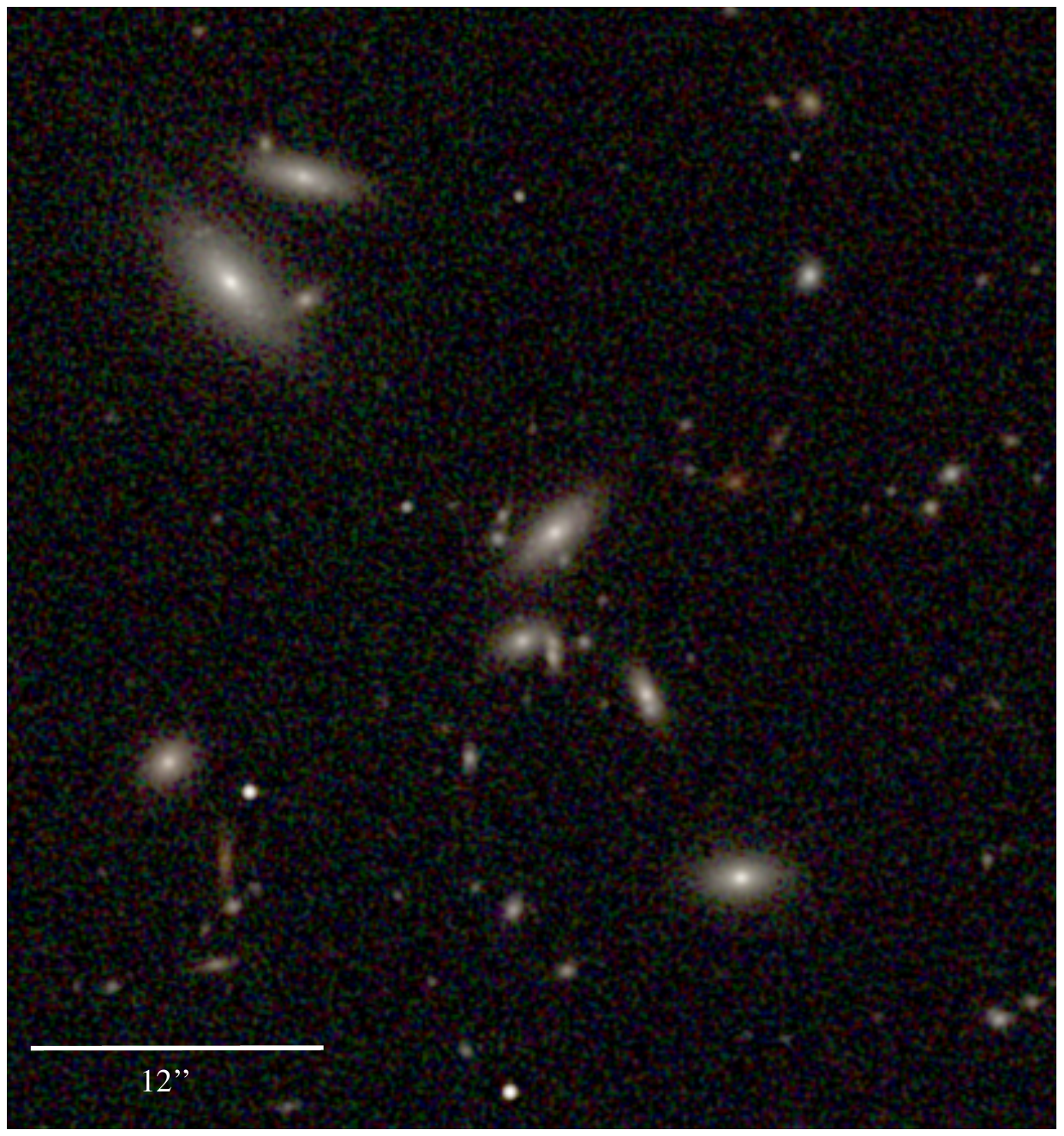}
\includegraphics[width=1.01\columnwidth]{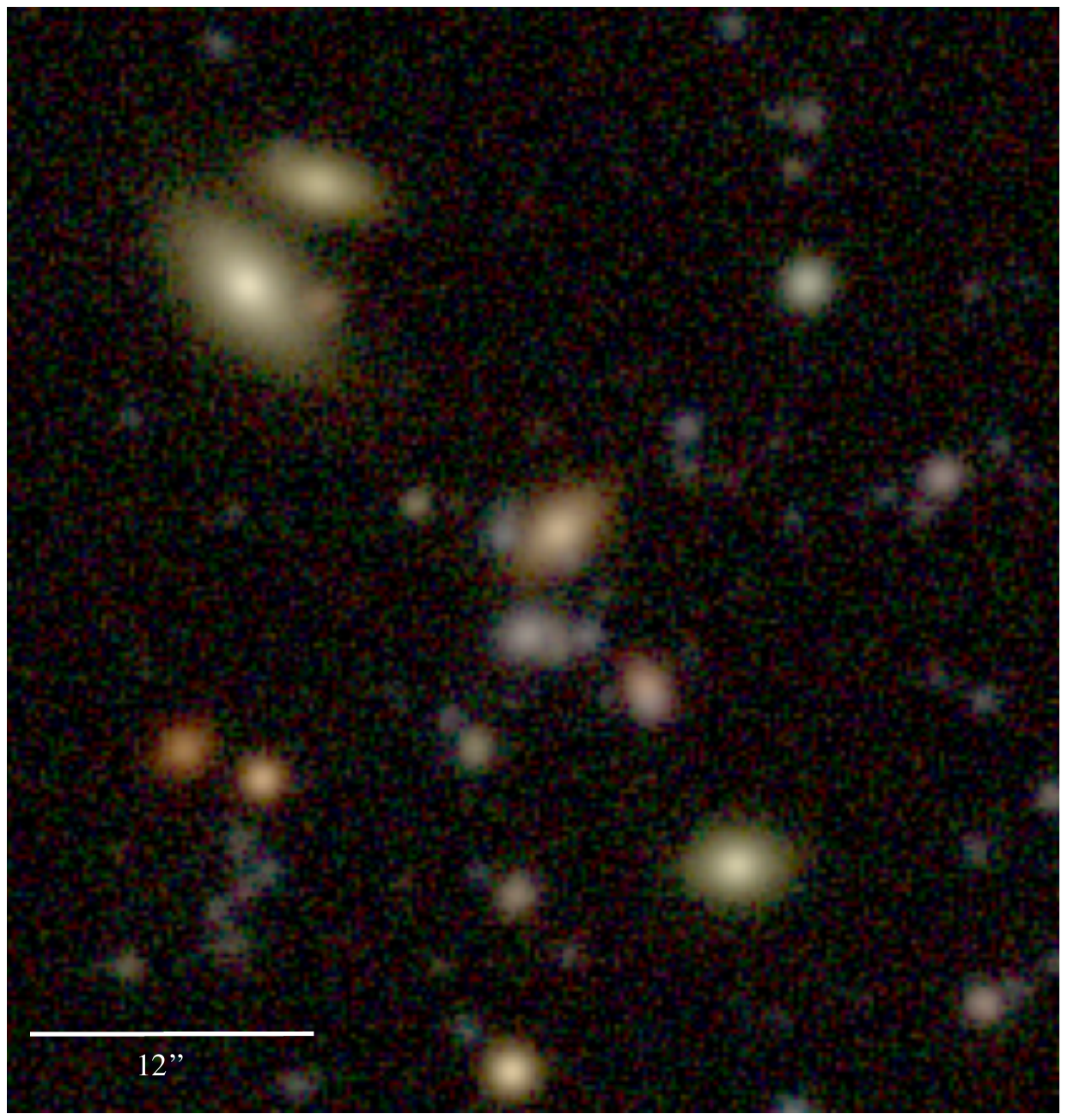}
\vspace{-1cm}
\caption{\label{fig:blend1}A 46 arcsec cutout from Fig.~\ref{fig:coaddimages} (Roman: left, Y106/J129/H158; LSST: right, g/r/i) that shows the power of combining LSST and Roman data to inform deblending the larger LSST data set. The Roman cutout both differentiates several blended systems in this region of the sky, but can also resolve very faint extended objects like the vertical red galaxy above the distance scale in the lower left. The difference in the coadd PSF is clearly visible in the star in the bottom central edge of the images.}
\end{figure*}

\begin{figure*}
\includegraphics[width=\textwidth]{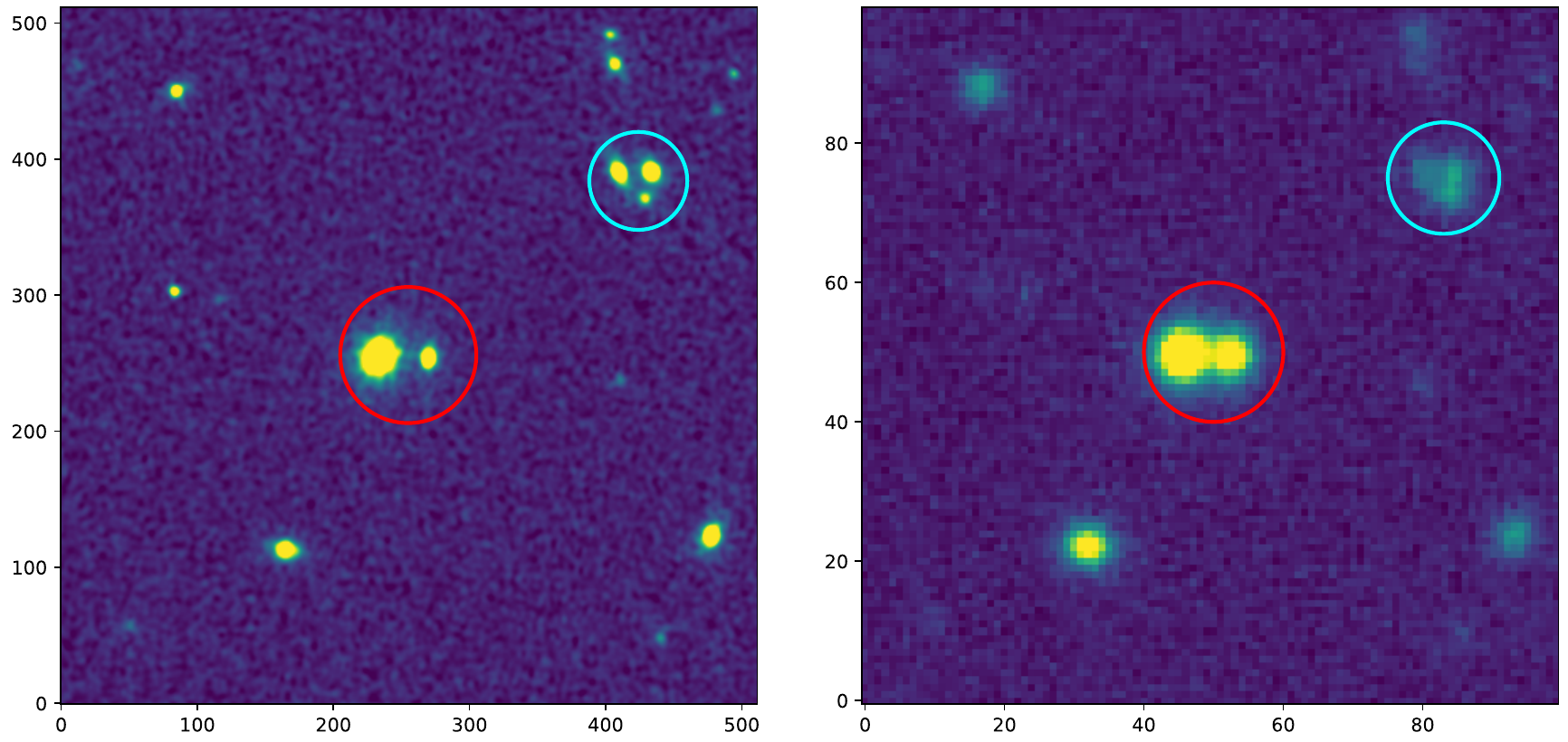}
\caption{\label{fig:blend2}A similar blending comparison as Fig.~\ref{fig:blend1}, with two blended systems indicated by circles, but monochromatic to emphasize photometric profile differences. Left: Roman H158. Right: LSST $r$. Each cutout is 20 arcmin on a side, and the pixel grid is indicated by the axes.}
\end{figure*}

The primary supported science goals for this set of simulations were intended to expand previous joint-survey simulations (e.g., \citealt{troxel23}) to better support exploring joint transient and other deep-field science. Thus, the largest component of the simulations is complete fields of the LSST ELAIS-S1 DDF and the Roman HLTDS. We include updated and expanded overlapping sections of wide-area surveys for each mission as well. The two surveys' complementarity is most easily seen in contrasts of wavelength coverage and resolution. We show the complementary wavelength coverage of the two surveys in Fig.~\ref{fig:bpass}. In Figs.~\ref{fig:blend1} \& \ref{fig:blend2}, we show example cutouts from Fig.~\ref{fig:coaddimages} that better demonstrate the difference in resolution and ability to identify (unrecognized) blends in the LSST imaging using Roman imaging. We describe each survey component as it is simulated in the following subsections.

\subsection{LSST Wide-Fast-Deep \& Deep-Drilling Field}

The Rubin Observatory LSST observation sequence is simulated to match the Baseline v3.2 results\footnote{\url{https://s3df.slac.stanford.edu/data/rubin/sim-data/sims_featureScheduler_runs3.2/}} from the Operations Simulator \citep{2014SPIE.9150E..15D,2016SPIE.9910E..13D,2016SPIE.9911E..25R}.\footnote{\url{https://github.com/lsst/rubin_sim}} These include observations spanning five of the ten-year LSST survey: 1) a region of the WFD survey overlapping the 70 deg$^2$ of the sky simulated for the {\tt Diffsky} galaxy catalog; and 2) the ELAIS-S1 DDF. This version of the LSST observation sequence implements a model of a ``rolling'' cadence for observations, which was not present in the LSST DESC DC2 simulations. This survey strategy prioritizes stripes of the sky for repeat observation in a ``rolling'' cadence over a several year period to boost transient discovery and light-curve sampling. However, this strategy results in challenges to recovering a homogeneous depth across the survey for static science, like weak lensing or galaxy clustering, which relies on comparing summary statistics to a statistical ``average'' model of the Universe. 

\begin{figure*}
\includegraphics[width=\textwidth]{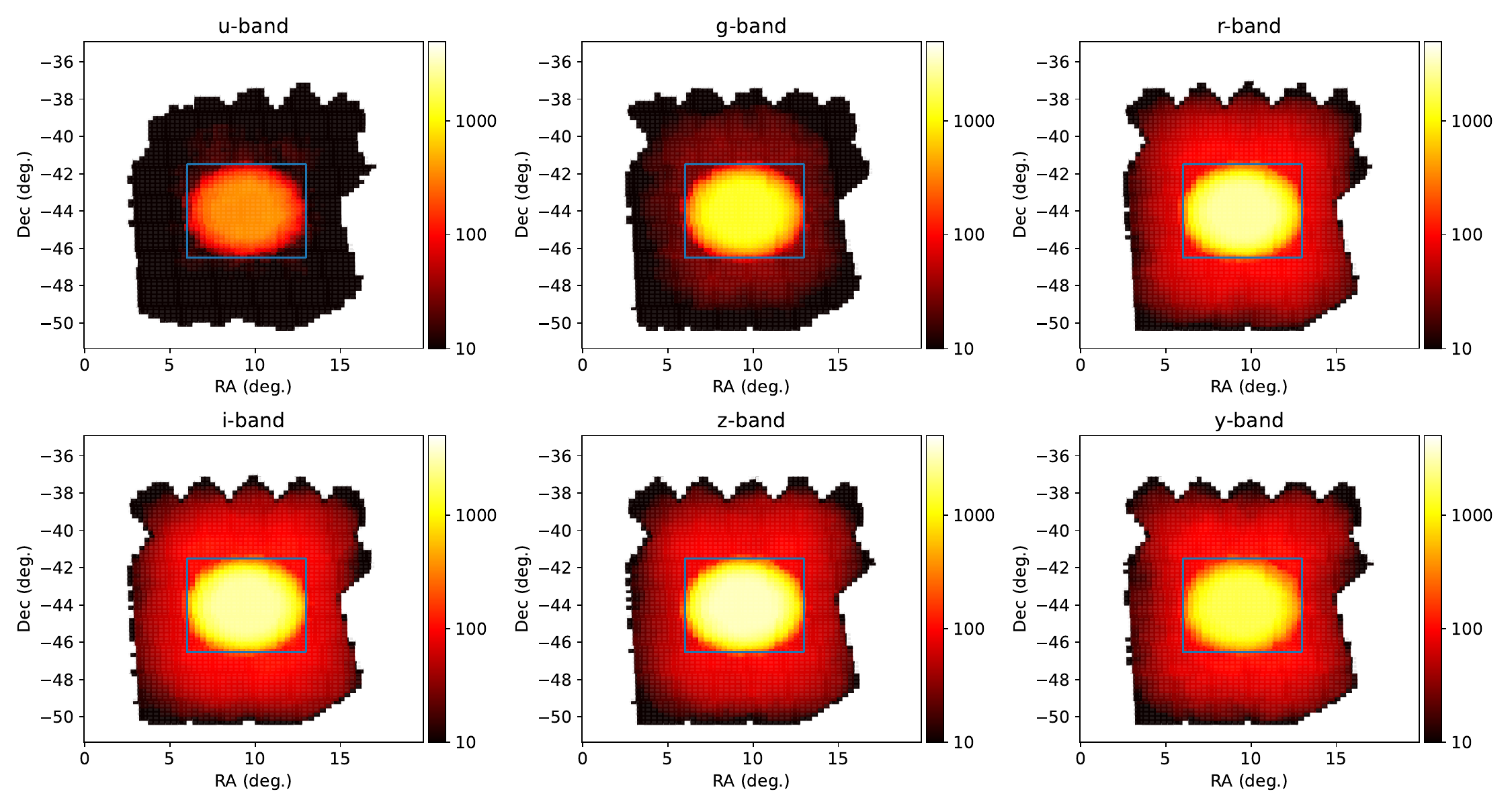}
\caption{\label{fig:lsstcoverage}Integrated exposure count maps for each bandpass of the simulated LSST survey surrounding the ELAIS-S1 DDF, including surrounding WFD exposures. The central box indicates the central 5$\times$5 deg$^2$ region of the DDF.}
\end{figure*}

Dithering of the pointings of the telescope and rotations are pseudo-random in the LSST observing sequences, unlike the Roman survey which has a regular scanning pattern. Since the specific pointings do not have strong structure in their position on the sky, we show in Fig.~\ref{fig:lsstcoverage} a simple exposure count map for each bandpass as a function of position on the sky for the simulations including both the central DDF region and overlapping WFD region.

\subsection{Roman High-Latitude Wide-Area Survey}

The Roman High-Latitude Wide-Area Survey (WAS) strategy simulated here is a subset of the Reference Survey described in \citet{troxel23}, Appendix A, with the addition of two filters (K213 and Wide). The actual survey strategy that will be used is being defined as part of the Roman Core Community Survey definition process, but the simulated strategy here and subsets thereof are representative of options that have been considered.

\begin{figure*}
\includegraphics[width=0.8\textwidth]{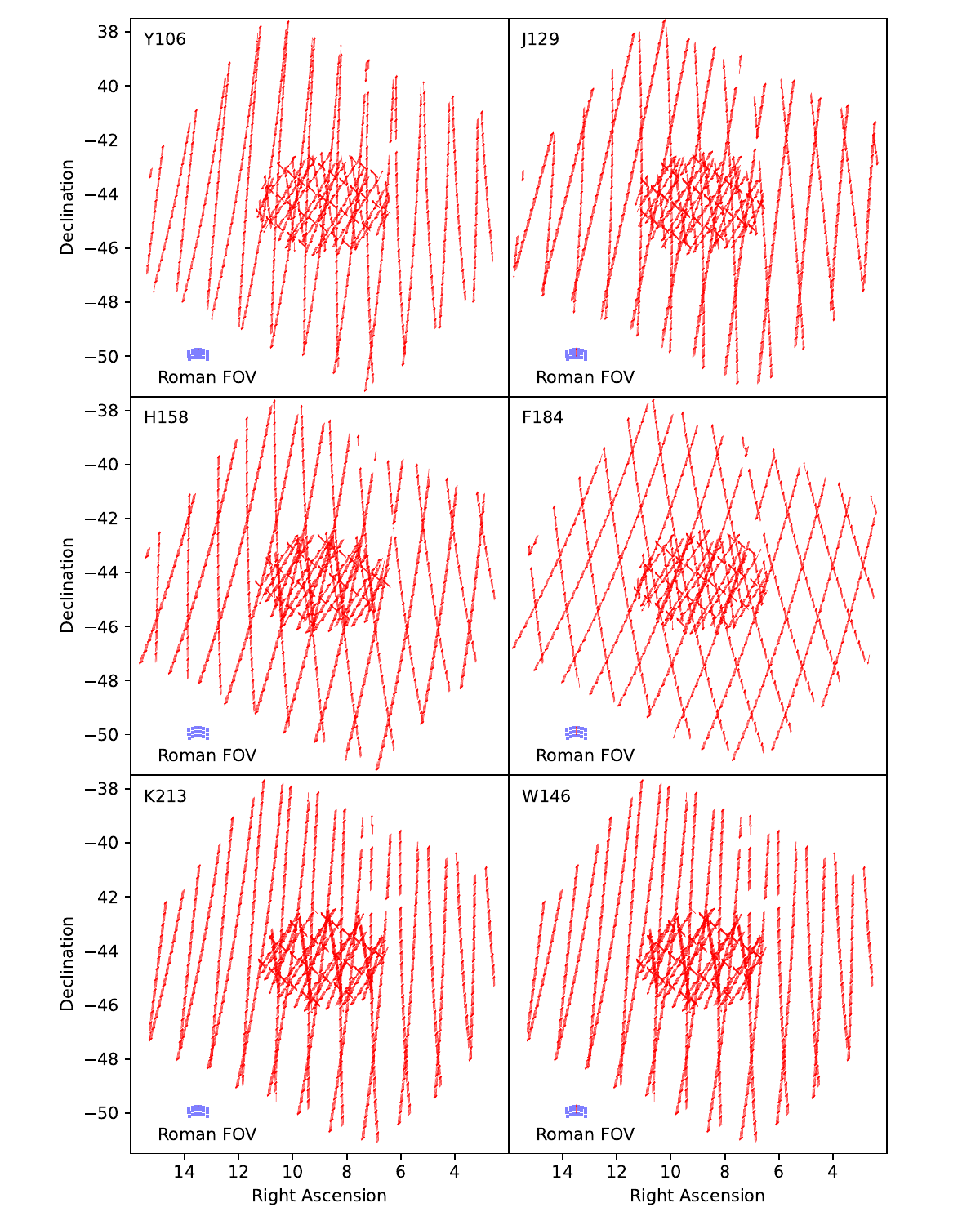}
\caption{\label{fig:romansurvey}Roman Wide-Area Survey observations simulated in this project. The six panels show the simulated region in each of the filters. Red arrows mark the positions and orientations of the observations, and are compared to scale within the Roman field-of-view (FOV) in the bottom-left inset in each panel. One can see the two wide-area passes (with slight offsets for dithering over chip gaps), as well as the deep region covered multiple times. The simulated survey includes gaps in survey coverage at the real-Universe bright stars $\alpha$ Phe (RA 6.6$^\circ$, Dec $-42.3^\circ$) and HD 2490 (RA $7.1^\circ$, Dec $-39.9^\circ$), where Roman would not observe in WAS mode, and thus pointings do not exist in the Reference survey design.}
\end{figure*}

The survey strategy is implemented as a nested ``{\tt for}'' loop. The innermost loop consists of several dither steps (3 or 4 positions) with small diagonal steps to cover the chip gaps. The next loop contains a $\approx 0.4$ degree step along the short axis of the Roman FOV so as to make a strip along the sky. These strips are then tiled to cover the 2D sky, and the outer {\tt for} loop contains 2 ``passes'' over each region of sky, at two different roll angles. The tiling strategy skips over bright stars. In addition, there are Deep Fields that are covered to greater depth by doing more than 2 passes. The $\approx 100$ deg$^2$ region considered in this simulation is centered on one of these Deep Fields. A graphical depiction of all of the pointings simulated is shown in Fig.~\ref{fig:romansurvey}.

The simulation schedules the Roman observations within the 5-year primary science mission period (MJD 61444.00--63270.25). Observations are scheduled when the fields are 54--126 degrees from the Sun, and with the direction to the Sun $120\pm15$ degrees clockwise from the field orientation (arrow in Fig.~\ref{fig:romansurvey}), as required by the Roman Field of Regard (see \citealt{troxel23}, Appendix A). The simulated region contains a total of 16,956 exposures, each of a duration of 140 seconds.

\subsection{Roman High-Latitude Time-Domain Survey}

The final Roman High-Latitude Time-Domain Survey (TDS) strategy is still being decided, however, several reference survey designs have been developed to study how well we can achieve mission requirements \citep{Hounsell2018,Rose2021,Hounsell2023}. In this simulation, we use a new variant on the TDS; one that would maximize the science goals of pre-survey simulated data rather than overall supernova cosmology figure-of-merit. 
Compared to the most recent reference survey \citep{Rose2021}, this simulation focuses on a large ``deep tier'' with the addition of K213 observations. The simulations are 12~deg$^2$ compared to the 4.2~deg$^2$ deep tier and 19~deg$^2$ wide tier. This simulation also has a sub-optimal square shape, rather than the more circular-like tiling patterns previously presented.

The observations are centered on the LSST ELAIS-S1 DDF, however, this field is i) outside the Roman continuous viewing zone and ii) has significantly higher zodiacal background than any proposed field \citep{Rose2021,Rose2023}. Thus, the real Roman TDS would not be carried out in this field. The choice of ELAIS-S1 was somewhat arbitrary, but is sufficiently far from the galactic plane and representative of the more ``typical'' LSST DDFs that overlapping observations from LSST and Roman are useful for pre-survey science and pipeline development. To compensate for (i) and (ii) we did two things. First we ignored solar proximity to observatory pointing assuming this field is always visible. Second, we simulated the expected zodiacal background of a high ecliptic field rather than the true zodiacal of ELAIS-S1. We use a fixed zodiacal background from RA~$=61.24$, decl~$=-48.42$ (Euclid Deep Field, South). 

\begin{figure*}
\includegraphics[width=0.7\textwidth]{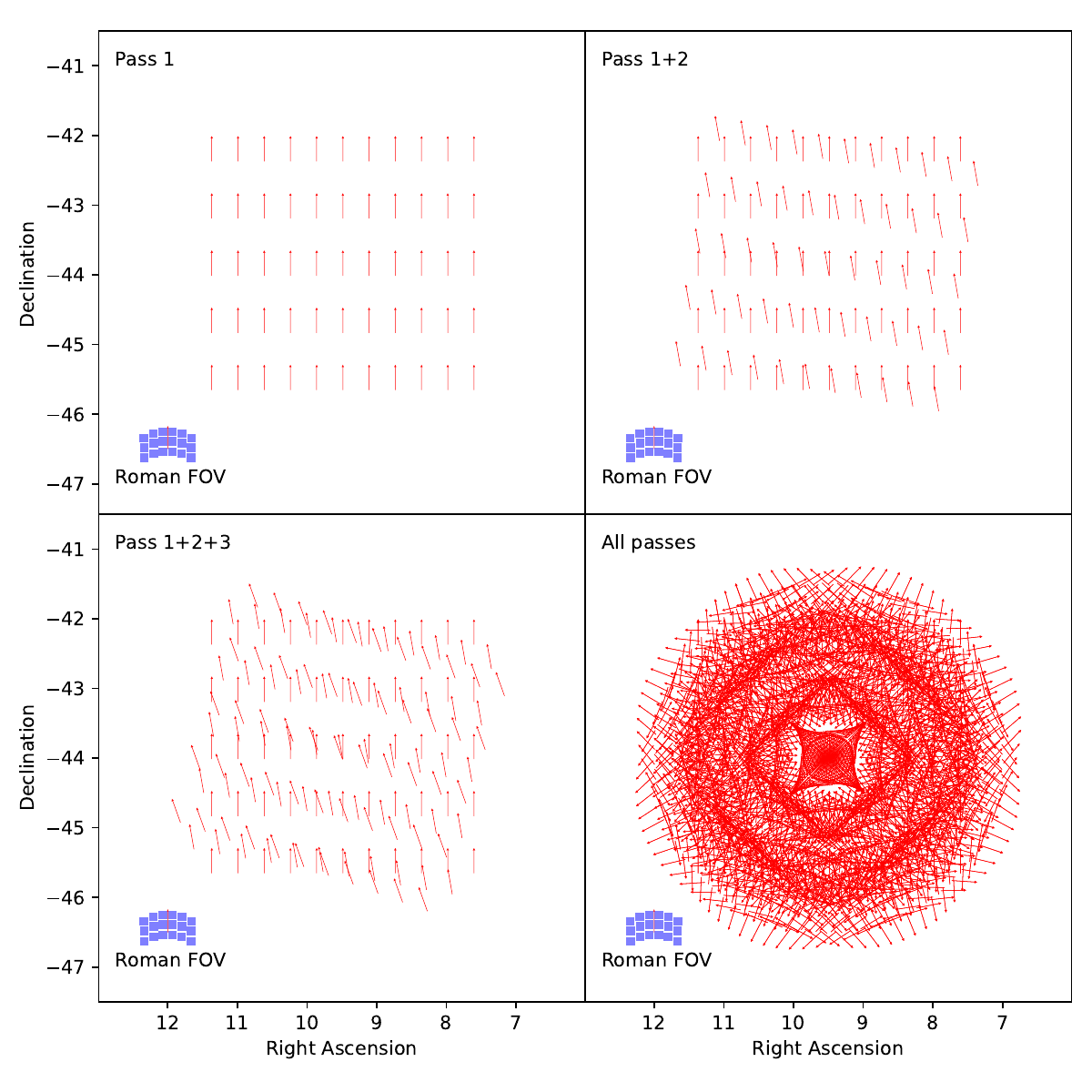}
\caption{\label{fig:romansurveytds}Roman Time-Domain Survey observations simulated in this project. Arrows mark the positions and orientations of the observations; one can see the two wide-area passes (with slight offsets for dithering over chip gaps), as well as the deep regions covered multiple times. The Roman field of view is shown in the inset. }
\end{figure*}

We simulate a single survey tier, with a combination of the previous wide and deep tiers that has an approx.~survey area of 12~deg$^2$. This is done with a 5-by-11 grid of pointings, which can be seen in the first panel of Fig.~\ref{fig:romansurveytds}.
Observations were taken every 5 days from MJD 62000.02--63563.06.
Each visit included a rotation of the field by 4.93~deg, to match the natural roll of the observatory. Due to a sign bug in interpreting the orientation of the observatory, this 4.93~deg rotation occurs opposite of the intended direction in the simulation. The build-up of these passes and the complete cycle are shown as they were simulated in the other panels of Fig.~\ref{fig:romansurveytds}. 

\begin{table}
\begin{center}
\caption{\label{tab:hltds-exp}Roman HLTDS Exposure Times}
\begin{tabular}{lc}
\hline
\hline
Filter & Exp. Time (s)\\ [0.2cm]
\hline
R062 & 161.025 \\
Z087 & 101.700 \\
Y106 & 302.275 \\
J129 & 302.275 \\
H158 & 302.275 \\
F184 & 901.175 \\
K213 & 901.175 \\
\hline
\end{tabular}
\end{center}
\end{table}

We observed in all 7 wide bands, which adds the K213 filter to the previous reference survey. We use the wide-tier exposure time for the R062 and Z087 filters. For the Y106, J129, H158, and F184 filters, we use the deep-tier exposure times. We also set the K213 filter to match the exposure of the F184 filter. These exposure times can be seen in Table~\ref{tab:hltds-exp}.

\section{Image Simulation Features}\label{sec:imagesims}

One of the main challenges of producing precisely matched simulations between telescopes or instruments, for which simulation packages typically have been (and in the case of LSST and Roman are) developed independently, is ensuring that objects are being simulated at all steps in the pipeline in exactly the same way. In the past, this has required immense work even when both simulation packages are based on the same underlying tool like GalSim.\footnote{\url{https://github.com/GalSim-developers/GalSim}} Most of this work comes in the form of: 1) reformatting input catalogs and 2) ensuring assumptions about the meaning of the input information are the same and that it is used in identical ways when modeling the objects in the images. For our previous joint simulation papers \cite{dc2a,dc2b,troxel23}, this was a process that spanned over several years. Many tools and packages have since been developed and updated to make this process much simpler, which turned a years-long process into only months for the current simulations described in this paper. 

The two main changes that are now uniformly utilized in both the LSST DESC and Roman image simulation packages used in these simulations, which are each described separately in the following subsections, include 1) utilizing the configuration framework for simulating images in GalSim \cite{2015A&C....10..121R} and 2) using a common framework for interpreting input information about astrophysical objects called SkyCatalogs.\footnote{\url{https://lsstdesc.org/skyCatalogs}} The SkyCatalogs format and associated API unify the representation and interpretation of input information, and within a configuration file for GalSim, can be used to construct object models in a self-consistent way across simulations. 

\subsection{LSST image simulation}
LSST images are simulated using the LSST DESC imSim framework,\footnote{\url{https://github.com/LSSTDESC/imSim}}
which is based on GalSim.  imSim has been developed to produce highly realistic simulated images for the LSST camera. In particular, a summary of the features implemented in the simulation is available as part of the imSim documentation.\footnote{\url{http://lsstdesc.org/imSim/features.html}} imSim was previously used to generate the LSST DESC DC2 image simulations \cite{dc2a,dc2b}, which were later used for the Rubin Observatory Data Preview 0 (DP0).\footnote{\url{https://dp0-2.lsst.io}} 

Since the LSST DESC DC2 simulation was completed, a major reorganisation of the code base has been carried out, and imSim has been refactored into a set of modules that are integrated into the GalSim configuration framework and incorporate simulation inputs from the SkyCatalog framework. Separate simulation modes for the LSST ComCam and LSSTCam have also been integrated. Updates to the realism of the simulation of LSST include: 1) optical ray tracing with the Batoid package,\footnote{\url{https://github.com/jmeyers314/batoid}} resulting in emergent vignetting and diffraction spikes; 2) sensor and optics throughputs; 3) airmass-dependent atmospheric transmission; 4) expanded controls of the telescope optical system that allow us to simulate the Active Optics System. 

\subsection{Roman image simulation}

Since \cite{troxel23}, the Roman image simulation package\footnote{\url{https://github.com/matroxel/roman_imsim/}} has been rewritten to utilize the GalSim configuration framework, which allows for a much simpler definition of simulation settings in a single yaml file. This makes reproducing and modifying the image simulation much simpler for users. Modules for interpreting a standardized observation sequence definition file and Roman instrument components (e.g., bandpases, WCS, PSF, and SCA models) have been updated within the GalSim configuration framework. Example configuration files for both survey modes are provided with the data release.

In addition, specific model details for the Roman WCS, PSF, and bandpasses have also been updated since \cite{troxel23} to match a snapshot of known parameters of the Roman telescope in November 2023, with specific details about sources of information provided in the \textsc{galsim.roman} documentation. The parameters of models of physical effects related to the conversion of photons to output counts in the Roman SCAs is unchanged from \cite{troxel23}, except for the additional inclusion of charge diffusion. The charge diffusion is modelled as a discrete photon operator applied to the image at the same stage as the PSF. The model is a fit to measurements using SCA 21536 in double-slit tests at the NASA GSFC Detector Characterization Laboratory (DCL) in October 2023. The measured modulation transfer function is best fit as a $\textit{sech}$, which we approximate as the sum of three 2D Gaussians that can then be analytically Fourier transformed:
\begin{equation}
\texttt{sech}(2\pi\sigma u) \approx \sum_{i=1}^3 w_i e^{-2\pi^2(c_i\sigma)^2 u^2}.
\end{equation}
For each component of this sum, the scaling of the Gaussian widths are $c_i=\{0.4522, 0.8050, 1.4329\}$, the relative weights $w_i=\{0.17519, 0.53146, 0.29335\}$, and $\sigma=0.3279$. 

\subsection{Known issues}

In this section we summarize some known issues associated with the simulation products, which are not possible to correct by regenerating the simulation due to the size of the simulated data products and associated computational cost. Rerunning even a significant portion of the time-domain surveys of either LSST or Roman is infeasible following the special allocation provided by ALCF. None of these issues invalidate the scientific utility of the simulations, but are features that users of the simulated data should be aware of.

In the generation of the LSST imaging:
\begin{enumerate}
\item The version of the LSST bandpasses (specifically v1.7 of the LSST throughputs) that was used to compute the fluxes for the non-variable objects, i.e., the stars and galaxies, differed from the version (v1.9) used to compute the fluxes from the transient objects.  Since the reference catalogs were built from the simulated stars, the transient object fluxes will have systematic offsets. These are largest in $i$ and $z$, and will be approx.~$\le -0.28$ in mag, but can be corrected by comparing the two effective bandpasses.
\item There was also a bug in how the normalization of the SEDs for each object was computed.  The SED normalizations are specified by the monochromatic magnitude at 500 nm, hereafter \texttt{mag\_norm}, in the rest frame of the object.  This allows the simulations to use a library of SEDs and scale those SEDs as desired via the single \texttt{mag\_norm} value.  However, because we used a very narrow bandpass at 500 nm instead of scaling the flux density at 500nm directly, the normalizations were sensitive to the local structure of the SEDs and resulted in systematic shifts in magnitude as large as $-0.4$ for a small fraction of objects.
\item We had intended each LSST exposure to have 30 second duration, but because of an error in interpreting the content of the Rubin observing cadence database, each LSST exposure was only 15 seconds long. This reduces the effective integrated exposure time of the simulated WFD survey by a factor of two from what was intended, and causes a difference in single visit depth of approx.~0.37 mag. 
\item Since very bright stars will saturate in the LSST CCDs and will take a lot of cputime to render, we decided to omit stars with fluxes $>3 \times 10^7$ photons per exposure.  Unfortunately, this cut was applied per exposure rather than at the catalog level for any band.  As a result, there are some bright stars that appear in exposures taken in, e.g., g-band, where the rendered flux is less than that limit, but do not appear in r- or i-band images covering the same sky location.  Since these stars would still be saturated anyway, they would be omitted from the final source or object catalogs.
\end{enumerate}

In the generation of the Roman imaging: 
\begin{enumerate}
\item The TDS observation sequence was implemented with a reflection in focal plane position angle, leading to an unintended rotation sequence between passes over the field. This is non-optimal for survey strategy, but otherwise does not impact the validity of any resulting images.
\item The Roman images were simulated with a GalSim bug in how the random seed for, e.g., drawing fluxes from a Poisson distribution was interpreted. This led to each object in an image using the same random seed, which means the resulting distribution of fluxes across objects in an image does not exactly follow a Poisson distribution and the mean flux is slightly biased from the truth. The mean bias in each image will be accounted for in any empirically derived zeropoint and further average out across images, but the skew in the noise distribution is something to be aware of for precise photometry applications. The standard deviation of the mean fractional bias in flux within a single image due to this bug is approx.~1--2$\times10^{-3}$, but can grow to a percent for the faintest objects.
\end{enumerate}

\section{Available Data Sets}
\label{sec:products}

\subsection{SkyCatalog truth tables}
\label{sec:object}

The truth information for the simulation suite is provided primarily in the form of a SkyCatalog. This is a set of compact Parquet and HDF5 files that contain all information about objects necessary to reproduce the simulated images. The files are intended to be interacted with via the SkyCatalogs Python module. More information about the table structure of SkyCatalog files, as well as simulation and catalog access, is outlined in Appendix~\ref{sec:access}.

\subsection{LSST image processing and catalogs}
\label{sec:processing}
The LSST imaging is processed using the LSST Science Pipelines code,\footnote{The version of the pipeline used in this work corresponds to weekly snapshot \textsc{w\_2024\_22}, which is close to version \textsc{v27\_0\_0}.} produced by the Rubin Data Management (DM) team. The overall processing steps are very similar to those performed for the DC2 data \citep{dc2a,dc2b} and more detailed descriptions have been covered in \cite{2018PASJ...70S...5B,2019ASPC..523..521B}. In addition, an online technical manual is maintained by the Rubin DM team (https://pipelines.lsst.io/). Therefore, we will only give brief descriptions of the processing steps for these Rubin simulations, focusing on those steps that relate to the available data products.

The visit level processing includes instrument signature removal, source detection, sky background and estimation, shape measurement, photometric and astrometric calibration, and initial source catalog generation.  The data products generated at this level are the "calibrated exposures" (also known as "calexps"), and the visit level source catalogs.

The next step is warp generation, i.e., resampling the PVIs onto a common pixel grid on the sky, and coadd generation. This grid is known as a "skymap" in Rubin and is defined in terms of "tracts" and "patches". A single world coordinate system (WCS) is used over a given tract and for the skymap used for the Rubin processing, each tract is divided into 10$\times$10 patches and each patch has a pixel scale of $0.2''$ (the same pixel scale as the LSST CCDs) and is 2000$\times$2000 pixels in size.  Available data products from these steps include the skymap and the coadds in each band.

After coadds are generated in all six LSST bands, multiband processing is performed.  This includes steps for source detection, deblending, source measurement (fluxes and shapes), and merging of the sources across the six bands to produce an object catalog.  After that step, we measure forced photometry at both the visit and coadd level for the objects.  The resulting data products are the final object catalogs with measurements in the six bands.

For difference imaging analysis (DIA), template images of the static sky are generated from the top one-third of visits that have the best seeing, and difference images are created for each visit.  Source detection, measurement, object association and forced photometry are then performed to generate the DIA object catalogs.

\subsection{Roman image processing and catalogs}
\label{sec:romanprocessing}

The primary Roman imaging is provided in a form that is close to what one would expect from a calibrated image. Versions of the imaging with a more complete Roman detector model, which would not be useful for science without an accurate calibration pipeline, can be produced upon request for small sections of the survey. The Roman component of the simulations may be updated in the future to include processing through the Roman Science Operations Center (SOC) pipeline\footnote{\url{https://github.com/spacetelescope/romancal}} when it is more mature. Roman data (and some of the OpenUniverse2024 simulated imaging accessible through the Roman Science Platform) will be provided in standard ASDF format. Since the OpenUniverse2024 simulated imaging is not currently compatible with the still under-development Roman pipeline, we provide the bulk imaging in standard FITS format for ease of use.

As Roman pixels are inherently Nyquist undersampled, an essential step in the image processing pipeline will be image coaddition: combining multiple undersampled images together to reconstruct fully sampled ones. The input images for coaddition contain differences including their pixel scales, permanent pixel masks, and PSFs. Image coaddition uses linear algebra techniques to homogenize the input image sample through transformations like resampling pixels onto a common grid, interpolating over masked regions, and homogenization of output PSFs onto a target PSF.

The most likely candidate software for this process is {\sc PyImcom} \citep{2024arXiv241005442C}, a Python-based implementation of Imcom \citep{2024MNRAS.528.2533H,2011ApJ...741...46R} developed to be both user-friendly and computationally efficient. Image combination occurs hierarchically; on the pixel scale, the $0.11''$ input pixels are transformed via linear combination into $0.039''$ output pixels. These pixels are then grouped into Postage Stamps ($32\times32$ output pixels), which are grouped into Blocks ($80\times80$ Postage Stamps), which make up full Mosaic images ($36\times36$ blocks). As an additional data product of the OpenUniverse2024 simulation suite, we have coadded a one deg$^2$ ($36\times36$ block) region of the Roman WAS in each filter except W146. The target PSF for this set of images was a Gaussian, with $\sigma_{\rm Gauss}$ ranging from $0.093 - 0.112$ for filters Y106 through K213. \citep{2024arXiv241011088L}.

\section{Example applications}
\label{sec:science}

Most of the OpenUniverse2024 simulation suite production resources and data volume are dedicated to simulating an overlapping LSST ELAIS-S1 DDF and Roman TDS to enable studying joint science applications related to transients, a use-case that was not supported in the previous set of joint simulations presented in \cite{troxel23}. OpenUniverse2024 has since spurred significant transient pipeline development, particularly within the Roman Alerts Promptly from Image Differencing (RAPID) and Roman Supernova Project Infrastructure Teams (PITs). The simulation of the Roman TDS is being extensively used to undertake end-to-end testing of the RAPID pipeline, including difference imaging algorithm selection, quantifying transient recovery statistics, and testing at full-scale before Roman launches. Similarly, difference imaging and photometry pipeline development has been ongoing in the Supernova PIT, including work to extend the simulations to include simulated Roman images using the Prism filter element. 

The simulations also include updated wide-field components for both LSST and Roman, and are much better suited than previous large LSST or Roman simulations for applications that require realistic optical and infrared galaxy colors. However, the limited area of the wide-field imaging in both simulated surveys, while a factor of 3-4 larger than the overlapping area in \cite{troxel23}, may still limit large-scale structure and weak lensing applications that require precise measurements of the two-point function or power spectrum. Detailed studies of sample selection, joint deblending, photometric redshift recovery, and other calibration applications are also ongoing. These and other specific pipeline development, science applications, and other studies using the OpenUniverse2024 simulations will be detailed in separate publications.

\begin{figure}
\includegraphics[width=0.5\columnwidth]{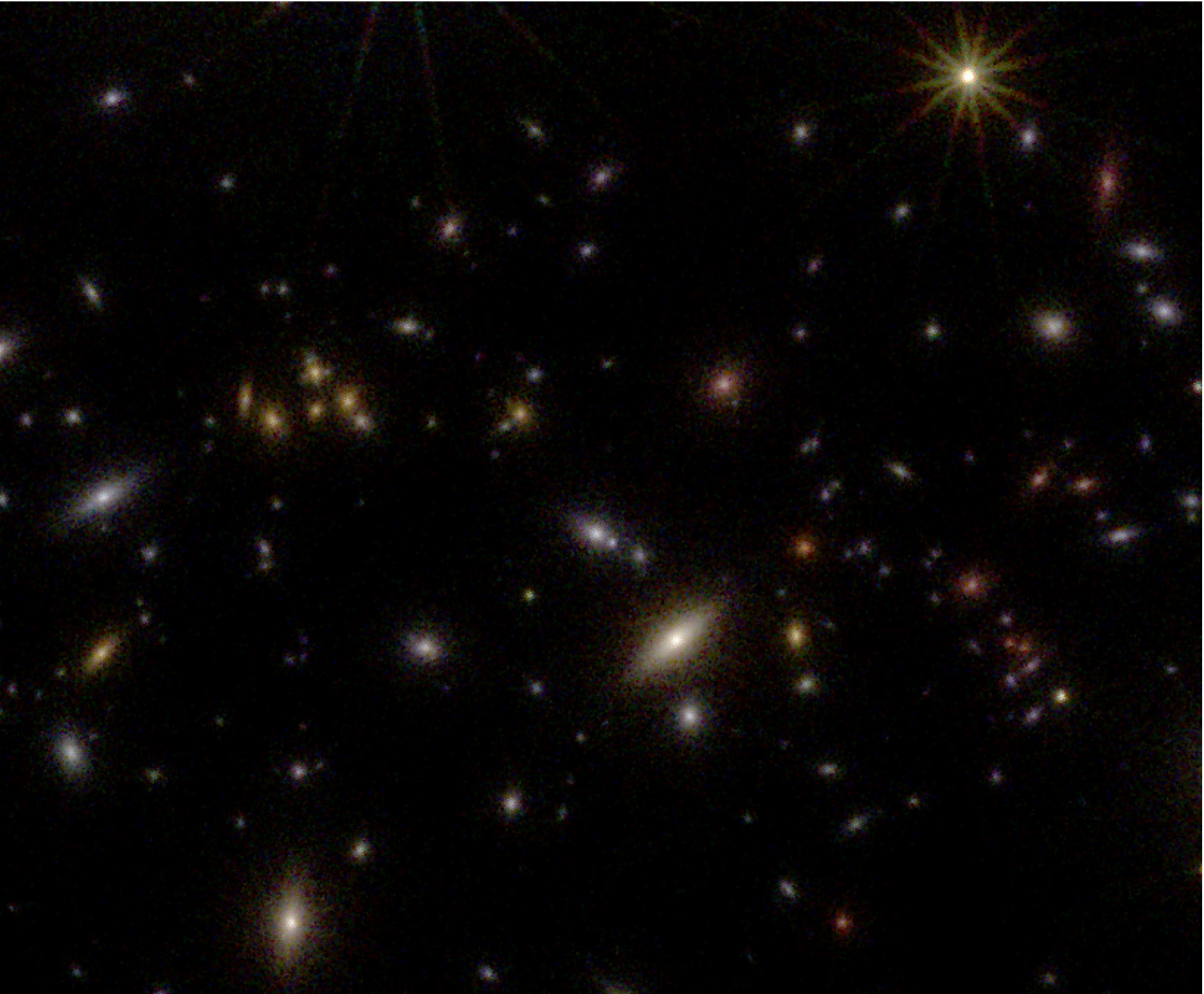}
\begin{overpic}[percent,width=0.5\columnwidth]{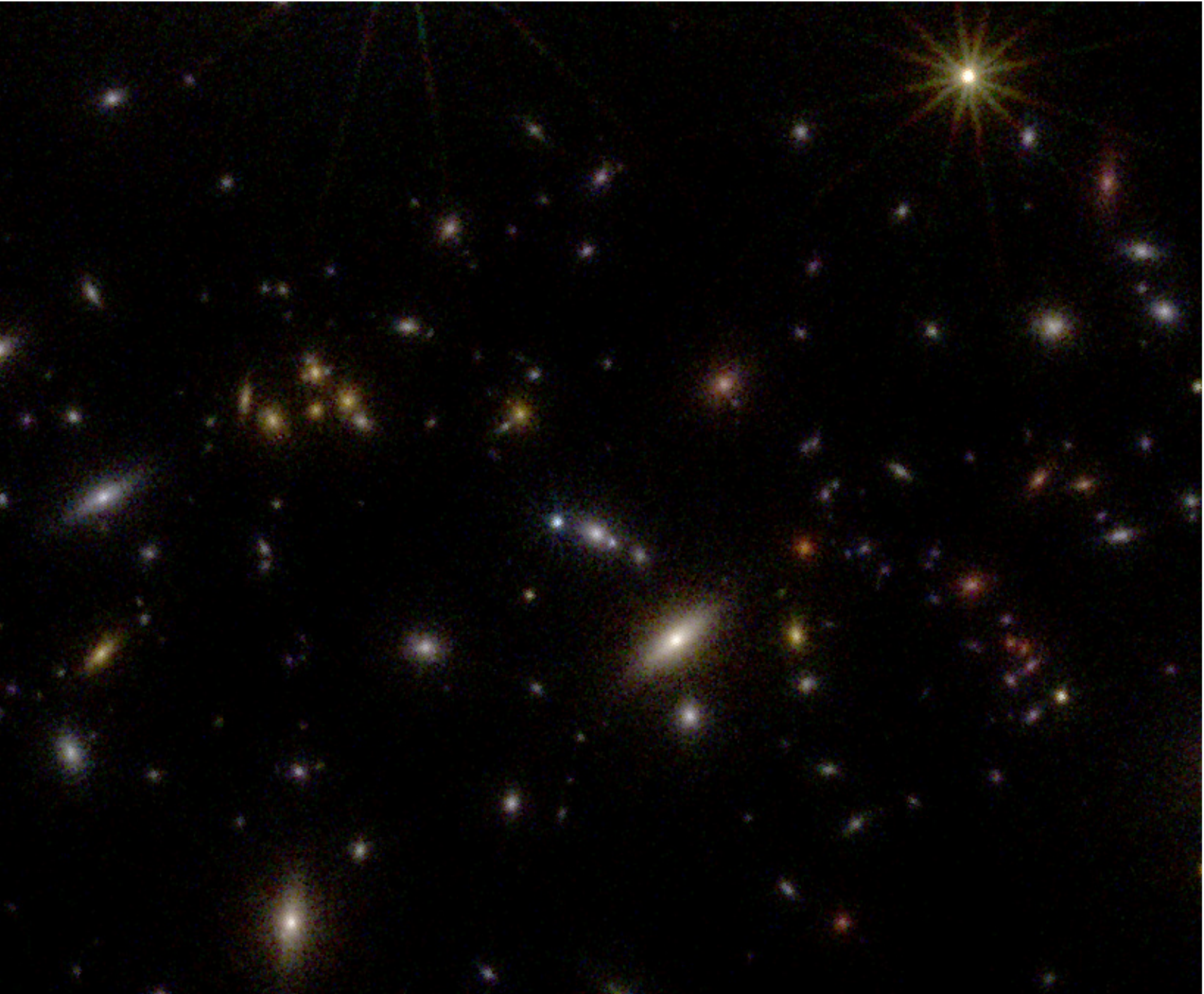} 
\linethickness{2pt}
\put(53,53){\color{red}\vector(-1,-2){6}}
\end{overpic}
\includegraphics[width=0.5\columnwidth]{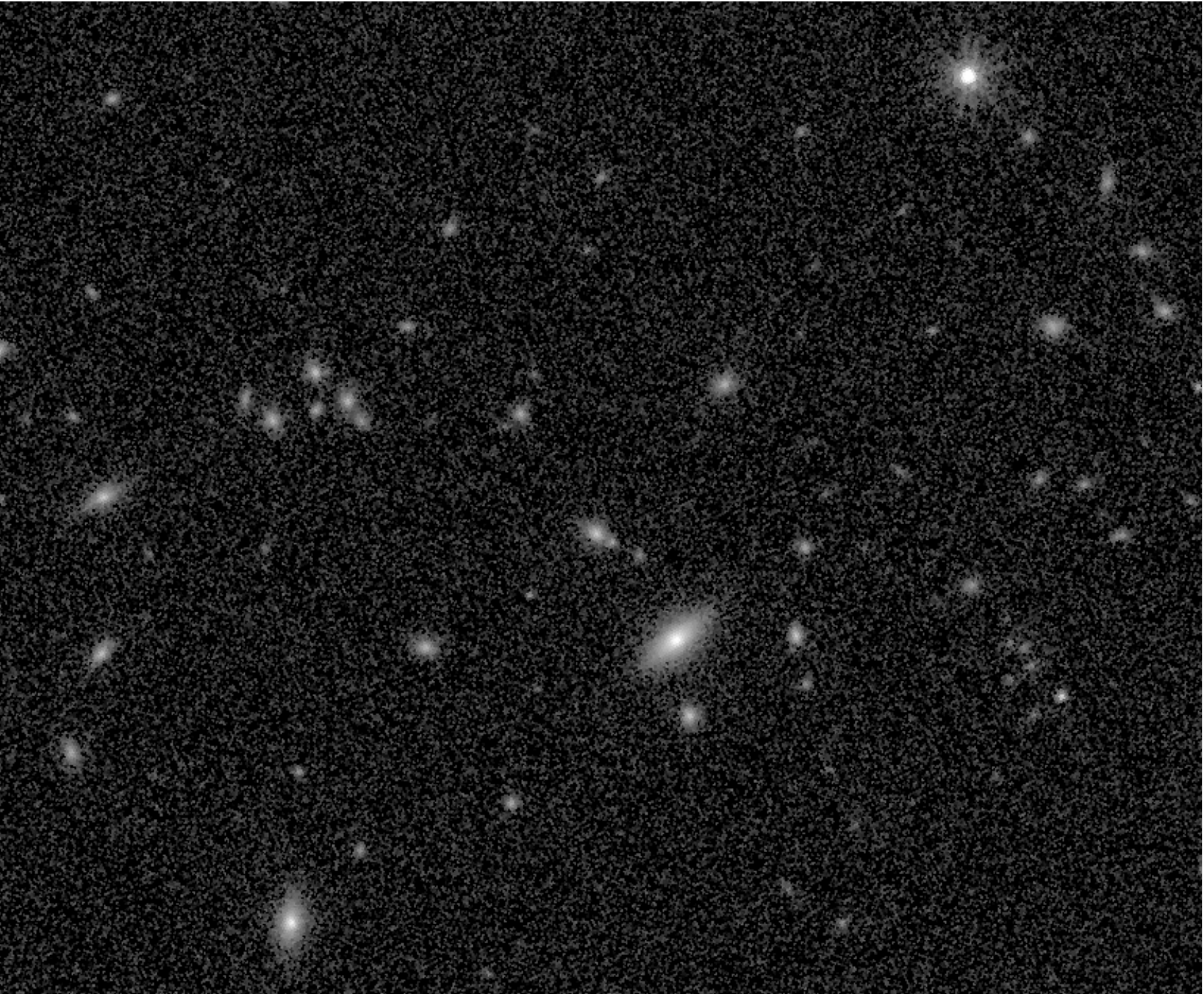}
\begin{overpic}[percent,width=0.5\columnwidth]{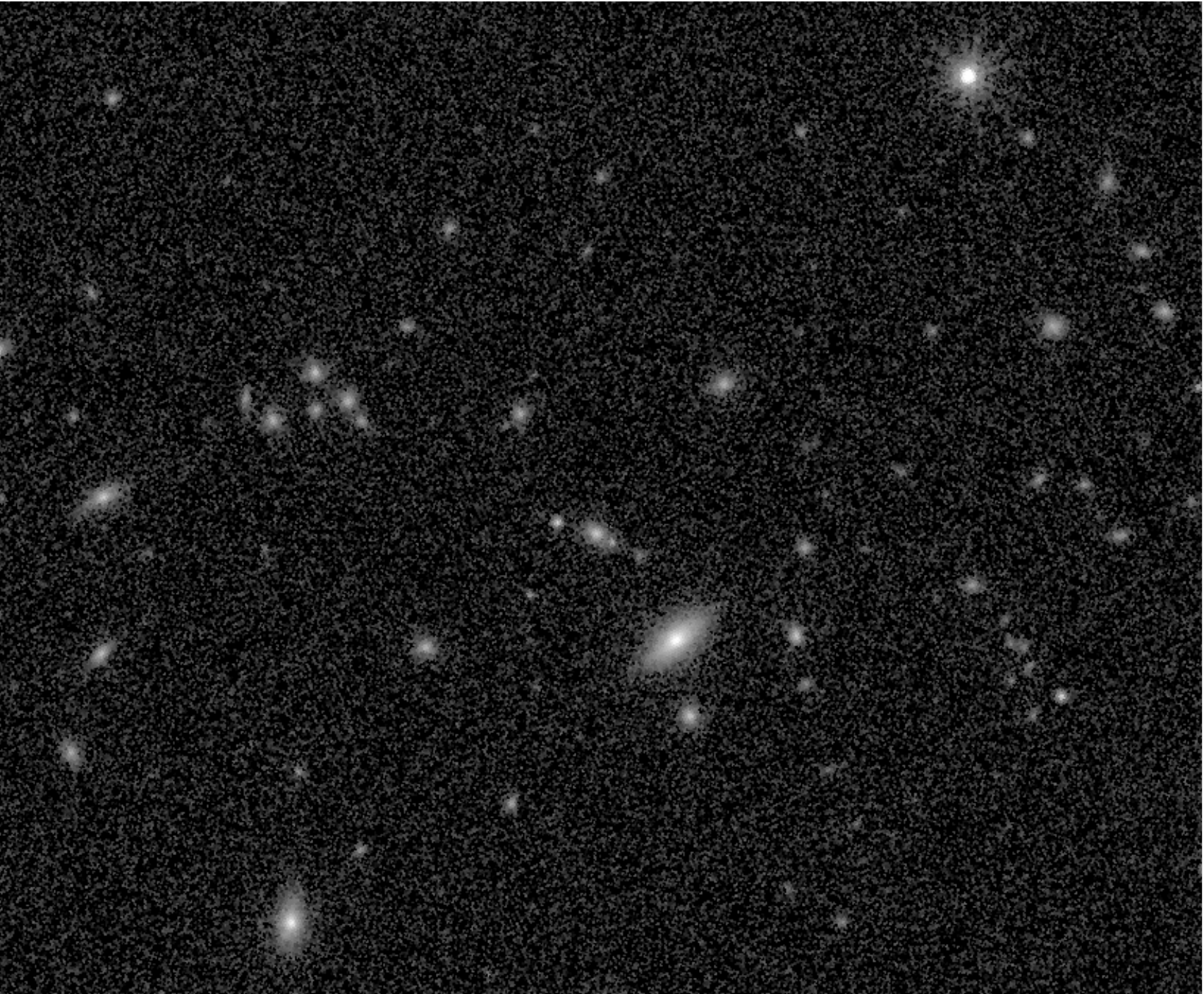} 
\linethickness{2pt}
\put(53,53){\color{red}\vector(-1,-2){6}}
\end{overpic}
\caption{Roman image cutouts around transient 20172782 at (RA, Dec) = (7.551093$^\circ$, $-$44.807181$^\circ$) showing two epochs with and without the SN Ia explosion. Each panel cutout is 71.4 arcsec wide. \textit{Left:} image scene before explosion, near MJD 62106. \textit{Right:} image scene around the time of peak brightness of the explosion, near MJD 62471. \textit{Top:} R062/Y106/H158 color cutouts built from overlapping images taken near the same MJD. Colour images show the scene built from the `true' images without backgrounds or detector effects to highlight the true models. \textit{Bottom:} H158 single-epoch cutouts built from the ``calibrated'' images. }
\label{fig:sn}
\end{figure}

\begin{figure*}
\includegraphics[width=\textwidth]{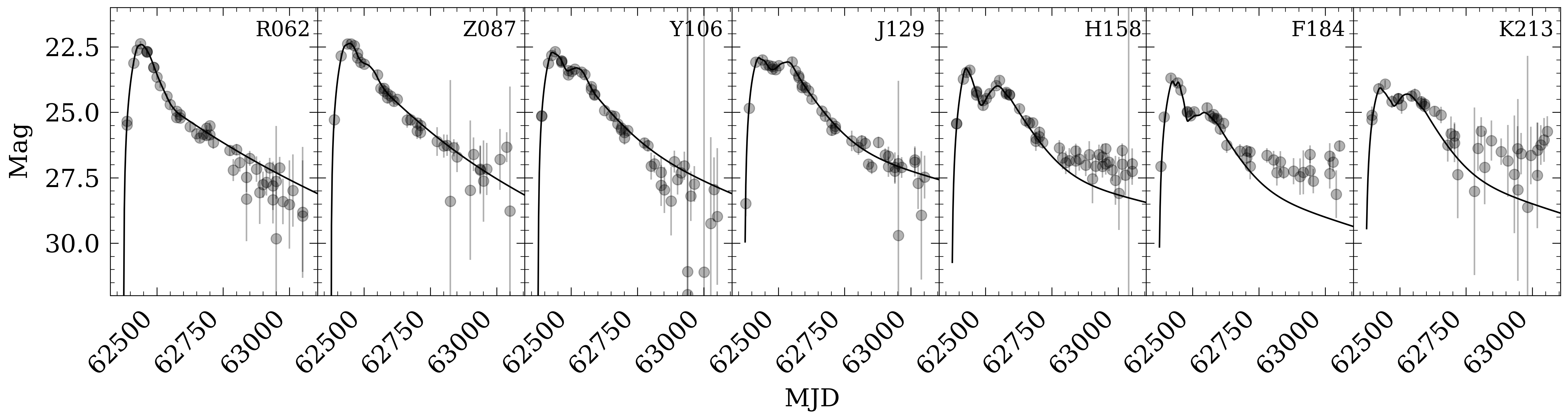}
\caption{Light curves for transient 20172782. The bandpass is indicated in the upper right of each panel. The solid line is the true light curve simulated by SNANA. The points are preliminary photometric measurements from single-epoch image subtractions using a proto-pipeline from the Roman SN PIT.}
\label{fig:lc}
\end{figure*}

\subsection{Transient identification and photometry}

A key use case of the OpenUniverse2024 simulations is to prepare the transient identification and photometry pipeline.  Transient pipelines must be ready as soon as repeat observations are available so that the survey can begin discovering transients with the start of operations; additionally, the simulations are critical for diagnosing various characteristics of the observations that may inform observing strategy.  

One initial discovery and photometry pipeline for Roman, called \textit{phrosty}\footnote{\url{https://github.com/Roman-Supernova-PIT/phrosty}} is being developed, particularly for the use case of measuring the photometry of the light curves of SNe Ia.  This pipeline will be presented in Aldoroty et al. in prep.  The pipeline uses the Saccadic Fast Four Transform (SFFT) method developed in \cite{Hu2022} for image subtraction and the SExtractor software \cite{Bertin1996} for detection.  The pipeline identifies `science' images in which the SN light can be detected, and pre-explosion `template' images that can be used for subtraction.  The pipeline can apply both aperture and PSF photometry, and the PSF photometry \citep{Bradley2016} can be performed using a previously determined PSF or empirically from stars in the image. 

Example cutouts of template and science images in which SN 20172782 is found are shown in Figs.~\ref{fig:sn}.  The cutouts show the color-combined images with R062/Y106/H158 using the `true' images without backgrounds or detector effects as well as the calibrated images from H158 single-epoch cutouts.  SN 20172782 is a bright SN Ia at $z=0.3$, and well-separated, so can be used  
for early diagnostic tests of the detection and photometry pipeline.

We show initial light-curve results of the pipeline run at a forced position of SN 20172782 in Fig.~\ref{fig:lc}. The points in the diagram show the recovered photometry and the curve is from the SN Ia model light curve injected as a time series on the images.  Overall, the recovered photometry is consistent to $1\%$ with the true photometry.  Deviations in the light curve seen later are due to single templates being used currently for the pipeline, and there are small flux offsets that propagate to differences in magnitudes at the faint end.  These offsets are not found when averaging over multiple template images.  There are no errors found yet in the point sources of the simulated photometry and current studies already show validation at the 5 mmag level.  Current best precision comes from aperture instead of PSF photometry due to limits of proper PSF fitting for the undersampled images.

\begin{figure}
    \centering 
    \includegraphics[width=0.47\textwidth]{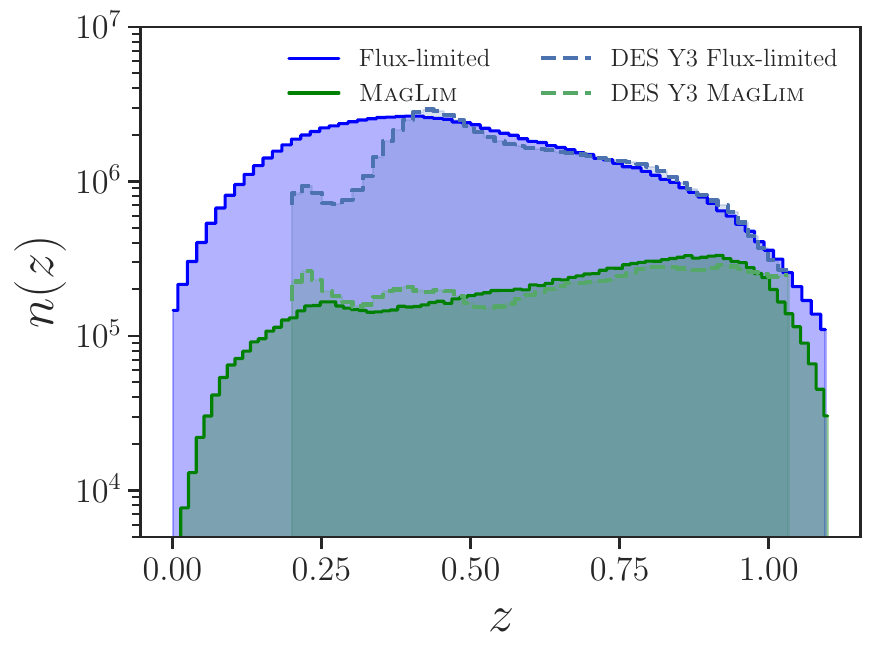}
    \caption{The galaxy number counts as a function of photometric redshift are shown for both the samples. The number count obtained here is re-normalized to match the area of DES \(\sim 5000\ \text{deg}^2\). The solid and dashed lines correspond to the lens samples obtained using the galaxy catalog from the OpenUniverse2024 simulation in this study, and the lens samples obtained by \citet{Porredon_2021} using the DES Year 3 catalog.}
    \label{fig:lens_samples}
\end{figure}

\subsection{Lens sample selection}

One of the aims of LSST DESC is to perform a cosmological analysis by studying cosmic shear in conjunction with galaxy-galaxy lensing and galaxy clustering, known as $3\times2$ point analysis. The inferred cosmological result is sensitive to the choices of the lens catalogs as it can directly impact the galaxy-galaxy lensing and galaxy clustering signals. 
There is a trade-off between selecting the largest galaxy samples which help reduce shot noise and choosing samples with high redshift accuracy, typically involving a smaller subset of galaxies. Balancing these factors is essential for an effective analysis. Hence it is important to find optimum lens samples for better cosmological constraints. OpenUniverse2024 enables the exploration of different lens samples with realistic colors for LSST DESC applications, in particular the selection of lens samples from both LSST optical and Roman near-infrared photometry due to substantially improved realism in colors across optical and infrared wavelengths relative to LSST DESC DC2.  In this section, we demonstrate that a preliminary study of lens selection produces results consistent with previous literature \cite{Porredon_2021}. 

We explore two lens samples based on the magnitude cuts in \textit{i}-band as a function of (photometric) redshift using the methodology described in \cite{Porredon_2021}. This approach is typically used because the \textit{i}-band provides the best signal-to-noise ratio per object across the relevant redshift range. Flux-limited samples are characterized by a constant apparent magnitude threshold in the \textit{i}-band, specifically \( i < a \), where \( a \) is a constant. 
One potential drawback of selecting all galaxies up to a fixed limiting magnitude is that, at low redshift, this approach tends to include a larger number of less luminous (primarily blue) galaxies, which can degrade the accuracy of photo-\textit{z} estimates. This can be alleviated by selecting samples known as $\textsc{MagLim}$ with a limiting magnitude that varies with redshift, expressed as \( i < az_{\text{photo}} + b \), where \( a \) and \( b \) are arbitrary constants and \( z_{\text{photo}} \) represents the photo-\textit{z} estimated via the introduction of a Gaussian photo-\textit{z} scatter defined by \( \sigma_z(1+z_\text{true})\) to the corresponding true redshift values. We have chosen  \(\sigma_z = 0.03, \text{and}\ 0.05\) for $\textsc{MagLim}$ and flux-limited samples, respectively. 
This method effectively prioritizes brighter galaxies at low redshift while allowing for the inclusion of fainter galaxies as redshift increases. Additionally, we exclude the brightest objects by applying a cutoff of \( i > 17.5 \).

Following \cite{Porredon_2021}, we choose flux-limited and $\textsc{MagLim}$ samples with the definition \(i<22.2\) and \(i<18+4z_\text{photo}\), respectively. They used the Dark Energy Survey (DES) imaging survey which covers \(\sim 5000\ \text{deg}^2\) of the southern sky. The number count of galaxies in the lens samples obtained here is re-normalized to match the area of DES. The galaxy counts as a function of photometric redshift are shown in Fig.~\ref{fig:lens_samples} for both samples. The solid and dashed lines correspond to the lens samples obtained using the galaxy catalog from the OpenUniverse2024 LSST WFD simulation, and the lens samples obtained by \cite{Porredon_2021} using the DES Year 3 catalog. We find a close agreement in the number counts of galaxies at $z_{\rm{true}}<1$ in Fig.~\ref{fig:lens_samples}, which also serves as a high-level validation of the realism of the galaxies being simulated in OpenUniverse2024. 

In the future, we plan to investigate different choices of lens samples for $3\times 2$ point analysis for the LSST DESC science applications. This will include an optimization in terms of number density and photometric redshift errors. Apart from these flux-limited lens samples, we can also create lens samples with accurate redshift estimates of the luminous red galaxies using the 
$\textsc{redMaGiC}$ algorithm \citep{Rozo_2016}. We will address all of this in future studies dedicated to the lens samples and their impact on cosmological inference.

\section{Summary and Outlook}
\label{sec:outlook}

Upcoming wide-area photometric surveys like the Rubin Observatory LSST and Roman Space Telescope High-Latitude surveys will enable us to view the Universe like never before. The complementarity of these ground- and space-based missions allows for an even wider and better-calibrated range of science than each mission individually. The OpenUniverse project provides simulation resources to maximize scientific discovery jointly among these new surveys. The OpenUniverse2024 simulation suite described in this paper provides a glimpse into what this new world of astrophysics and cosmology will look like by providing hundreds of terabytes of synthetic overlapping LSST and Roman survey imaging. 

We have produced simulated imaging for approx.~70 deg$^2$ of the Rubin Observatory LSST Wide-Fast-Deep survey and the Roman Space Telescope High-Latitude Wide-Area Survey, as well as overlapping versions of the ELAIS-S1 Deep-Drilling Field for LSST and the High-Latitude Time-Domain Survey for Roman. This simulation suite was made uniquely possible by: i) a special ALCF Discretionary award using all of the Theta cluster before the system was retired, and ii) the combined resources and technical talent among the participating collaborations and teams. 

The resulting simulations and their inputs are documented in this paper, which include: i) an early version of the updated extragalactic model called {\tt Diffsky}, which substantially improves the realism of optical and infrared photometry of objects; ii) updated transient models that extend through the wavelength range probed by Roman and Rubin; and iii) improved survey, telescope, and instrument realism. The tools that we have developed to enable efficient and accurate joint simulations of multiple surveys are also described. Both the simulated data products and the code packages and tools used to create them are available publicly with this paper to maximize the science impact of the simulations in the broader community. 

The simulations have been instrumental in providing updated realistic synthetic data for use in the participating experiments, particularly in driving forward early pipeline development for several of the Roman Project Infrastructure Teams. We anticipate a range of scientific publications will follow from these teams utilizing these simulated surveys. Some examples of the validation work undertaken to characterize the simulations and potential science use cases have been described and we have documented issues uncovered during this period of early discovery.

Continued development of joint simulation tools and ongoing work using the current OpenUniverse2024 simulations to study the joint science gain from these surveys, test the science pipelines for LSST and Roman, and develop strategies to maximize science return from the surveys will continue within OpenUniverse and the collaborating teams. We hope to collaborate with more teams to expand the overlapping simulated survey data available to the community -- in particular, overlapping Euclid-like VIS survey imaging is already in advanced development, and will also be released in 2025

The incredible statistical power these next generation of surveys will enable us to harness brings with it an incredible challenge to control data calibration and systematics. These highly realistic simulations of LSST and Roman data enable us to continue to work on addressing these challenges now, both individually and jointly using both these surveys' data together, to support maximizing science potential once the surveys begin.

\section*{Acknowledgments}
\phantomsection
\addcontentsline{toc}{section}{Acknowledgments}

This work was supported in part by the OpenUniverse effort, which is funded by NASA under JPL Contract Task 70-711320, "Maximizing Science Exploitation of Simulated Cosmological Survey Data Across Surveys". 

Argonne National Laboratory's work was supported under the U.S. Department of Energy contract DE-AC02-06CH11357. 

The DESC acknowledges ongoing support from the Institut National de Physique Nucl\'eaire et de Physique des Particules in France; the Science \& Technology Facilities Council in the United Kingdom; and the Department of Energy, the National Science Foundation, and the LSST Corporation in the United States.  DESC uses resources of the IN2P3 Computing Center (CC-IN2P3--Lyon/Villeurbanne - France) funded by the Centre National de la Recherche Scientifique; the National Energy Research Scientific Computing Center, a DOE Office of Science User Facility supported by the Office of Science of the U.S.\ Department of Energy under Contract No.\ DE-AC02-05CH11231; STFC DiRAC HPC Facilities, funded by UK BIS National E-infrastructure capital grants; and the UK particle physics grid, supported by the GridPP Collaboration.  This work was performed in part under DOE Contract DE-AC02-76SF00515.

Work in the Roman HLIS PIT is supported by NASA grant 22-ROMAN11-0011, "Maximizing Cosmological Science with the Roman High Latitude Imaging Survey." 

Work in the Roman SN PIT is supported by NASA under award number 80NSSC24M0023, "A Roman PIT to Support Cosmological Measurements with Type Ia Supernovae". 

Work in the RAPID PIT is supported by NASA under the award number 80NSSC24M0020, "RAPID: Roman Alerts Promptly from Image Differencing". 

The SLAC National Accelerator Laboratory, operated under DOE Contract DE-AC02-76SF00515, is the host laboratory for DESC and manages support for the DESC pipeline scientist and computing infrastructure teams.

This research used resources of the Argonne Leadership Computing Facility, which is a DOE Office of Science User Facility supported under Contract DE-AC02-06CH11357. This work was done through a special ALCF Discretionary award using all of Theta before the system was retired. This research used resources of the National Energy Research Scientific Computing Center (NERSC), a Department of Energy Office of Science User Facility using NERSC award ERCAP0026324. This research used resources at the Duke Compute Cluster.

This paper makes use of software developed for Vera C. Rubin Observatory. We thank the Rubin Observatory for making their code available as free software at \url{http://pipelines.lsst.io/}.

The work of A.K. was carried out at the Jet Propulsion Laboratory,
California Institute of Technology, under a contract with
NASA. L.G. acknowledges financial support from AGAUR, CSIC, MCIN and AEI 10.13039/501100011033 under projects PID2023-151307NB-I00, PIE 20215AT016, CEX2020-001058-M, ILINK23001, COOPB2304, and 2021-SGR-01270. The material is based upon work supported by NASA under award number 80GSFC24M0006.

This paper has undergone internal review by the LSST Dark Energy Science Collaboration. The internal reviewers were Arun Kannawadi and Chris Walter.

Author contributions follow, in the alphabetical order of the author list:
A.A.~led development and calibration of the SFH model. 
L.A.~contributed to validation of transients and Roman images. 
G.B-M.~led calibration of the Diffsky model. 
A.B.~contributed to lens sample selection and paper writing. 
J.B.~contributed to development of extragalactic catalogs. 
J.B.~led development of the SkyCatalogs software used by the image simulations. 
G.B.~implemented the optical raytracing component of the LSST imSim image simulation. 
A.B.~contribued to generation of calibration products for the ISR steps (ptc/brighter-fatter). 
K.C.~led development of the {\sc PyImcom} software used for IMCOM coadd production and contributed to paper writing. 
J.C.~led simulation of LSST images and paper writing. 
E.C.~contributed to lens sample selection. 
V.D.~coordinated public archiving of the simulation. 
Y.F.~contributed to Roman image simulation validation. 
L.G.~reviewed and provided comments on paper. 
A.H.~led development of Diffsky model. 
K.H.~led development of the Outer Rim simulation and PI of ALCF Discretionary allocation.  
C.H.~led IMCOM coadd production and paper writing. 
R.H.~contributed to proposal preparation and TDS design. 
B.J.~provided comments on the paper.
M.J.~contributed to LSST and Roman image simulations and validation. 
J.J.~contributed to transient modelling and validation. 
A.K.~reviewed the paper and contributed to Roman modules in GalSim. 
M.K.~contributed to transient modelling and validation. 
R.K.~led transient simulation and paper writing. 
A.K.~is PI of the OpenUniverse team. 
R.K.~contributed to transient simulation and paper writing. 
E.K.~contributed to the development and production of the extra-galactic catalog that underlies the image simulations, and paper writing. 
R.L.~contributed to transient modelling and validation. 
K.L.~contributed the Roman observing sequence and paper writing. 
C.L.~contributed to Roman image simulation development and validation. 
I.L.~contributed to the lens sample selection. 
E.M.~developed the charge diffusion model for Roman.
A.M.~contributed to transient modelling and validation. 
R.M.~contributed to Roman image simulation development and validation. 
J.M.~contributed to transient modelling and validation. 
S.M.~contributed to the integration of DSPS SEDs with GalSim and to validation of the extragalactic catalog. 
C.M.~contributed to Roman image simulation validation. 
J.M.~implemented the optical raytracing component of the LSST imSim image simulation. 
B.M.~contributed to the lens sample selection. 
R.P.~contributed to transient modelling and validation. 
A.P.~led development of the optimization algorithms used to calibrate Diffsky.
A.PM.~contributed to the LSST Science Pipelines code and algorithms used in this work.
B.R.~worked on designing the Roman HLTDS observation sequence and paper writing. 
D.R.~contributed to Roman TDS survey definition and paper writing. 
B.R.~contributed to transient modelling and validation. 
A.S.~contributed to the lens sample selection. 
N.S.~contributed to the lens sample selection. 
D.S.~contributed to validation of transient injection and paper writing. 
J.S.~contributed to creating blending figures in paper. 
M.T.~led the OpenUniverse2024 project and paper writing. 
N.VA.~contributed to development of extragalactic catalogs. 
S.VD.~contributed to transient modelling and validation. 
C.W.~coordinated addition of needed features and inputs to the DESC imSim simulation package and reviewed the paper.  
T.W.~contributed to validation of Roman simulation output. 
M.Y.~contributed to validating the simulation framework, and developing and validating IMCOM.  
L.Y.~contributed to transient modelling and validation. 
T.Z.~contributed to the GalSim Roman PSF model and its validation.

\section*{Data Availability}

The data underlying this article are available through the NASA/IPAC Infrared Science Archive (IRSA) at \url{https://irsa.ipac.caltech.edu/data/theory/openuniverse2024/overview.html}.\footnote{\url{https://doi.org/10.26131/IRSA596}} The transient models are also available through Zenodo at \url{https://zenodo.org/records/14749318}.


\phantomsection
\addcontentsline{toc}{section}{References}

\bibliographystyle{aasjournal}
\bibliography{bibliography}

\appendix

\section{Data Products and Access at IPAC}
\label{sec:access}

A 10TB data preview\footnote{\url{https://doi.org/10.26131/IRSA569}} of the OpenUniverse2024 simulations was made publicly available by the NASA/IPAC Infrared Science Archive (IRSA; irsa.ipac.caltech.edu) in summer 2024. The full approx.~400TB simulated survey data is being made available with this paper. The data is hosted both on-premises at IPAC and in the cloud via Amazon Web Services (AWS) through the Open Data Repository program. There is a persistent landing page for all data access at IPAC.\footnote{\url{https://irsa.ipac.caltech.edu/data/theory/openuniverse2024/overview.html}}. This includes links and scripts for downloading the data from on-premises, as well as tutorials for accessing the data from Python. In the sections below, we provide additional details about the individual data products.

\subsection{Input SkyCatalog truth files}

There are 7 classes of catalogs that make up the SkyCatalog truth information.\footnote{\url{https://lsstdesc.org/skyCatalogs/roman-rubin-1.1.2/diffsky_galaxy.html}}\footnote{\url{https://lsstdesc.org/skyCatalogs/roman-rubin-1.1.2/UW_stars.html}}\footnote{\url{https://lsstdesc.org/skyCatalogs/roman-rubin-1.1.2/snana.html}}
In each filename listed below, "\textsc{<hpixid>}" refers to the nside 32 healpixel\footnote{\url{https://healpix.sourceforge.io}} area of the sky the file corresponds to. They can be interacted with directly, or by using the SkyCatalog interface by referencing the provided \textsc{skycatalog.yaml} file.
\begin{itemize}
\item \textsc{galaxy\_<hpixid>.parquet} -- Table of galaxy properties in Parquet format, described in Table \ref{tab:galaxy}.
\item \textsc{galaxy\_flux\_<hpixid>.parquet} -- Table of galaxy fluxes in relevant Roman and Rubin bandpasses, in Parquet format, described in Table \ref{tab:galaxy_flux}.
\item \textsc{galaxy\_sed\_<hpixid>.hdf5} --   Low-resolution galaxy spectral energy distributions (SEDs) in HDF5 format. The wavelength grid is contained under the 'meta' group. The 'galaxy' group contains a tree of object IDs split into subgroups of 10k objects, each containing the SED for each component of the object (bulge, disk, and star-forming knots). The original resolution SED for each object component can be obtained via the routines in the \textsc{lsstdesc-diffsky} package.\footnote{\url{https://github.com/LSSTDESC/lsstdesc-diffsky/}} The SEDs in the HDF5 file are intended to be accessed via SkyCatalogs, which has a class to interpret them.
\item \textsc{pointsource\_<hpixid>.parquet} -- Table of Milky Way star properties in Parquet format, described in Table \ref{tab:stars}.
\item \textsc{pointsource\_flux\_<hpixid>.parquet} -- Table of Milky Way star fluxes in Parquet format, similar in format to galaxy Table \ref{tab:galaxy_flux}.
\item \textsc{snana\_<hpixid>.parquet} --  Table of transient properties in Parquet format, described in Table \ref{tab:transients}.
\item \textsc{snana\_<hpixid>.hdf5} -- Table of transient SEDs.  There is a group in the HDF5 file for each transient, named for the corresponding \textsc{id} of the transient from the parquet file.  Each group has the following datasets: \\\noindent 1) \textbf{mjd} holds the times (in MJD) at which the transient flux density is simulated, and is an array of length $n_t$;\\\noindent   2) \textbf{lamba} holds wavelengths in Angstroms, and is an array of length $n_\lambda$;\\\noindent  3) \textbf{flambda} holds flux densities in $\mathrm{erg}\,\mathrm{s}^{-1}\,\mathrm{\Angstrom}^{-1}\,\mathrm{cm}^{-2}$ at each simulated time and each wavelength, and is a 2d array of size $(n_t, n_\lambda)$; and \\\noindent    4) an additional 42 datasets, each arrays of length $n_t$, holding simulated magnitudes and magnitude corrections. These are named \textbf{mag\_\textit{j}}, \textbf{synmag\_\textit{j}}, and \textbf{magcor\_\textit{j}}, where \textbf{\textit{j}} for each of $R$, $Z$, $Y$, $J$, $H$, $F$, $K$, $W$, $u$, $g$, $r$, $i$, $z$, $y$ Roman and Rubin bandpasses. The capital letters correspond to Roman filters, and the lower-case letters correspond to Rubin filters.  The purpose of these columns is to provide a correction for the integrated flux from the low-resolution spectra in this file (see Sec.~\ref{magcor} for more details).

\end{itemize}

\begin{table*}
\begin{center}
\caption{\label{tab:galaxy} Contents of the SkyCatalog galaxy parquet tables. \textsc{<component>} represents the spin-2 component of ellipticity. }

\begin{tabular}{lccl}
\hline
\hline
Column name & Datatype & Units & Description \\ [0.2cm]
\hline
\textsc{galaxy\_id} & int64 & -- & Unique galaxy object id.\\ 
\textsc{ra} & float64 & Degrees & Right Ascension.\\ 
\textsc{dec} & float64 & Degrees & Declination.\\ 
\textsc{redshift} & float64 & -- & Object redshift (inc. peculiar velocity). \\ 
\textsc{redshiftHubble} & float64 & -- & Object redshift (w/o peculiar velocity). \\ 
\textsc{peculiarVelocity} & float64 & km/sec & Object peculiar velocity.\\ 
\textsc{shear<component>} & float64 & -- & Shear ($\gamma$) experienced by object.\\ 
\textsc{convergence} & float64 & -- & Convergence ($\kappa$) experienced by object.\\ 
\textsc{spheroidHalfLightRadiusArcsec} & float32 & arcsec & Bulge component half-light radius.\\ 
\textsc{diskHalfLightRadiusArcsec} & float32 & arcsec & Disk component half-light radius.\\ 
\textsc{diskEllipticity<component>} & float64 & -- & Intrinsic ellipticity of the disk component. \\ 
\textsc{spheroidEllipticity<component>} & float64 & -- & Intrinsic ellipticity of the bulge component.\\ 
\textsc{um\_source\_galaxy\_obs\_sm} & float32 & $M_{\sun}$ ($h=0.7$) & Stellar mass\\ 
\textsc{MW\_rv} & float32 & -- & Milky Way dust model $R_v = 3.1$.\\ 
\textsc{MW\_av} & float32 & -- & Milky Way dust model $A_v=R_v E(B-V)$.\\ 

\hline
\end{tabular}
\end{center}
\end{table*}

\begin{table*}
\begin{center}
\caption{\label{tab:galaxy_flux} Contents of the SkyCatalog galaxy flux parquet tables. Bandpasses (\textsc{<band>}) for LSST include u, g, r, i, z, and y. Bandpasses for Roman include R062, Z087, Y106, J129, H158, F184, K213, and W146.}
\begin{tabular}{lccl}
\hline
\hline
Column name & Datatype & Units & Description \\ [0.2cm]
\hline
\textsc{galaxy\_id} & int64 & -- & Unique galaxy object id. \\ 
\textsc{lsst\_flux\_<band>} & float32 & photons/sec/cm$^2$ & Total flux of composite object through bandpass.\\ 
\textsc{roman\_flux\_<band>} & float32 & photons/sec/cm$^2$ & Total flux of composite object through bandpass.\\ 
\hline
\end{tabular}
\end{center}
\end{table*}

\begin{table*}
\caption{\label{tab:stars} Contents of the SkyCatalog point source parquet tables. }

\begin{tabular}{lccl}
\hline
\hline
Column name & Datatype & Units & Description \\ [0.2cm]
\hline
\textsc{object\_type} & string & -- & Type of point source. Only stars in this catalog.\\ 
\textsc{id} & string & -- & Unique object id. \\ 
\textsc{ra} & float64 & Degrees & Right Ascension.\\ 
\textsc{dec} & float64 & Degrees & Declination.\\ 
\textsc{host\_galaxy\_id} & int64 & -- & Not applicable in this simulation.\\ 
\textsc{magnorm} & float64 & -- & Not applicable in this simulation.\\ 
\textsc{sed\_filepath} & string & -- & Path to file containing SED for object.\\ 
\textsc{MW\_rv} & float32 & -- & Milky Way dust model $R_v = 3.1$.\\ 
\textsc{MW\_av} & float32 & -- & Milky Way dust model $A_v=R_v E(B-V)$.\\ 
\textsc{mura} & float64 & milliarcsec/year & Not applicable in this simulation.\\ 
\textsc{mudec} & float64 & milliarcsec/year & Not applicable in this simulation. \\ 
\textsc{radial\_velocity} & float64 & km/s & Radial velocity \\ 
\textsc{parallax} & float64 & milliarcsec & Parallax\\ 
\textsc{variability\_model} & string & -- & Not applicable in this simulation.\\ 
\textsc{salt2\_params} & string & -- & Not applicable in this simulation.\\ 
\hline
\end{tabular}
\end{table*}

\begin{table*}
\caption{\label{tab:SNe} Contents of the SkyCatalog SNe parquet tables. }

\begin{tabular}{lccl}
\hline
\hline
Column name & Datatype & Units & Description \\ [0.2cm]
\hline
\textsc{id} & int64 & -- & Unique object id. \\ 
\textsc{ra} & float64 & Degrees & Right Ascension.\\ 
\textsc{dec} & float64 & Degrees & Declination.\\ 
\textsc{host\_id} & int64 & -- & Host galaxy id. \\ 
\textsc{gentype} & int16 & -- & Type of object simulated.\\ 
\textsc{model\_name} & object & -- & Name of object type.\\ 
\textsc{start\_mjd} & float32 & days & Earliest visibility.\\ 
\textsc{end\_mjd} & float32 & days & Latest visibility.\\ 
\textsc{z\_CMB} & float32 & -- & Redshift.\\ 
\textsc{mw\_EBV} & float32 & -- & Milky Way extinction E(B-V).\\ 
\textsc{mw\_extinction\_applied} & bool & -- & Always False in this simulation.\\ 
\textsc{AV} & float32 & -- & Host galaxy dust model (random in this simulation).\\ 
\textsc{RV} & float32 & -- & Host galaxy dust model.\\ 
\textsc{v\_pec} & float32 & km/s & Peculiar velocity\\ 
\textsc{host\_ra} & float64 & Degrees & Host galaxy RA.\\ 
\textsc{host\_dec} & float64 & Degrees & Host galaxy Dec.\\ 
\textsc{host\_mag\_g} & float32 & -- & Host magnitude.\\ 
\textsc{host\_mag\_i} & float32 & -- & Host magnitude.\\ 
\textsc{host\_mag\_F} & float32 & -- & Host magnitude.\\ 
\textsc{host\_sn\_sep} & float32 & arcsec & Separation from host.\\ 
\textsc{peak\_mjd} & float32 & days & Time of peak flux.\\ 
\textsc{peak\_mag\_g} & float32 & -- & Transient peak magnitude.\\ 
\textsc{peak\_mag\_i} & float32 & -- & Transient peak magnitude.\\ 
\textsc{peak\_mag\_F} & float32 & -- & Transient peak magnitude.\\ 
\textsc{lens\_dmu} & float32 & -- & Distance modulus shift from lensing.\\ 
\textsc{lens\_dmu\_applied} & bool & -- & Always False in this simulation.\\ 
\textsc{model\_param\_names} & object & -- & Model parameter names.\\ 
\textsc{model\_param\_values} & object & -- & Model parameter values.\\ 
\textsc{MW\_av} & float32 & -- &Milky Way dust model. \\ 
\textsc{MW\_rv} & float32 & -- &Milky Way dust model. \\ 

\hline
\end{tabular}
\end{table*}

\subsection{Simulated Roman survey data} 

Both the Roman Wide-Area Survey (WAS) and Time-Domain Survey (TDS) follow the same data structure, and include the following classes of files. Image and per-SCA truth files for each survey are organized in a directory tree first by bandpass, then observing sequence or pointing id, then individual files for each SCAs of the pointing that were simulated.

\begin{itemize}
\item Per-SCA truth files, which contain information on the position of the object centroid in image coordinates, the type of object, and the as-drawn, composite object flux values. Files have naming convention: \textsc{roman\_<survey>\_index\_<bpass>\_<pointing>\_<sca>.txt}.
\item Simulated "True" images include the appropriate bandpass and PSF/charge diffusion model, but otherwise no sources of noise, backgrounds, or other non-idealities of the detectors, except for object poisson noise. Files have naming convention: \textsc{roman\_<survey>\_truth\_<bpass>\_<pointing>\_<sca>.fits.gz}.
\item Simulated "Calibrated" images include relevant backgrounds and major sources of noise, but otherwise lack detector non-idealities that would prevent treating the images as final calibrated products. Files have naming convention: \textsc{roman\_<survey>\_simple\_model\_<bpass>\_<pointing>\_<sca>.fits.gz}.
\item In the Roman WAS, coadds over a small portion of the survey are simulated using \textsc{imcom}. More information on the \textsc{imcom} coadds can be found in Sec.~\ref{sec:processing}. The subdirectory structure corresponds to sub-images (blocks) as described in Fig. 4 of \cite{2024MNRAS.528.2533H}.
\item Simulation metadata including the driver configuration file for the image simulation (\textsc{<survey>.yaml}), the observation sequence information for each pointing in the simulated survey (\textsc{roman\_<survey>\_obseq\_*.fits}), and tables of RA and Dec for each SCA center for each pointing (\textsc{roman\_<survey>\_obseq\_*\_radec.fits}). The observation sequence table is described in Table \ref{tab:romanobseq}.
\end{itemize}

\begin{table*}
\begin{center}
\caption{\label{tab:romanobseq}Description of the Roman observation sequence table data.}
\begin{tabular}{lccl}
\hline
\hline
Column name & Datatype & Units & Description \\ [0.2cm]
\hline
\textsc{date} & float64 & Days & Observation timestamp (MJD). \\ 
\textsc{exptime} & float64 & s & Exposure time.\\ 
\textsc{ra} & float64 & Degrees & Right ascension (J2000) of WFI field center.\\ 
\textsc{dec} & float64 & Degrees & Declination (J2000) of WFI field center.\\ 
\textsc{pa} & float64 & Degrees & Position angle (J2000) of the Roman WFI +Y direction.\\ 
\textsc{filter} & char[4] & - & Filter name. \\
\hline
\end{tabular}
\end{center}
\end{table*}

\subsection{Simulated Rubin survey data}

There are 8 kinds of simulated data products in this distribution. Aside from the simulated input images, i.e., the raw data files, these simulated data products are all produced by the Rubin Science Pipelines code and are intended to be accessed via the Rubin Data Butler \citep{2022SPIE12189E..11J}.  For more detailed descriptions of the Rubin code and their outputs, see \cite{2018PASJ...70S...5B,2019ASPC..523..521B}.\footnote{\url{https://pipelines.lsst.io/index.html}} 

Each type of simulated Rubin data is described below.

\begin{itemize}
\item LSSTCam raw exposures consist of the raw, pixel data, one file per exposure per CCD, simulated as if produced by LSSTCam observations.

\item Spatial SkyMap\footnote{\url{https://github.com/lsst/skymap}} partitioning used for generating the coadded images.

\item Files containing the reference catalog stars that are used for calibrating the simulated LSSTCam images.

\item Calibrated exposures include detrended, background-subtracted image, and mask and variance plane data, the PSF model for that image, WCS, and zero-point information. 

\item The catalog of detected sources in each calexp and their measured properties.

\item Calibrated coadded image dataset with the same info as the per-exposure calexps.

\item  FITS image containing the number of exposures contributing to each pixel in the coadd.

\item A catalog of forced photometry measurements for the objects identified in the multiband coadds.
\end{itemize}

\bsp
\label{lastpage}
\end{document}